\documentclass[a4paper,12pt]{article}
\usepackage{latexsym,amssymb}
\usepackage{cite}
\usepackage[arrow,matrix,curve]{xy}
\usepackage{amsmath}

\usepackage{mathrsfs}

\usepackage{graphicx}
\usepackage{showidx}
\usepackage{cite}

\usepackage{amssymb}
\usepackage{amscd}
\usepackage{amsfonts}
\usepackage{amsthm}

\numberwithin{equation}{section}

\DeclareMathAlphabet{\mathpzc}{OT1}{pzc}{m}{it}

\global\arraycolsep=2pt 
\def\s[#1,#2]{[#1\stackrel{{\displaystyle\star}}{,}#2]}

 1

\newcommand{\eq}{\begin{equation}}
\newcommand{\eqa}{\begin{eqnarray}}
\newcommand{\en}{\end{equation}}
\newcommand{\ena}{\end{eqnarray}}
\newcommand{\enn}{\nonumber \end{equation}}

\def\sk{\vskip .4cm}
\def\noi{\noindent}

\def\de{\delta}
\def\epsi{{\varepsilon}}
\def\st {\star}

\def\Fh{\widehat F}

\def\D/h{\widehat{\fmslash D}}

\def\Tr{{\rm{Tr}}}

\def\th{\theta}
\def\om{\omega}
\def\upomega{{{\omega}}}
\def\Om{\Omega}
\def\al{\alpha}
\def\la{\lambda}
\def\be{\beta}
\def\ga{\gamma}
\def\Ga{\Gamma}

\def\de{\delta}

\def\5bar{{\overline 5}}

\def\sma#1{\mbox{\footnotesize #1}}

\def\RR{{\mathcal R}}

\def\FF{\mathcal F}
\def\GG{\mathcal G}

\def\s'O{\stackrel{_{{\displaystyle\st \footnotesize '}}}{_{^{^{\displaystyle\otimes}}}}}

\def\La{\Lambda }

\def\HH{{\mathcal H}}
\def\LL{{\mathcal L}}
\def\CL{{\mathcal L}}
\def\ll{{\mathcal L}}
\def\D{\Delta}
\def\1s{{1_\st }}
\def\3s{{3_\st }}
\def\2s{{2_\st }}
\def\ef1{{1_\FF}}
\def\ef2{{3_\FF}}
\def\ef3{{2_\FF}}

\def\hbar{\lambda}

\def\le{\langle}
\def\re{\rangle}

\def\AA{\mathcal A}

\def\F{{\mathfrak{F}}}
\def\th{\mathsf{\Theta}}

\def\cc{\mathbb{C}}
\newcommand\rr{\mathbb{R}}
\def\zz{\mathbb{Z}}

\def\lg{{\rm log}}

\newcommand{\alh}{\widehat\al}
\newcommand{\beh}{\widehat\be}

\newcommand{\asf}{\mathsf{a}}
\newcommand{\bsf}{\mathsf{b}}
\newcommand{\csf}{\mathsf{c}}
\newcommand{\dsf}{\mathsf{d}}

\newcommand{\Asf}{\mathsf{A}}
\newcommand{\Bsf}{\mathsf{B}}
\newcommand{\Csf}{\mathsf{C}}
\newcommand{\Dsf}{\mathsf{D}}

\newcommand{\K}{K\"ahler }
\newcommand{\KK}{{\cal K}}
\newcommand{\VV}{{V}}
\newcommand{\VVV}{{\cal V}}
\newcommand{\del}{\partial}
\newcommand{\pa}{\partial}
\newcommand{\eqn}[1]{(\ref{#1})}
\newcommand{\rref}[1]{(\ref{#1})}
\newcommand{\nn}{\nonumber}
\newcommand{\Omm}{\Omega}
\newcommand{\N}{{\cal N}}
\newcommand{\cN}{{\cal N}}
\newcommand{\M}{{\cal M}}
\newcommand{\cM}{{\cal M}}

\newcommand{\A}{{\cal A}}
\newcommand{\im}{{\rm Im}\,}
\newcommand{\rea}{{\rm Re}\,}
\newcommand{\ns}{{{\mbox{\tiny{$\N$}}}}}
\newcommand{\Ps}{{{\mbox{\small{$\cal P$}}}}}
\newcommand{\nt}{{{\mbox{\tiny{$\N$}}}}}
\newcommand{\rmi}{i}

\newcommand{\matc}{\begin{array}{c}}
\newcommand{\matcc}{\begin{array}{cc}}
\newcommand{\matccc}{\begin{array}{ccc}}
\newcommand{\matcccc}{\begin{array}{cccc}}
\newcommand{\emat}{\end{array}}

\newcommand{\VBH}{{\mathscr{V}}_{\!BH}}
\newcommand{\SBH}{{\mathscr{S}}}
\newcommand{\HHH}{{H}}
\newcommand{\LLL}{{\mathfrak{L}}}
\newcommand{\Bv}{\boldsymbol{B}}
\newcommand{\Hv}{\boldsymbol{H}}
\newcommand{\Dv}{\boldsymbol{D}}
\newcommand{\Ev}{\boldsymbol{E}}
\newcommand{\ppi}{{\varphi}}
\newcommand{\ddel}{{\Delta}}

\newcommand{\pn}{n'}
\newcommand{\nm}{n}

\newcommand{\beqn}{\eq}
\newcommand{\eeqn}{\en}
\newcommand{\Lag}{\LL}
\newcommand{\ap}{{\alpha'}}

\newcommand{\CF}{{\cal F}}
\newcommand{\CK}{{\cal G}}
\newcommand{\CLh}{{\widehat{\CL}}}

\begin{document}


\begin{titlepage}

\hfill {DISTA-UPO/08}

{\vspace{-2.8em}}

\hfill {CERN-PH-TH/2008-144}

\sk\sk

\hfill {LBNL-69756}

\begin{center}{\bf{\large {Duality Rotations in Nonlinear Electrodynamics and 
in\\[.2em] Extended Supergravity}}}

\sk
{\bf Paolo Aschieri$^{1,2,3}$, Sergio Ferrara$^{4,5,6}$ and Bruno Zumino$^{7,8}$}
\sk

{\it $^{1}$Centro Studi e Ricerche ``Enrico Fermi'' Compendio Viminale, 00184 Roma, Italy}\\
  {\it $^2$Dipartimento di Scienze e Tecnologie
 Avanzate, Universit\`{a} del
 Piemonte Orientale,}\\  {\it $^3$INFN, Sezione di Torino, gruppo collegato di Alessandria }\\
{\it Via Bellini 25/G, 15100 Alessandria, Italy}\\
{\small{\texttt{aschieri@to.infn.it}}}\\[.5em]

        {\it $^4$Physics Department,Theory Unit, CERN, 
        CH 1211, Geneva 23, Switzerland}\\
        {\it $^5$INFN - Laboratori Nazionali di Frascati, 
        Via Enrico Fermi 40,I-00044 Frascati, Italy}\\
{\it $^6$Department of Physics and Astronomy, University of California,\\ 
Los Angeles, CA USA}\\
{\small\texttt{sergio.ferrara@cern.ch}}\\[.5em]

{\it $^7$Department of Physics, University of California 
\\ $^8$Theoretical Physics Group, Bldg. 50A5104,
Lawrence Berkeley National Laboratory \\ Berkeley,
CA 94720 USA\\}
{\small\texttt{bzumino@lbl.gov}}\\[.5em]
\sk\sk\sk

{ Invited contribution to Rivista del Nuovo Cimento in occasion of the\\[.15em] 2005 Enrico Fermi Prize of the Italian Physical Society}
\sk\sk\sk
{\small{ DEDICATED TO THE MEMORY OF JULIUS WESS}}
\sk\sk
\begin{abstract} 
 We review the general theory of duality rotations which, in four dimensions,
exchange electric with magnetic fields.
Necessary and sufficient conditions in order for a theory 
to have duality symmetry are established.
A nontrivial example is Born-Infeld theory with $n$ abelian 
gauge fields and with $Sp(2n,\rr)$ self-duality.
We then review duality symmetry in supergravity theories. In the case of $N=2$
supergravity duality rotations are in general not a symmetry of the theory but 
a key ingredient in order to formulate the theory itself. This is due to the 
beautiful relation between the geometry of special K\"ahler manifolds and 
duality rotations.
\end{abstract}
\sk\sk\sk\sk

\setcounter{page}{0}

\end{center}
\end{titlepage}

\renewcommand{\thepage}{\arabic{page}}

\def\thefootnote{\arabic{footnote}} \setcounter{footnote}{0}

\tableofcontents

\section{Introduction}
It has long been known that the free Maxwell's equations are invariant under the rotation of the electric field into the magnetic fields; this is also the case if electric and magnetic charges are present. In 1935, Schr\"odinger \cite{Sch}  showed that the nonlinear electrodynamics of Born and Infeld 
\cite{BI}, (then proposed as a new fundamental theory of the electromagnetic field and presently relevant in describing the low energy
effective action of $D$-branes in open string theory), 
has also, quite remarkably, this property. 
Extended supergravity theories too, as first pointed out in  
\cite{FSZ77, crju}
exhibit electric-magnetic duality symmetry.  Duality symmetry 
thus encompasses photons self-interactions, gravity interactions 
and couplings to spinors (of the magnetic moment type, not minimal couplings).

Shortly after \cite{FSZ77, csf, crju} the general theory of duality invariance with abelian gauge fields  coupled to fermionic and bosonic matter was developped 
in \cite{GZ, BZ}. 
Since then the duality symmetry of extended supergravity theories has been 
extensively investigated \cite{bibbia, cdfv, solv, adf96}, 
and examples of Born-Infeld type 
lagrangians with electric-magnetic duality have been presented, in the case
of one abelian gauge field \cite{GR1,GR2,GZ1,GZ2,araki} 
and in the case of many abelian gauge fields \cite{BMZ, 
ABMZ, ABMZ2, Ivanov}.
Their supersymmetric generalizations have been considered in 
\cite{APT, RT} and with different scalar couplings and noncompact 
duality group in \cite{BMZ, ABMZ, Kuzenko:2000tg, Kuzenko:2000uh, 
Kuzenko:2002vk}. 

We also mention that duality symmetry can be generalized 
to arbitrary even dimensions by using antisymmetric tensor 
fields such that the rank of their field strengths 
equals half the dimension of space-time, see \cite{tanii, Ferrara}, and 
\cite{fresissa, adf96, Julia, Julia2, araki, ABMZ, Kuzenko:2000uh, 
Kuzenko:2002vk}. 

\sk
We provide a rigorous formulation of the general theory of four-dimensional electric-magnetc duality in lagrangian field theories where many abelian vector fields are coupled to scalars, fermions and to gravity. When the scalar fields lagrangian is described by a non-linear sigma model with a symmetric space $G/H$ where $G$ is noncompact and $H$ is its maximal compact subgroup, the coupling of the
scalars with the vector fields is uniquely determined by a symplectic representation of $G$ (i.e. where the representation space is equipped with an invariant 
antisymmetric product). 
Moreover fermions coupled to the sigma model, which lie in representations of $H$, must also be coupled to vectors through particular Pauli terms as implied by electric-magnetic duality.

This formalism is realized in an elegant way in extended supergravity theories in four dimensions and can be generalized to dyons \cite{schwinger} in $D$-dimensions, which exist when $D$ is even and the dyon is a $p$-brane with $p=D/2-2$. In the context 
of superstring theory or $M$ theory electric-magnetic dualities can arise from many sources, namely $S$-duality, $T$-duality or a combination thereof called $U$-duality \cite{huto}. From the point of view of a four dimensional observer such dualities manifest as some global symmetries of the lowest order Euler-Lagrange equations of the underlying four dimensional effective theory. 

The study of the relations between the symmetries of higher dimensional 
theories  and their realization in four dimension is rich and fruitful, 
and duality rotations are an essential ingredient.
Seemingly different lagrangians with different elementary 
dynamical fields can be shown to describe equivalent equation 
of motions by using duality.
An interesting example is provided by 
the $N=8$, $D=4$ supergravity lagrangian  
whose duality group is $G=E_{7,(7)}$, this is the formulation of Cremmer and 
Julia \cite{crju}.
An alternative formulation obtained from dimensional reduction of the $D=5$ 
supergravity, exhibits an action that is invariant under a different group 
of symmetries.
These two theories can be related only after a proper duality rotation of 
electric and magentic fields which involves a suitable Legendre 
transformation (a duality rotation that is not a symmetry transformation).

Let us also recall that duality rotation symmetries can be further enhanced to 
local symmetries  (gauging of duality groups). The corresponding 
gauged supergravities appear as string compactifications in the presence 
of fluxes and as generalized compactifications of (ungauged) higher 
dimensional supergravities.

As a main example consider again 
the $N=8$, $D=4$ supergravity lagrangian  of Cremmer and Julia, it is invariant
 under $SO(8)$ (compact subgroup of $E_{7,(7)}$). The gauging of $SO(8)$  
corresponds to the gauged $N=8$ supergravity of 
De Witt and  Nicolai \cite{dwni}. As shown in \cite{Andrianopoli:2002mf} the 
gauging of a different subgroup, that is the natural choice in the equivalent 
formulation of the theory obtained from dimensional reduction of $D=5$ 
supergravity,
corresponds to the gauging of a flat group in the sense of 
Scherk and Schwarz dimensional reduction \cite{Scherk:1979zr},
and gives the massive deformation of the $N=8$ supergravity as obtained by 
Cremmer, Scherk and Schwarz \cite{Cremmer:1979uq}.

\sk
Electric-Magnetic duality is also the underlying symmetry  which encompasses 
the physics of extremal black holes and of the ``attractor mechanism''
\cite{fekast, feka, strom3}, for recent reviews on the attractor mechanism 
see \cite{Bellucci:2007ds, Andrianopoli:2006ub, Ferrara:2008hw}.
Here the Bekenstein-Hawking entropy-area formula
\[
S={1\over 4} A
\]
is directly derived by the  evaluation of a certain black hole potential $\VBH$
at its attractive critical points \cite{fegika}
\[
S=\pi \,\VBH\big|_C
\]
where the critical points $C$ satisfy $\partial\VBH|_C=0$. The potential $\VBH$
is a quadratic invariant of the duality group; it depends on both 
the matter and the gauge fields configuration.  
In all extended supersymmetries with $N>2$, the entropy $S$ can also be
computed via a certain duality invariant combination of the magnetic and electric charges $p,q$ of the fields configuration \cite{kako,FM}
\[
S=\pi \SBH(p,q)~.
\]

\sk\sk
In the remaining part of this introduction  we present the structure of 
the paper by
summarizing its different sections.
\sk
In Section 2 we give a pedagogical introduction to $U(1)$ duality
rotations in nonlinear theories of electromagnetism.   The basic
aspects of duality symmetry are already present in this simple case with just one abelian gauge field:
the hamiltonian is invariant  (duality rotations are canonical 
transformations that commute with the hamiltonian); the lagrangian 
is not invariant but must transform in a well defined way.
 The Born-Infeld theory is the main example of duality invariant nonlinear 
theory.

\sk
In Section 3 the general theory is formulated with many abelian 
gauge fields interacting with bosonic and fermionic matter.
Necessary and for the first time sufficient conditions in order 
for a theory to have duality symmetry are established. 
The maximal symmetry group in a theory with $n$ abelian gauge 
fields includes $Sp(2n, \rr)$. If there are no scalar fields the 
maximal symmetry group is $U(n)$. The geometry of the symmetry 
transformations on the scalar fields is that of the coset space 
$Sp(2n,\rr)/U (n)$ that we study in detail.  
The kinetic term for the scalar fields is constructed by using this coset 
space geometry. 
In Subsection 3.6 we present the Born-Infeld lagrangian with $n$ abelian gauge 
fields and $Sp(2n,\rr)$ duality symmetry \cite{ABMZ}. The self-duality of this 
lagrangian is proven by studying another example:
the Born-Infeld lagrangian with $n$ complex gauge fields
and $U(n,n)$ duality symmetry. Here $U(n,n)$ is the group 
of holomorphic duality rotations.
 We briefly develop the theory 
of holomorphic duality rotations.

The Born-Infeld lagrangian with  $U(n,n)$ self-duality is per se interesting,
the scalar fields span the coset space $U(n,n)\over U(n)\times U(n)_{}$,
in the case $n=3$ this is the coset space of the scalars of $N=3$
supergravity with$^{}$ $3^{}$ vector multiplets. 
This Born-Infeld lagrangian is then a natural candidate for 
the nonlinear generalization of $N=3$ supergravity.

We close this sections by presenting, in a formulation with auxiliary fields,
the supersymmetric version of this Born-Infeld Lagrangian \cite{BMZ, ABMZ}. 
 We also present the form without 
auxiliary fields of the supersymmetric Born-Infeld Lagrangian with a single 
gauge field and a scalar field; this theory is invariant under $SL(2, \rr)$ 
duality, which reduces to $U(1)$ duality if the value of the scalar field is 
suitably fixed.  Versions of this theory without the scalar field were 
presented in \cite{DP, CF, BG}.

\sk
In Section 4 we apply the general theory of duality rotation to 
supergravity theories with $N>2$ supersymmetries. In these supersymmetric 
theories the duality group is always a subgroup $G$ of $Sp(2n,\rr)$, where
$G$ is the isometry group of the sigma model $G/H$ of the scalar fields.
Much of the geometry underlying these theories is in the (local) 
embedding of $G$ 
in $Sp(2n,\rr)$. 
The supersymmetry transformation rules, the structure of the central and matter charges and the duality invariants associated to the entropy and the potential
of extremal black holes configurations are all expressed in terms of the
embedding of $G$ in $Sp(2n,\rr)$ \cite{adf96}. We thus present a unifying formalims.
We also explicitly construct the symplectic bundles (vector 
bundles with a symplectic product on the fibers) associated to these theories,
and prove that they are topologically trivial; this is no more the case
for generic $N=2$ supergravities. 

\sk
In Section 5 we introduce special \K geometry as studied in
differential geometry, we follow in particular the work of Freed \cite{Freed}, see also \cite{LledoV} (and \cite{Cortes}) 
and then develop the mathematical definition
up to the construction of those explicit flat symplectic sections used in 
$N=2$ supergravity. We thus see for example that the flat 
symplectic bundle of a rigid special \K manifold $M$ is just the tangent bundle $TM$ with symplectic product given by the \K form. A similar construction applies in the case of local
special geometry (there the flat tangent bundle is
not of the \K manifold $M$ but is essentially the tangent bundle 
of a complex line bundle $L\rightarrow M$). 
This clarifies the global aspects of special geometry
and the key role played by duality rotations in the formulation of $N=2$ supergravity with scalar fields taking value in the target space $M$. Duality rotations
are needed for the theory to be globally well defined.


\sk
In Section 6 duality rotations in nonlinear electromagnetism are considered 
on a noncommutative spacetime, $[x^\mu,x^\nu]=i\th^{\mu\nu}$. 
The noncommutativity tensor $\th^{\mu\nu}$
must be light-like. 
A nontrivial example of nonlinear 
electrodynamics on commutative spacetime is presented and 
using Seiberg-Witten map between commutative and 
noncommutative gauge theories noncommutative $U(1)$ Yang Mills theory
is shown to have duality symmetry. This theory formally is nonabelian,
$\widehat F_{\mu\nu}=\partial_\mu \widehat{A}_\mu-\partial_\nu \widehat{A}_\mu- 
i[\widehat{A}_\mu,\widehat{A}_\nu]$,
its self-duality is in this respect remarkable. One can also enhance the duality group to $Sp(2,\rr)$ and couple 
this noncommutative theory  to axion,
dilaton and Higgs fields, these latter via minimal couplings.
Duality in noncommutative spacetime allows to relate 
space-noncommutative magnetic monopoles to space-noncommutative 
electric monopoles \cite{Aschieri:2001zj, Aschieri:2001gf}. 

A special kind of noncommutative spacetime is a lattice space
(it can be studied with noncommutative geometry techniques). Duality rotations on a lattice have been studied in \cite{Aschieri:2003}.

\sk
In  Appendix 7
we prove some fundamental properties 
of the symplectic group $Sp(2n,\rr)$ and of the coset space $Sp(2n,\rr)/U(n)$.
We also collect for reference some main formulae and definitions.
\sk 
In Appendix 8 a symmetry property of the trace of a solution of a 
polynomial matrix equation is proven. This allows the explicit formulation of
the Born-Infeld lagrangian with $Sp(2n,\rr)$ duality symmetry presented in  
Section 3.7.

\section{$U(1)$ gauge theory and duality symmetry}
Maxwell theory is the prototype of electric-magnetic 
duality invariant theories. 
In vacuum the equations of
motion are
\eqa
{\pa}_{\mu}
{F}^{\mu\nu}  &=&0~,\nn\\
{\pa}_{\mu}
\tilde{F}^{\mu\nu}&=&0~, 
\ena
where ${\tilde{F}}^{\mu\nu}
\equiv 
\frac{1}{2}\epsilon^{\mu\nu\rho\sigma} F_{\rho\sigma}$.
They are invariant under rotations 
$\big({}^F_{\tilde F} \big)$ 
$\mapsto \big({}^{\cos\al}_{\sin\al} {}^{~-\sin\al}_{~\;\cos\al}\big) \big({}^F_{\tilde F}\big)$, or using vector notation under rotations
$\big({{}^{\Ev}_{\Bv}}\big)$
$\mapsto \big({}^{\cos\al}_{\sin\al} {}^{~-\sin\al}_{~\;\cos\al}\big) \big({{}^{\Ev}_{\Bv}} \big)$.
This rotational symmetry, called duality symmetry, and also duality invariance
or self-duality, is reflected in the invariance of the hamiltonian 
${\cal H}={1\over 2}(\Ev^2+\Bv^2)$,
notice however that the lagrangian ${\cal L}={1\over 2}(\Ev^2-\Bv^2)$ is not
invariant. This symmetry is not an internal symmetry because it rotates 
a tensor into a pseudotensor.

We study this symmetry for more general electromagnetic theories. 
In this section and the next one conditions on the lagrangians of 
(nonlinear) elecromagnetic theories will be found that guarantee the duality symmetry (self-duality) of the equations of motion. 
\sk
The key mathematical point that allows to establish
criteria for self-duality, thus avoiding  the explicit check of the symmetry 
at the level of the equation of motions, is that the equations of motion 
(a system of PDEs) can be conveniently split in a set of equations that 
is of degree 0 (no derivatives on the field strengths $F$), the so-called 
constitutive relations  (see e.g. \eqn{constitutive}, or \eqn{defK}), and 
another set of degree 1 (see e.g. \eqn{max2}, \eqn{max1} or \eqn{max22}, 
\eqn{max11}).
Duality rotations act as an obvious symmetry of the set of equations of degree
1, so all what is left is to check that they act as a symmetry
on the set of equations of degree 0. It is therefore plausible 
that this check can be equivalently formulated as a specific transformation 
property of the lagrangian under duality rotations (and independent from the spacetime dependence $F_{\mu\nu}(x)$ of the fields), indeed both the 
lagrangian and the equations of motions of degree 0 are functions of the field 
strength $F$ and not of its derivatives. 

\subsection{Duality symmetry in nonlinear electromagnetism}
Maxwell equations read
\eqa
\partial_t \Bv=-\nabla\times \Ev~~,~~~\nabla \cdot\Bv=0\label{max2}
\\
\partial_t \Dv=\nabla\times \Hv~~,~~~\nabla \cdot\Dv=0\label{max1}
\ena
they are complemented by the relations between the electric field $\Ev$,
the magnetic field $\Hv$, the electric displacement $\Dv$ and the 
magnetic induction $\Bv$. In vacuum we have  
\eq
\Dv=\Ev~,~~\Hv=\Bv~.\label{constlin}
\en
In a nonlinear theory we still have the equations \eqn{max2}, \eqn{max1}, but 
the relations $\Dv=\Ev,~\Hv=\Bv$ are replaced by the nonlinear constitutive relations
\eq
\Dv=\Dv(\Ev,\Bv)~~,~~~\Hv=\Hv(\Ev,\Bv)~\label{constitutive}
\en
(if we consider a material medium with electric and magnetic properties 
then these equations are the constitutive relations of the material, and
\eqn{max2} and \eqn{max1} are the macroscopic Maxwell equations). 

Equations \eqn{max2}, \eqn{max1}, \eqn{constlin} are invariant under 
the group of general linear transformations  
\eq\label{rotBD}
\left(\begin{array}{c}
\Bv' \\
\Dv'
\end{array}\right)=
\left(\begin{array}{cc}
A & B\\
C & D
\end{array}\right)
\left(\begin{array}{c}
\Bv  \\
\Dv
\end{array}\right)
~~~,~~~~
\left(\begin{array}{c}
\Ev' \\
\Hv'
\end{array}\right)=
\left(\begin{array}{cc}
A & B\\
C & D
\end{array}\right)
\left(\begin{array}{c}
\Ev  \\
\Hv
\end{array}\right)~~.
\en
\sk
\noi We study under which conditions also the nonlinear constitutive 
relations \eqn{constitutive} are invariant. We find constraints on 
the relations 
\eqn{constitutive} as well as  on the transformations 
\eqn{rotBD}. 

We are interested in nonlinear theories that admit a lagrangian formulation 
so that relativistic covariance of the equations \eqn{max2}, \eqn{max1}, 
\eqn{constitutive} and their inner consistency is automatically ensured.
This requirement is fulfilled if the constitutive relations \eqn{constitutive}
are of the form 
\eq\label{const2}
\Dv={{ {\del \LL(\Ev,\Bv)}}\over {\del \Ev}}~~,~~~\Hv=-{{\del { \LL(\Ev,\Bv)}}\over \del \Bv}~~,
\en
where $\LL(\Ev,\Bv)$ is a Poincar\'e invariant function of $\Ev$ and $\Bv$. 
Indeed if we consider $\Ev$ and $\Bv$ depending on a gauge potential $A_\mu$
and vary the lagrangian
 $\LL(\Ev,\Bv)$ with respect to $A_\mu$,
we recover \eqn{max2}, \eqn{max1} and \eqn{const2}. 
This property is most easily shown by using four component notation. 
We group the constitutive relations \eqn{const2} in the constitutive relation\footnote{a practical convention 
is to define ${ {\partial F_{\rho\sigma}}\over{\partial F_{\mu\nu}}  } 
=\de^{\mu}_{\rho}\de^{\nu}_{\sigma} $ rather than 
${{ {\partial F_{\rho\sigma}}\over{\partial F_{\mu\nu}}}}
=\de^{\mu}_{\rho}\de^{\nu}_{\sigma}-
\de^{\nu}_{\rho}\de^{\mu}_{\sigma}~$. This explains the factor $2$ in 
\eqn{defK}.}
\eq
{\tilde G}^{\mu\nu}~=~2{\del \LL(F)\over 
\del F_{\mu\nu}}~;
\label{defK}
\en
we also define  $G_{\mu\nu}=-{1\over 2}\epsilon_{\mu\nu\rho\sigma}{\tilde G^{\rho\sigma}}$, so that  
${{\tilde{G}}}^{\mu\nu}=
\frac{1}{2}\epsilon^{\mu\nu\rho\sigma} G_{\rho\sigma}$ ($\epsilon^{0123}=-\epsilon_{0123}=1$).
If we consider the field strength $F_{\mu\nu}$ as a function of 
a (locally defined) gauge potential $A_\mu$, then equations \eqn{max2} and 
\eqn{max1} are respectively the Bianchi identities for
$F_{\mu\nu}=\partial_\mu A_\nu-\partial_\nu A_\mu$ and the 
equations of motion for $\LL(F(A))$,
\eqa
{\pa}_{\mu}
{\tilde F}^{\mu\nu}  &=&0~,\label{max22}\\
{\pa}_{\mu}
\tilde{G}^{\mu\nu}&=&0~ \label{max11}~.
\ena

In our treatment of duality rotations we study the symmetries of the equations
\eqn{max22}, \eqn{max11} and \eqn{defK}. The lagrangian $\LL(F)$ is always a 
function of the field strength $F$; it is not seen as a function of the 
gauge potential $A_\mu$; accordingly the Bianchi identities for $F$ are considered part of the equations of motions for $F$.

Finally we consider an action $S=\int {\!\cal L} \,d^4x$
with lagrangian density $\LL=\LL(F)$ that depends on $F$
but not on its partial derivatives; it also depends on a
spacetime metric $g_{\mu\nu}$ that we generally omit writing explicitly\footnote{Notice that \eqn{max22}, \eqn{max11} are also 
the equation of motions in the presence of
a nontrivial metric. Indeed $S=\int {\!\cal L} \,d^4x
=\int {\! L} \,\sqrt{g}d^4x$. The equation of motions are
$
\partial_{\mu}(\sqrt{g}\, F^{*\,\mu\nu} )=
{\partial}_{\mu}
\tilde{F}^{\mu\nu}  =0  ~,~
{\partial}_{\mu}(\sqrt{g}\,  G^{*\,\mu\nu} )=
{\partial}_{\mu}
\tilde{G}^{\mu\nu}  =0~,
$
where the Hodge dual of a two form 
$\Omega_{\mu\nu}$ is defined by 
$\Omega^*_{\mu\nu}\equiv \frac{1}{2}\sqrt{g}\,
\epsilon_{\mu\nu\rho\sigma}
\Omega^{\rho\sigma}$
.},
 and on at least one dimensionful constant in order to allow for 
nonlinearity in the constitutive relations \eqn{defK} (i.e. \eqn{constitutive}). 
We set this dimensionful constant to 1.

\sk
The duality rotations \eqn{rotBD} read 
\eq\label{rotFG}
\left(\begin{array}{c}
F' \\
G'
\end{array}\right)=
\left(\begin{array}{cc}
A & B\\
C & D
\end{array}\right)
\left(\begin{array}{c}
F  \\
G
\end{array}\right)~.
\en
Since by construction equations \eqn{max22} and \eqn{max11} are invariant 
under \eqn{rotFG}, these duality rotations are a symmetry of
the system of equations \eqn{max22}, \eqn{max11}, \eqn{defK}
(or \eqn{max2}, \eqn{max1}, \eqn{constitutive}),
iff on shell the constitutive relations are invariant in form, i.e., iff
the functional dependence of $\tilde G'$ from $F'$ is the same as 
that of $\tilde G$ from $F$, i.e. iff 
\eq
{{\tilde G}'}{}^{\mu\nu}~=~{2}\frac{\del {\LL(F')}}{\del F'_{\mu\nu}}~,
\label{defK'}
\en
where $F'_{\mu\nu}$ and $G'_{\mu\nu}$ are given in \eqn{rotFG}.
This is the condition that constrains the lagrangian $\LL(F)$ and the 
rotation parameters in \eqn{rotFG}.
This condition has to hold on shell of \eqn{defK}-\eqn{max11}; however
\eqn{defK'} is not a differential equation and therefore has to hold just using 
\eqn{defK}, i.e., off shell of \eqn{max22} and \eqn{max11}
(indeed if it holds for constant field strengths $F$ then it holds for 
any $F$).
\sk
In order to study the duality symmetry condition \eqn{defK'}
let $\big({}^{A~B}_{C~D}\big)=\big({}^{1~0}_{0~1}\big)+\epsilon\big({}^{a~b}_{c~d}\big)+\ldots $, and consider infinitesimal $GL(2,\rr)$ rotations $G\rightarrow G + \epsilon\ddel
G,~F\rightarrow F+ \epsilon\ddel F$,
\eq
 \ddel \left( \matc F \\
G \emat \right)
=
\left( \matcc a & b \\ c & d \emat
\right) \left( \matc F \\ G \emat \right)~,\label{FGtrans}
\en
so that the duality condition reads
 \eq
\tilde{G}^{}+\ddel\tilde{G}^{}= 2{\del \LL(F+\ddel
F)\over \del (F_{}+ \ddel F_{})} \label{constr}~.
\en
The right hand side simplifies to\footnote{here and in the following we 
suppress the spacetime indices
so that for example $F\tilde G=F_{\mu\nu}\tilde G^{\mu\nu}$; notice that
$F\tilde G=\tilde F G$, $\tilde{\tilde F}=-F$,  and $\tilde F\tilde G=-FG$
where $FG=F^{\mu\nu}G_{\mu\nu}$.} 
 \eqa
  {\del \LL(F+\ddel F)\over
\del (F_{}+ \ddel F_{})} &=& {\del
\LL(F+\ddel F)\over \del F_{} } {\del F_{} \over \del
(F_{}+ \ddel F_{})}\nonumber\\[.3em] &=&{\del \LL(F+\ddel F)\over
\del F_{}}-  {}{\del \LL(F)\over \del
F_{} } {\del (\ddel F_{}) \over \del F_{}}\nonumber
\ena
 \noi then, using (\ref{FGtrans}) and (\ref{defK}),
condition (\ref{constr}) reads
 \eq\label{ABCD}
c\tilde{F}^{}+ d {}{}\tilde{G}^{}
=2 {\del (\LL(F+\ddel F)-\LL(F))\over \del F_{}}
-2a{\del \LL(F)\over \del F}
-b
 \tilde{G}^{}  {}{\de
G_{} \over{\del F_{}}} 
_{}~.
\en
In order to further simplify this expression we write
$ 
2\tilde F={\del\over \del F_{}} F_{} {}{}\tilde{F}^{} 
$
and we factorize out the partial derivative $\del \over \del F_{\,}$.
We thus arrive at the equivalent condition 
\eq \LL(F+\ddel F)-\LL(F) -{c\over 4}  
F_{}{}{}\tilde{F}^{}-{b\over 4}  \label{16}
G_{}{}{}\tilde{G}^{}\,=\,(a+d)(\LL(F)-\LL_{F\!=0})~.
\en
The constant term $(a+d)\LL_{F\!=\!0}$,  nonvanishing for example in 
D-brane lagrangians,  is obtained by observing that when $F=0$ also 
$G=0$.

Next use $\LL(F+\ddel
F)-\LL(F)
= {\del \LL(F)\over \del
  F_{}}\ddel F_{}=
{1\over 2} a 
F_{}{}{}\tilde{G}^{}
+
{1\over 2} b 
G_{}{}{}\tilde{G}^{}
$
in order to rewrite expression \eqn{16} as 
\eq {b\over 4} 
G_{}{}{}\tilde{G}^{}-{c\over 4} 
F_{}{}{}\tilde{F}^{}= (a+d)(\LL(F)-\LL_{F\!=0})-{a\over 2}
 F_{}{}{}\tilde{G}^{}\label{eqmpar}~.
\en
If we require the nonlinear lagrangian $\LL(F)$ to reduce to the usual 
Maxwell lagrangian 
in the weak field limit, $F^4 <\!<  F^2$,
i.e.,  $\LL(F)=\LL_{F\!=\!0} -1/4\int \! FF d^4x +O(F^4)$, 
then $\tilde G = -F + O(F^3)$, and we obtain the constraint
(recall that $\tilde{\tilde G}=-G$)
$$b=-c~~~,~~~~a=d~,$$
the duality group can be at most $SO(2)$ rotations times dilatations. 
Condition (\ref{eqmpar}) becomes
\eq {b\over 4}
\Big(G_{}{}{}\tilde{G}^{}+
F_{}{}{}\tilde{F}^{}\Big)\,=\, 2a\Big( \LL(F)-\LL_{F\!=0}-
{1\over 2}F_{}{}{}{\partial \LL\over\partial F}^{}\Big)\label{eqmparloc}~.
\en
The vanishing of the right hand side holds only if 
either $\LL(F)-\LL_{F\!=0}$ is quadratic in $F$ (usual electromagnetism) 
or $a=0$. We are interested in nonlinear theories;
by definition in a nonlinear theory $\LL(F)$ is not quadratic
in $F$. This shows that dilatations alone cannot be a duality symmetry.
If we require the duality group to contain at least $SO(2)$ rotations then 
\eq
G_{}{}{}\tilde{G}^{}+
F_{}{}{}\tilde{F}^{}=0\label{thecondition}~,
\en
and $SO(2)$ is the maximal duality group.
Relation (\ref{eqmparloc}) is nontrivially satisfied iff 
$$a=d=0~,$$ and \eqn{thecondition} hold.
\sk
In conclusion equation \eqn{thecondition} is a 
necessary and sufficient condition for a nonlinear electromagnetic 
theory to be symmetric under $SO(2)$ duality rotations, and $SO(2)\subset GL(2,\rr)$ is the maximal connected Lie group of duality rotations of pure 
nonlinear electromagnetism\footnote{This symmetry cannot even extend 
to $O(2)$ because already in the case of usual electromagnetism the 
finite rotation $\big({}  _{\;0~~1} ^{-1~0}{} \big)$ does not satisfy 
the duality condition \eqn{defK'}. It is instructive to see the obstruction 
at the hamiltonian level.
The hamitonian itself is invariant under $D\rightarrow D$, $B\rightarrow -B$,
but this transformation is not a canonical transformation:
the Poisson bracket \eqn{poisbra} is not invariant.}.
\sk
This conclusion still holds if we consider a
nonlinear lagrangian $\LL(F)$ that in the weak field limit $F^4 <\!<  F^2$  
(up to an overall normalization factor) reduces to the most general linear lagrangian
$$
\LL(F)=\LL_{F\!=\!0} -{1\over 4} FF +{1\over 4}\theta F\tilde F+O(F^4)~.
$$
In this case $G = \tilde F+\theta F + O(F^3)$. We substitute in \eqn{eqmpar}
and obtain the two conditions 
(the coefficients of the scalar 
$F^2$ and of the speudoscalar $F\tilde F$ have to vanish separately) 
\eq
c=-b(1+\theta^2) ~~~,~~~~~d-a=2\theta b~.
\en 
The most general infinitesimal duality transformation is therefore
\eq
\left( \matcc a & b \\ -b(1+\theta^2)\; & \;a+2\theta b \emat
\right) ~=~
\left( \matcc a+\theta b & 0 \\ 0 & a+\theta b \emat
\right) +\Theta \left( \matcc  0 & b \\ -b & 0 \emat
\right) \Theta^{-1}\label{trasfmatgen}
\en
where $\Theta=\left( \matcc  1 & 0 \\ \theta & 1 \emat
\right)$. We have dilatations and $SO(2)$ rotations, they act on the vector
$\left( \matc  F \\ G \emat \right)$ via the conjugate representation 
given by the matrix $\Theta$.
Let's now remove the weak field limit assumtion
 $F^4 <\!<  F^2$. We proceed as before.
>From \eqn{defK'} (or from \eqn{eqmpar}) we immediately obtain 
that dilatations alone are not a duality symmetry of the nonlinear equations of motion.  
Then if $SO(2)$ rotations are a duality symmetry we have that they are the maximal duality symmetry group. This happens if
\eq
G\tilde G + (1+\theta)^2 F\tilde F=2\theta F\tilde G~.\label{condgenggfft}
\en

\sk
Finally we note that the necessary and sufficient conditions 
for $SO(2)$ duality rotations  \eqn{condgenggfft} (or  \eqn{thecondition})  
can be equivalently expressed as invariance of
\eq
\LL(F)-{1\over 4}F\tilde G~.\label{invariance}
\en
Proof:
the variation of  expression \eqn{invariance} under $F\rightarrow F+\Delta F$
is given by $\LL(F+\Delta F)-\LL(F)
-{1\over 4}\Delta F\,\tilde G-{1\over 4}F\Delta\tilde G~.$ Use of \eqn{16}
with $a+d=0$ (no dilatation) shows that this variation vanishes.

%
%

%
%

\subsection{Legendre Transformations}
In the literature on gauge theories of abelian $p$-form potentials, 
the term duality transformation denotes a 
different transformation from the one we have introduced,
a Legendre transformation, that is not a symmetry transformation.
 In this section
we relate these two different notions, see \cite{GZ2} for further 
applications and examples.

Consider a theory of nonlinear electrodynamics ($p=1$)
with lagrangian $\LL(F)$. The equations of motion and
the Bianchi identity for $F$ can be derived from the Lagrangian 
$\LL(F, F_{\rm D})$ defined by
\eq
\LL(F, F_{\rm D}) = \LL(F) -{1\over 2} \, 
F  \tilde{F}_{\rm D}~,
~~~~~~~F_{\rm D}{}^{\mu\nu} = \partial^\mu A_{\rm D}{}^\nu 
- \partial^\nu A_{\rm D}{}^\mu~,\label{LFFD}
\en
where $F$ is now an unconstrained antisymmetric tensor 
field, $A_{\rm D}$ a Lagrange multiplier field  
and $F_{\rm D}$ its electromagnetic field.
[Hint: varying with respect to $A_{\rm D}$ gives
the Bianchi identity for $F$, varying with respect to
$F$ gives $G^{\mu\nu}=F_{\rm D}{}^{\mu\nu}$ 
that is equivalent to the initial equations of motion
$\partial_\mu\tilde G^{\mu\nu}=0$ because
$F_{\rm D}{}^{\mu\nu} = \pa^\mu A_{\rm D}{}^\nu -\pa^\nu A_{\rm D}{}^\mu $ 
(Poincar\'e lemma)].

Given the lagrangian \eqn{LFFD} one can also 
first consider the equation of motion for $F$,
\eq
G(F) ~=~ F_{\rm D}~,\label{mapdual}
\en
that is solved by expressing $F$ as a function of the dual 
field strength, $F = F (F_{\rm D})$. 
Then inserting this solution into $\LL(F, F_{\rm D})$, 
one gets the dual model
\eq
\LL_{\rm D} (F_{\rm D}) 
\equiv  \LL(F(F_{\rm D}))
-{1\over2}F(F_{\rm D})\cdot \tilde F_{\rm D}\,
\label{legendre}~.
\en 
Solutions of the \eqn{legendre} equations of motion are, tothether with 
\eqn{mapdual}, solutions of the 
\eqn{LFFD} equations of motion. Therefore solutions to the \eqn{legendre}
equations of motion are via \eqn{mapdual} in  1-1  correspondence 
with solutions of the $\LL(F)$ equations of motion.
\sk
One can always perform a Legendre transformation and describe the physical 
system with the new dynamical variables $A_{\rm D}$ and the new lagrangian 
$\LL_{\rm D}$ rather than $A$ and $\LL$.

The relation with the duality rotation symmetry (self-duality)
of the previous section is that if the system admits duality 
rotations then the solution  $F_{\rm D}$ of the $\LL_{\rm D}$ equations of motion
is also a solution of the $\LL$ equations of motion, we have a symmetry because 
the dual field  $F_{\rm D}$ is a solution of the 
original system.
This is the case because 
for any solution $\LL$ of the self-duality equation, 
its Legendre transform $\LL_{\rm D}$ satisfies:
\eq
\LL_{\rm D} (F) ~=~\LL(F)~.
\label{L=L}
\en
This follows from considering a finite $SO(2)$ duality rotation 
with angle $\pi/2$. Then $F\rightarrow F'=G(F)=F_{\rm D}$, and 
invariance of 
\eqn{invariance},  i.e.
$
\LL(F')-{1\over 4}F'\tilde G'=
\LL(F)-{1\over 4}F\tilde G~,
$
implies $\LL_{\rm D} (F_{\rm D}) ~=~\LL(F_{\rm D})$, i.e., $\eqn{L=L}$.
\sk

In summary, a Legendre transformation is a duality rotation 
only if the symmetry condition \eqn{defK} is met. If the self-duality condition 
\eqn{defK} does not hold, a Legendre transformation leads to a dual formulation of the theory in terms of a dual Lagrangian $\LL_{\rm D}$, 
not to a symmetry of the theory.

\subsection{Hamiltonian theory}
The symmetric energy monentum tensor of a nonlinear theory of electromagnetism 
(obtained via Belinfante procedure or by varying with respect to the metric) 
is given by\footnote{symmetry of $T^{\mu\nu}$ 
follows immediately by observing that the tensor structure of
$\tilde G^{\mu\nu}$ implies $\tilde G^{\mu\nu}=f_s(F)F^{\mu\nu}+f_p(F)\tilde F^{\mu\nu}$ with scalars $f_s(F)$ and $f_p(F)$  depending on $F$, 
the metric $\eta=diag(-1,1,1,1)$ and the completely antisymmetric 
tensor density $\epsilon_{\mu\nu\rho\sigma}$. (Actually, if the lagrangian is parity even, $f_s$ 
is a scalar function while $f_p$ is a pseudoscalar function).}
\eq\label{enmom}
T^\mu_{~\nu}=\tilde G^{\mu\la} F_{\nu\la} +\del^{\mu}_{~\nu}\LL ~.
\en
The equations of motion \eqn{max11} and \eqn{max22} imply its conservation, 
$\del_\mu T^\mu_{~\nu}=0$. 
Invariance of the energy momentum tensor under duality rotations is easily proven by observing that for a generic antisymmetric tensor $F_{\mu\nu}$ 
\eq
\tilde F^{\mu\la}F_{\nu\la} =-{1\over 4}\del^\mu_{~\la}\tilde F^{\rho\sigma}F_{\rho\sigma}~,
\en
and then by recalling the duality symmetry condition \eqn{thecondition}.
\sk
In particular the hamiltonian 
\eq\label{hamiltonian}
{\HH}= T^{00}=\Dv\!\cdot\!\Ev-\LL
\en
of a theory that has duality rotation symmetry is invariant.
\sk
In the hamiltonian formalism duality rotations are canonical transformations, 
since they leave the hamiltonian invariant they are usual symmetry transformations. 
We briefly describe the hamiltonian formalism of (nonlinear) electromagnetism by 
avoiding to introduce the vector potential $A_\mu$; this is appropriate since 
duality rotations are formulated independently from the notion of vector potential.
Maxwell equations \eqn{max2}, \eqn{max1} and the expression of the hamiltonian suggest to consider $\Bv$ and $\Dv$ as the analogue of canonical coordinates and momenta 
$q$ and $p$, while $\Ev$, that enters the lagrangian togheter with $\Bv$, is
the analogue of $\dot q$. 

Recalling the constitutive relations in the lagrangian form \eqn{const2} 
we obtain that the hamiltonian $\HH=\HH(\Dv,\Bv)$ is just given by the Legendre 
transformation \eqn{hamiltonian}. Moreover $\Hv={\partial\HH\over \del \Bv}$ and 
$\Ev={\partial\HH\over \del \Dv}$. The equations of motion are
\eqa\label{ham22}
\partial_t \Bv&=&-\nabla\times {\delta \HH\over\delta \Dv}~,
\\[.3em]
\partial_t \Dv&=&\nabla\times {\delta \HH\over\delta \Bv}~\label{ham11}~.
\ena
The remaning equations $\nabla \cdot \Bv=0$, $\nabla \cdot \Dv=0$ are 
constraints that imposed at a given time are satisfied at any other time.
The Poisson bracket between two arbitrary functionals $U$, $V$ of the 
canonical variables is 
\eq\label{poisbra}
\{U,V\}=\int
{\del U\over\del \Dv}\!\cdot\! 
\left(\nabla \times {\del V\over\del \Bv}\right)
-
{\del V\over\del \Dv} \!\cdot\!
\left(\nabla \times {\del U\over\del \Bv}\right) ~ d^3r~,
\en
in particular the only nonvanishing parenthesis between 
the canonical variables $\Bv$ and $\Dv$ are 
$\{\Bv^i(r),\Dv^j(r')\}=\epsilon^{ijk}\del_k\del^3(r-r')$. 
The equations of motion \eqn{ham22} and \eqn{ham11} assume then the canonical 
form $\partial_t \Bv=-\{\Bv,\HHH\}~,~~
\partial_t \Dv=\{\Dv,\HHH\}$
where $\HHH=\int \HH \,d^3r $ is the hamiltonian ($\HH$ being the hamiltonian density). We see that
 $\HHH$ as usual is the generator of time evolution.
The consitency and the hidden Poincar\'e invariance of the
present formalism is proven in \cite{birulabook}. 

\sk 
In the canonical formalism the generator of duality rotations  is
the following nonlocal integral \cite{birula}, \cite{DeserT}
\eq
\La={1\over 8\pi}\int \!\int ~{\Dv_1\!\cdot\! (\nabla\times \Dv_2) +\Bv_1\!\cdot\! (\nabla\times \Bv_2) \over  |r_1-r_2|}\, d^3r_1d^3r_2
\en
where the subscripts indicate that the fields are taken at the points $r_1$ and $r_2$. We have $\{\Dv,\La\}=\Bv$ and $\{\Bv,\La\}=-\Dv$~.
\sk 

Finally we remark  that it is straighforward  
to establish duality symmetry in the hamiltonian 
formalism. Indeed there are three independent scalar combinations of the canonical 
fields $\Bv$ and $\Dv$, they can be taken to be: $\Dv^2+\Bv^2$, $\Dv^2-\Bv^2$ and $(\Dv\times \Bv)^2$. The last two scalars are duality invariant and therefore any hamiltonian that depends just on them leads to a 
theory with duality symmetry. The nontrivial problem in this approach in now
to constrain the hamiltonian so that the theory is Lorentz invariant 
\cite{Deser}, \cite{birula}.  The condition is again \eqn{thecondition} i.e., $\Dv\!\cdot\!\Hv=\Ev\!\cdot\!\Bv$. 

\subsection{Born-Infeld lagrangian}

A notable example of a lagrangian whose equations of motion 
are invariant under duality rotations is given by the  Born-Infeld one 
\cite{BI}
\eqa
\LL_{\rm{BI}}&=&
1-\sqrt{-{\rm{det}}(\eta+{F})}\\[.3em] 
&=&
1- {\sqrt{1+ {1\over 2} F^{\,2}-\frac{1}{16}(F\tilde F)^2}}\\[.3em]
&=& 1-\sqrt{1-\Ev^2+\Bv^2-(\Ev\!\cdot\!\Bv)^2}~.
\ena
In the second line we have simply expanded the 4x4 determinant and 
espressed the lagrangian in terms of the only two independent Lorentz 
invariants associated to the electromagnetic field: 
$F^{\,2}\equiv F_{\mu\nu} F^{\mu\nu}\,,~
F\tilde F \equiv F_{\mu\nu}\tilde F^{\mu\nu}$. 

The explicit expression of $G$ is 
\eq\label{Geasy}
{G}_{\mu\nu}=
\frac{\tilde F_{\,\mu\nu}+{1\over 4} F\tilde F\,F_{\mu\nu}}{
\sqrt{1+{1\over 2}F^2-\frac{1}{16}(F\tilde F)^2}}~,
\en
and the duality condition \eqn{thecondition} is  readily seen to hold.
The hamiltonian is
\eq
\HH_{\rm{BI}}=\sqrt{1 +\Dv^2+\Bv^2+(\Dv\times \Bv)^2}\,\,-1~.
\en
Notice that while the $\Ev$ and $\Bv$ variables are constrained by the
reality of the square root in the lagrangian, the hamiltonian 
variables $\Dv,\Bv$ are unconstrained.
By  using the equations of motion and \eqn{thecondition} it can be explicitly
verified that the generator of duality rotations is time independent, $\{\La, H\}=0$.

\subsection{Extended duality rotations}
The duality symmetry of the equations of motion
of nonlinear electromagnetism can be  extended to $SL(2,\rr)$.
We observe that the definition of duality symmetry we used can be relaxed
by allowing the $F$ dependence of $G$ to change by a linear term: 
$G=2{\partial \LL\over\partial F}$ and $G=2{\partial \LL\over\partial F}+\vartheta F$ togheter with the  Bianchi identities for $F$ give equivalent 
equations of motions for $F$. 
Therefore the transformation
\eqa
\left(\begin{array}{c}
F' \\
G'
\end{array}\right)&=&
\left(\begin{array}{cc}
1 &0\\
 \vartheta & 1
\end{array}\right)
\left(\begin{array}{c}
F  \\
G
\end{array}\right)\label{Etrans}
\ena
is a symmetry of any nonlinear electromagnetism. It corresponds to the lagrangian change $\LL\rightarrow \LL+{1\over 4}\vartheta F\tilde F$.
This symmetry alone does not act on $F$, but it is useful if the nonlinear 
theory has $SO(2)$ duality symmetry.
In this case \eqn{Etrans} extends duality symmetry from $SO(2)$ to 
$SL(2,\rr)$ (i.e. $Sp(2,\rr)$). 
Notice however that the $SL(2,\rr)$ 
transformed solution, contrary to the $SO(2)$ one, has a different 
energy and energy momentum tensor (recall \eqn{enmom}).
On the other hand, as we show in Section \ref{enmomten},  if the constant 
$\vartheta$ is promoted 
to a dynamical field we have invariance of the 
energy momentum tensor under $SL(2,\rr)$ duality.

\section{General theory of duality rotations}
We study in full generality the conditions in order to have theories with
duality rotation symmetry. By properly introducing scalar fields 
(sigma model on coset space) we 
enhance theories with a compact duality group
to theories with an extended noncompact duality group.
A Born-Infeld lagrangian with $n$ abelian field strengths  
and $U(n)$ duality group (or $Sp(2n,\rr)$ in the presence of scalars)  
is constructed.

\subsection{General nonlinear theory}
We consider a theory of $n$ abelian gauge fields  
possibly coupled to other bosonic and fermionic fields that we denote $\ppi^\al$,
($\al=1,...p$).
We assume that the $U(1)$ gauge potentials 
enter the action $S=S[F,\ppi]$ only trough the 
field strengths $F^\La_{\mu\nu}$ ($\La=1,\dots,n$), and that the action
does not depend on partial derivatives of the field strengths.
Define ${\tilde G}_{\!\La}^{\;\mu\nu}=~2{\del \LL
\over\del F^\La_{\mu\nu}}$, i.e,
\eq
{\tilde G}_{\!\La}^{\;\mu\nu}=~2{\delta S[F,\ppi]
\over\delta F^\La_{\mu\nu}}~;
\label{defKK}
\en
then the Bianchi identities and the equations 
of motions for $S[F,\ppi]$ are
\eqa
{\pa}_{\mu}
{\tilde F}^{\La\,\mu\nu}  &=&0~,\label{max222}\\[.3em]
{\pa}_{\mu}
\tilde G_{\!\La}^{~\mu\nu}&=&0~, \label{max111}\\[.3em]
{\delta S[F,\ppi]\over\delta \ppi^\al}&=&0\label{chieq}~.
\ena
The field theory is described by the system of 
equations \eqn{defKK}-\eqn{chieq}.
Consider the duality transformations
\eqa\label{rotFGs}
\left(\begin{array}{c}
F' \\
G'
\end{array}\right)&=&
\left(\begin{array}{cc}
A & B\\
C & D
\end{array}\right)
\left(\begin{array}{c}
F  \\
G
\end{array}\right)
\\[.5em]
\ppi'^\al&=&\Xi^\al (\ppi)\label{rotchi}
\ena
where $\big({} _{C~D} ^{A~B}{} \big)$ is a generic constant $L(2n,\rr)$ 
matrix and the $\ppi^\al$ fields transformation in full detail reads 
$\ppi'^\al=\Xi^\al(\ppi, \big({}  _{C~D} ^{A~B}{} \big))$,
with no partial derivative of $\ppi$ appearing in $\Xi^\al$.
\sk
These duality rotations are a symmetry of
the system of equations 
\eqn{defKK}-\eqn{chieq} iff, 
given $F$, $G$, and $\ppi$ solution of
\eqn{defKK}-\eqn{chieq} then
$F'$, $G'$ and $\ppi'$, that by construction satisfy 
$\pa_\mu{\tilde F'}{}^{\La\,\mu\nu}  =0$ and 
${\pa}_{\mu}{\tilde G'}_\La{}^{\!\mu\nu}=0$,
satisfy also
\eqa
{{\tilde G}'}_\La{}^{\!\mu\nu}&\,=\,&{2}\frac{\de {S[F',\ppi']}}{\de F'^\La_{\mu\nu}}~,
\label{defK's}\\[.3em]
{\delta S[F',\ppi']\over\delta\ppi'^\al}&\,=\,&0\label{chieq'}~.
\ena
We study these on shell conditions in the case of infinitesimal 
$GL(2n,\rr)$ rotations $$F\rightarrow F'=F+ \ddel F~~,~~~
G\rightarrow G'= G + \ddel G~,~$$
\eqa
 \ddel \left( \matc F \\
G \emat \right)
&=&
\left( \matcc a & b \\ c & d \emat
\right) \left( \matc F \\ G \emat \right)~,\label{FGtransgen}\\[.5em]
\ddel\ppi^\al&=&\xi^\al(\ppi)~.\label{ppitrans}
\ena
%
%
%
%
The right hand side of \eqn{defK's} can be rewritten as 
 \eq\label{47}
{\de  S[F',\ppi']\over
\de F'^\La} = \int_y{\de
S[F',\ppi']\over \de F^\Sigma(y) } {\de F^\Sigma(y) \over \de
F'^\La}~.
\en
We now invert the matrix $\left(\matcc {\delta F'\over\delta F} & 
{\delta F'\over\delta \ppi} \\[.3em] {\delta \ppi'\over\delta F} & 
{\delta \ppi'\over\delta \ppi} \emat \right)_{_{_{_{_{}}}}} \!$, recall that 
$F_1\tilde F_2 =\tilde F_1 F_2$ and observe that 
$$\int_y{\de S[F,\ppi]\over \de F(y) }\, b\, {\de G(y) \over \de F^\La}
={1\over 4}{\de \over \de F^\La} \int_y{\tilde G \,b \, G }+ 
{1\over 4}\int_y\tilde G (b-b^t) {\de G\over \de F^\La}~.$$
We thus obtain 
\eq
{\de  S[F',\ppi']\over
\de F'^\La} \;=\;{\de
S[F',\ppi']\over \de F^\La } -a^{\Sigma\La}{\de
S[F',\ppi']\over \de F^\Sigma } 
-{1\over 4}{\de \over \de F^\La} \int_y{\tilde G\, b\, G}
-{1\over 4}\int_y\tilde G (b-b^t) {\de G\over \de F^\La}~.
\en
Since the left hand side of  \eqn{defK's} is 
$\tilde G_\La+{1\over 2}{\delta\over\delta F^\La}\int_y 
\tilde F\,c\, F+ {1\over 2} (c-c^t)_{\La\Sigma}\tilde F^\Sigma+ 2d_{\La}^{\;\Sigma\,}{\delta S[F,\ppi]\over\delta F^\Sigma}$,  we rewrite  \eqn{defK's} as
\eqa\label{thecondgende}
{\de\over \de F^\La}\left(S[F',\ppi']-S[F,\ppi]-{1\over 4} \int_y (\tilde F\,c \,F +\tilde G\,b\, G)
\right) ~~~~~~~~~~~~~~~~~~~~~~~~~~~~~~~~~~~~~~~~~~~~~~\\[.3em]
~~~~~~~~~~~~~~~~~~~~~~~~~~~~~~~=~
  (a^t+d)_{\La}^{\;\Sigma}{\de\over\de F^\Sigma}S[F,\ppi]+{1\over 4}(c-c^t)_{\La\Sigma} \tilde F^{\Sigma} + {1\over 4}\int_y\tilde G (b-b^t) {\de G\over \de F^\La}~.\nn
\ena
%
%
Since this expression
does not contain derivatives of $F$, the functional variation becomes just a partial derivative, and \eqn{thecondgende} is equivalent to
\eqa\label{deldelF}
{\del\over\del F^\La} \left( 
\LL(F',\ppi')-\LL(F,\ppi)-{1\over 4}  \tilde 
F\,c\, F -{1\over 4}\tilde G\,b\, G \right)\;~~~~~~~~~~~~~~~~~~~~~~~~~~~~~~~~~~~~ ~~~~~~~~~~~~\\
~~~~~~~~~~~~~~~~~~~~~~~~~~~~=\;(a^t+d)_{\La}^{\;\Sigma}{\del\over\del F^\Sigma}\LL(F,\ppi)+{1\over 4}(c-c^t)_{\La\Sigma} \tilde F^{\Sigma}+
 {1\over 4}\tilde G (b-b^t) {\del G\over \del F^\La}\;.~\nn
\ena
Here $\LL(F,\ppi)$ is a shorthand notation for a lagrangian that depends on $F$,
$\ppi^\al$, $\partial\ppi^\al$ and eventually higher partial derivatives of the 
fields $\ppi^\al$, say up to order $\ell$. 
Equation \eqn{deldelF} has to hold on shell of \eqn{defKK}-\eqn{chieq}. 
Since this equation has no partial derivative of $F$ and at most  
derivatives of $\ppi^\al$ up to order $\ell$, if it holds on shell of 
\eqn{defKK}-\eqn{chieq} then it holds just on shell of \eqn{defKK}, 
and of the fermions fields equations, the scalar and vector partial differential 
equations being of higher order in derivatives of $F$ or $\ppi^\al$ fields.
In particular if no fermion is present \eqn{deldelF} 
holds just on shell of \eqn{defKK}. 

Since the left hand side of \eqn{deldelF} is a derivative with respect to $F^\La$ so must be the right 
hand side. This holds if we consider infinitesimal dilatations, parametrized by 
${\kappa\over 2}\in\rr$, and infinitesimal $Sp(2n,\rr)$ transformations
\eq
a^t+d=\kappa 1\!\!1~~,~~~~b^t=b~~,~~ c^t=c~.~~ \label{liedilsp}
\en 
We can then remove the  derivative $\del\over \del F^\La$ and 
obtain the equivalent condition
\eq\label{thecondgenf}
\LL(F',\ppi')-\LL(F,\ppi)-\kappa\LL(F,\ppi)
-{1\over 4}  \tilde 
F\,c\, F -{1\over 4}\tilde G\,b\, G =f(\ppi)
\en
where $f(\ppi)$ can contain partial derivatives of $\ppi$ up to the 
same order as in the lagrangian.
\sk
We now show that $f(\ppi)$ in \eqn{thecondgenf} is independent 
from  $\ppi$.
Consider the $\ppi$-equations of motion \eqn{chieq'},
\eqa\label{ppieqmot}
{\delta S[F',\ppi']\over\delta\ppi'^\al}&=&
\int_y{\delta S[F',\ppi']\over\delta\ppi^\be(y)}{\delta\ppi^\be(y)\over\delta\ppi'^\al}
+\int_y{\delta S[F,\ppi]\over\delta F(y)}{\delta F(y)\over\delta\ppi'^\al}\nn\\
&=&
{\delta S[F',\ppi']\over\delta\ppi^\al}
-{\delta S[F,\ppi]\over\delta\ppi^\be}{\del\xi^\be\over\del\ppi^\al}
-{1\over 4} {\delta\over \delta\ppi^\al} \int_y \tilde G\,b\, G\nn\\
&=&{\delta S[F,\ppi]\over\delta\ppi^\al}
-{\delta S[F,\ppi]\over\delta\ppi^\be}{\del\xi^\be\over\del\ppi^\al}
+{\delta\over\delta\ppi^\al}\left(S[F',\ppi']-S[F,\ppi]-{1\over 4}
\int_y \tilde G \,b\,G 
\right)\nn
\ena
where only first order infinitesimals have been retained, and where techniques similar to those used in the study of \eqn{47} have been applied. 
On shell the left hand side has to vanish; since the first two addends on the right hand side are proportional to the $\ppi$-equations of motion, this happens iff on shell 
\eq
{\delta\over\delta\ppi^\al}\left(S[F',\ppi']-S[F,\ppi]-\kappa S[F,\ppi]-{1\over 4}
\int_y (\tilde G\,b\, G  +\tilde F\,c\, F) \right)\;=\;0 ~.
\en
Comparison with \eqn{thecondgenf} shows that  on shell
\eq\label{f=0}
{\de\over\de\ppi^\al} f(\ppi)=0~.
\en
In this expression no field 
strength $F$ is present and therefore the equations of motion of our 
interacting system are of no use;
equation \eqn{f=0} holds also off shell and we conclude that $f(\ppi)$ is 
$\ppi$ independent, it is just a constant depending on the parameters $a,b,c,d$  
(it usually vanishes). We thus have the condition
\eq\label{thecondgen0}
\LL(F',\ppi')-\LL(F,\ppi)-\kappa_{}\LL(F,\ppi)
-{1\over 4}  \tilde 
F\,c\, F -{1\over 4} \tilde G\,b\, G ~=\; {\mbox{\it const}}_{a,b,c,d}
\en
If we expand $F'$ in terms of $F$ and $G$, we obtain the equivalent condition 
\eq\label{thecondgen}
\Delta_\ppi\LL(F,\ppi)={1\over 4}\tilde F \,c\, F - {1\over 4}  \tilde G\,b\, G+ \kappa_{}\LL(F,\ppi) -{1\over 2} \tilde G\,a\, F+ { \mbox{\it const}}_{a,b,c,d}~
\en
where $\,\Delta_\ppi\LL(F,\ppi)=\LL(F,\ppi')-\LL(F,\ppi)$.
\sk
Equation \eqn{thecondgen}, where
${\tilde G}_{\!\La}^{\;\mu\nu}=~2{\del \LL
/\del F^\La_{\mu\nu}}$, 
is a necessary and sufficient condition 
in order to have duality symmetry. This condition is on shell of the 
fermions equations of motion, in particular if no fermion is present 
this condition is off shell. 
In the presence of fermions,  
equation \eqn{thecondgen} off shell is a 
sufficient condition for duality symmetry.

The duality symmetry group is \eq
\rr^{>0}\times SL(2n,\rr)~,
\en
 the
group of dilatations times  symplectic transformation; it is the connected Lie group generated by the Lie algebra \eqn{liedilsp}. It is also the maximal group of duality rotations as the example (or better, the limiting case) studied in the next section shows.
\sk

We have considered dynamical fermionic and bosonic fields $\ppi^\al$.
If a subset $\chi^r$ of these fields is not dynamical 
the corresponding equations of motion are of the same order as those 
defining $G$, and thus \eqn{deldelF} and \eqn{thecondgen} hold on shell 
of all these equations. Moreover since no $\partial\chi^r$ appears in the lagrangian, the duality transformations for these fields can include the field strength $F$, i.e., $\chi^r\rightarrow \chi'^r=\Xi^r(F,\chi)$.  In this case there is an extra addend in \eqn{47}. The necessary and sufficient duality condition \eqn{thecondgen} does not change.
\sk

We also notice that condition \eqn{thecondgen0} in the absence of dilatations ($\kappa=0$), and for $const_{a,b,c,d}=0$  is equivalent to the invariance of 
\eq
\LL-{1\over 4}\tilde F G~.\label{invLFG}
\en

\noi
{\subsection{The main  example and the scalar fields fractional transformations}
Consider the Lagrangian
\eq\label{examplef2} 
{1\over 4}\N_{2\,\La\Sigma} F^\La F^\Sigma+{1\over 4} \N_{1\,\La\Sigma}F^\La\tilde F^\Sigma +{{\mathscr{L}}}(\phi)
\en
where the real symmetric matrices $\N_1(\phi)$ and $\N_2(\phi)$ and the lagrangian $\mathscr{L}(\phi)$ are just functions of the bosonic fields $\phi^i$, $i=1,\ldots m$, (and their partial derivatives).

Any nonlinear lagrangian in the limit of vanishing fermionic fields and of weak 
field strengths $F^4<<F^2$ reduces to the one in \eqn{examplef2}. 
A straighforward calculation shows that this lagrangian has 
$\rr^{>0}\times SL(2n,\rr)$ duality symmetry if the matrices $\N_1$ and $\N_2$ 
of the scalar fields transform as
\eqa\label{DN1}
\Delta_{\,} \N_1&=&c+d_{\,}\N_1-\N_1 a -\N_1\,b\,\N_1+\N_2\,b\,\N_2 ~~,\\
\Delta_{\,} \N_2&=&d_{\,}\N_2-\N_2 a -\N_1\,b\,\N_2-\N_2\,b\,\N_1 ~~,\label{DN2}
\ena
and 
\eq
\Delta_{\,}\mathscr{L}(\phi)=\kappa \mathscr{L}(\phi)~.\label{spdinvL}
\en
If we define $$\N=\N_1+i\N_2~,$$
i.e., $\N_1=\rea\N~,~~\N_2=\im \N$, the transformations \eqn{DN1}, \eqn{DN2} read
\eq
\Delta_{\,} \N\;=\;c+d_{\,}\N-\N a-\N\,b\,\N ~,\label{DN1N2}
\en
the finite version is the fractional transformation
\eq\label{sp2nrfract}
\N'=(C+D\N)\,(A+B\N)^{-1}~.
\en
Under \eqn{sp2nrfract} the imaginary part of $\N$ transforms as
\eq
\N_2'=(A+B\N)^{-\dagger}\N_2(A+B\N)^{-1}\label{Kahler}
\en
where ${-\dagger}$ is a shorthand notation for the  hermitian conjugate 
of the inverse matrix. 
\sk
The kinetic term ${1\over 4}\N_{2\,\La\Sigma} F^\La F^\Sigma$
is positive definite if the symmetric matrix $\N_2$ is negative definite.
In Appendix 7.2 we show that the  matrices $\N=\N_1+i\N_2$
with $\N_1$ and $\N_2$ real and symmetric, and $\N_2$ positive definite, 
are the coset space $Sp(2n,\rr)\over U(n)$. 

A scalar lagrangian that satisfies the variation \eqn{spdinvL} can
always be constructed using the geometry of the coset space
$Sp(2n,\rr)\over U(n)$, see Section \ref{csfndr}.
\sk
This example also clarifies the condition \eqn{liedilsp} that we have 
imposed on the $GL(2n,\rr)$ generators. It is a straighfoward calculation
to check that the equations
\eqn{max222}, \eqn{max111} and 
\eq
\tilde G= \N_2F+\N_1\tilde F\label{constexf2}
\en 
have duality symmetry under $GL(2n,\rr)$ transformations with
$\Delta\N$ given in \eqn{DN1N2}. 
However it is easy to see that equation \eqn{deldelF} implies, for the 
lagrangian \eqn{examplef2}, that 
condition  \eqn{liedilsp} must hold. The point is that we want the 
constitutive relations $G=G[F,\ppi]$ to follow 
from a lagrangian.
Those following from the lagrangian \eqn{examplef2} are 
\eqn{constexf2} with $\N_1$ and $\N_2$ necessarily {\sl symmetric}
matrices. Only if the transformed  matrices $\N'_1$ and $\N'_2$ are
again symmetric we can have $\tilde G'={\del \LL(F',\ppi')\over\del F'}$ as in 
\eqn{defK's}, (or more generally $\tilde G'={\del \LL'(F',\ppi')\over\del F'}$).
 The constraints $\N'_1={\N'_1}^t$, $\N'_2={\N'_2}^t$, 
reduce the duality group to $\rr^{>0}\times SL(2n,\rr)$.
\sk
In conclusion equation \eqn{thecondgen} is a necessary and sufficient 
condition for a theory of $n$ abelian gauge fields 
coupled to bosonic matter to be symmetric under 
$\rr^{>0}\times SL(2n,\rr)$  duality rotations, 
and  $\rr^{>0}\times SL(2n,\rr)$ is the maximal connected Lie group of duality 
rotations.

\subsection{A basic example with fermi fields}
Consider the Lagrangian with Pauli coupling
\eq
\LL_0=-{1\over 4}F_{\mu\nu}F^{\mu\nu}\label{FPSI'}
-{1\over 2}\overline\psi\partial\!\!\!\slash \psi-{1\over 2}\overline\xi
\partial\!\!\!\slash
\xi
+{1\over 2}\la F^{\mu\nu}\overline\psi\sigma_{\mu\nu}\xi
\en
where $\sigma^{\mu\nu}={1\over 4}[\gamma^\mu,\gamma^\nu]$ and $\psi, \xi$ 
are two Majorana spinors. 
We have 
\eq
\tilde G^{\mu\nu}=2 {\partial\LL_0\over\partial F_{\mu\nu}}=
-F^{\mu\nu}+  \la\overline\psi\sigma^{\mu\nu}\xi
\en
and the duality condition \eqn{thecondgen}  for an 
infinitesimal $U(1)$ duality rotation $\big( {}^{\,0}_{-b} {\,}^b_{0}\big)$ reads 
\eq\label{condFPSI}
\Delta_{\psi}\LL_0+\Delta_{\xi}\LL_0=
-{b\over 4}\la \tilde F\overline\psi\sigma \xi+{b\over 4}\la^2\,
\overline\psi\sigma^{\mu\nu}\xi\:\overline\psi\tilde\sigma_{\mu\nu}\xi~.
\en
It is natural to assume that the kinetic terms of the fermion fields 
are invariant under this duality rotation 
(this is also the case for the scalar lagrangian
${{\mathscr{L}}}(\phi)$ in \eqn{spdinvL}), then using 
$\gamma_5\sigma^{\mu\nu}=i\tilde\sigma^{\mu\nu}$
we see that the coupling of the fermions with the field strength
is reproduced if the fermions rotate according to 
\eqa
\Delta \psi&=&{i\over 2}b\gamma_5\psi~,\\[.3em]
\Delta \xi&=&{i\over 2}b\gamma_5\xi~;
\ena 
we also see that we have to add to the lagrangian $\LL_0$
a new interaction term quartic in the fermion fields.
Its coupling is also fixed by duality symmetry to be $-\la^2/8$.
\sk
The theory with $U(1)$ duality symmetry is therefore given by
the lagrangian \cite{FSZ77}
\eq
\LL=-{1\over 4}F_{\mu\nu}F^{\mu\nu}\label{FPSI}
-{1\over 2}\overline\psi\partial\!\!\!\slash \psi-{1\over 2}\overline\xi
\partial\!\!\!\slash
\xi
+{1\over 2}\la F^{\mu\nu}\psi\sigma_{\mu\nu}\xi
-{1\over 8} \la^2\,\overline\psi\sigma_{\mu\nu}\xi\:
\overline\psi\sigma^{\mu\nu}\xi~.
\en
Notice that fermions transform under the double cover of $U(1)$ indeed
under a rotation of angle $b=2\pi$ we have $\psi\rightarrow -\psi$,
$\xi\rightarrow -\xi$, this is a typical feature of fermions transformations
under duality rotations, they transform under the double cover of the maximal 
compact subgroup of the duality group. This is so because the interaction 
with the gauge field is via fermions bilinear terms.

\subsection{Compact and noncompact duality rotations}
\sk
\subsubsection{Compact duality rotations}
The fractional transformation \eqn{sp2nrfract} is also characteristic of
nonlinear theories. The subgroup of $Sp(2n,\rr)$ that leaves invariant a
fixed value of the scalar fields $\N$ is $U(n)$. This is easily seen by 
setting $\N=-i1\!\!1$. Then infinitesimally we have relations 
\eqn{liedilsp} with $\kappa=0$ and $b=-c$, $a=-a^t$, i.e. we have
the antisymmetric matrix 
$$\left(\matcc a & \,b \\[.3em] -b & \,a
 \emat \right)~,
$$
$a=-a^t$, $b=b^t$.
For
finite transformations the $Sp(2n,\rr)$ relations \eqn{sp2nr} are 
complemented by
\eq
A=D~~,~~B=-C~.
\en 
Thus  $A-iB$ is a unitary matrix (see also \eqn{UinUspSp}).
$U(n)$ is the maximal compact subgroup of $Sp(2n,\rr)$, it is the group of
orthogonal and symplectic $2n\times 2n$ matrices.   
\sk
More in general from Section 3.1
we easily conclude that a necessary and sufficient condition for a theory 
with just $n$ abelian gauge fields to have $U(n)$ duality symmetry is
(cf. \eqn{thecondgen})
\eqa\label{FFGG}
\tilde  F^\La F^\Sigma +\tilde G^\La G^\Sigma=0 \\[.3em]
\tilde G^\La F^\Sigma -\tilde G^\Sigma F^\La=0 \label{GFGF}
\ena
for all $\La, \Sigma$.
Moreover since any nonlinear lagrangian in the limit
of weak field strengths $F^4<<F^2$ reduces to the one in 
\eqn{examplef2} (with a fixed value of $\N$),
 we conclude that  $U(n)$ is the maximal duality group 
for a theory with only gauge fields.

\sk
Condition \eqn{GFGF} is equivalent to 
\eq
(F^\Sigma{\del\over\del F^\La}-
F^\La{\del\over\del F^\Sigma})\LL=0~,\label{soninv}
\en 
i.e. to the invariance of the 
Lagrangian under $SO(n)$ rotations of the $n$ field strengths $F^\Sigma$. 
Condition \eqn{FFGG} concerns on the other hand the invariance of the 
equations of motion under transformation of the electric field strengths
into the magnetic field strengths. 

In a theory with just $n$ abelian gauge fields 
the field strengths appear in the Lagrangian
only through the Lorentz invariant combinations
\eq
\alpha^{\La\Sigma} \equiv \frac{1}{4} F^{\La} {F}^{\Sigma},
~~\beta^{\La\Sigma} \equiv 
\frac{1}{4}  \widetilde{F}^{\La} {F}^{\Sigma},
\label{alphadef}
\en
and equation \eqn{soninv}, tell us that $\LL$ is a scalar under 
$SO(n)$ rotations; e.g.
$\LL$ is a sum of traces, or of products of traces, of monomials 
in $\alpha$ and $\beta$
(we implicitly use the metric $\delta_{\La\Sigma}$ 
in the $\al$ and $\be$  products).

If we define
\eq 
\LL_{\alpha} \equiv \frac{\partial \LL}{\partial \alpha^t}~,~~
\LL_{\beta} \equiv \frac{\partial \LL}{\partial \beta^t}~,
\en
then using the chain rule and the definitions \eqn{alphadef}
we obtain that \eqn{FFGG} is equivalent to
\eq
\LL_{\beta} \beta \LL_{\beta} -\LL_{\alpha} \beta \LL_{\alpha} +
\LL_{\alpha} \alpha \LL_{\beta} + \LL_{\beta} \alpha \LL_{\alpha} +
\beta = 0~.\label{bselfdual}
\en
If we define
\eq
p \equiv -\frac{1}{2}(\alpha +i\beta) ~,~~
q \equiv -\frac{1}{2}(\alpha - i\beta)~,
\en
then \eqn{bselfdual} simplifies and reads
\eq
 p - \LL_p\, p \,\LL_p =  q- \LL_q \, q \,\LL_q~.
\en

Condition \eqn{bselfdual} in  the case of a single gauge field
was considered in~\cite{GZ2} togheter with other equivalent 
conditions, in particular 
$\LL_u\LL_v=1$, where $u={1\over 2}(\al+(\al^2+\be^2)^{1\over 2})$, 
$v={1\over 2}(\al-(\al^2+\be^2)^{1\over 2})$, see also \cite{Ivanov}.

\subsubsection{Coupling to scalar fields and noncompact duality rotations}\label{csfndr} 
By freezing the values of the scalar fields $\N$ we have obtained a 
theory with only gauge fields and with $U(n)$ duality symmetry.
Vice versa (following \cite{araki} that
extends to $U(n)$ the $U(1)$ interacting theory discussed in~\cite{GZ1,GZ2}) 
we show that given a theory invariant under $U(n)$ duality rotations
it is possible to extend it via $n(n+1)$ scalar fields $\N$  to a theory 
invariant under $Sp(2n,\rr)$. 
Let $\ll(F)$ be the lagrangian of the theory with $U(n)$ duality. From
\eqn{thecondgen0} we see that under a $U(n)$ duality rotation
\eq
\LL(F')-\LL(F)
=-{1\over 4}  \tilde 
F\,b\, F +{1\over 4} \tilde G\,b\, G~.\label{inLUn}
\en
In particular $\LL(F)$ is invariant under the orthogonal subgroup 
$SO(n)\subset U(n)$ given by the matrix $\big({}^A_0 {}^{~0}_{-A^t}\big)$. 
This is the so-called
electric subgroup of the duality rotation group $U(n)$ 
because it does not mix
the electric fields  $F$ with the dual fields $G$. 
\sk
Define the new lagrangian 
\eq\label{extension}
\LLL(F,R,\N_1) = \ll(RF)+{1\over 4}\tilde F\N_1 F 
\en
where  $R=(R^\La_{~\Sigma})_{\La,\Sigma=1,...n}$ is an arbitrary nondegenerate real matrix and $\N_1$ is a real symmetric matrix.
Because of the $O(n)$ symmetry the new lagrangian depends only on the combination
\eq
\N_2=-R^tR~,
\en
rather than on $R$. Thus $\LLL(F,R,\N_1)=\LLL(F,\N)$ where $\N=\N_1+i\N_2$.
 
We show that $\LLL$ satisfies the duality condition \eqn{thecondgen},
\eq\label{condscalars}
(\Delta_F+\Delta_R +\Delta_{\N_1} )\LLL(F,R,\N_1)={1\over 4}\tilde F\,c\,F+{1\over 4}\tilde G\,b\,G
\en
where as always $\tilde G={2}{ \partial \LLL\over \partial F}$, and where
$\N_1$ transforms as in \eqn{DN1}
and 
\eq
\Delta R=-R(a+b\N_1)~~,~
\en
so that $\N_2=-R^tR$ transforms as in \eqn{DN2}. Notice that we could also 
have chosen the transformation $\Delta R=\La R- R(a+b\N_1)$ with $\La$ an 
infinitesimal $SO(n)$ rotation.

We first immediately check \eqn{condscalars} in the case of the rotation 
$\big( {}^0_c {}^0_0\big)$. Then in the case $\big( {}^a_0 {}^0_d\big)$, where $a=-d^t$.
Finally we consider the duality rotation $\big( {}^0_0 {}^b_0\big)$.
It is convenient to introduce the notation
\eq
{\FF}=RF~~,~~~\GG=2{\partial \LL(\FF)\over \partial \FF}~.
\en
We observe that $\LL(\FF)$ satisfies the $U(n)$ duality conditions
\eqn{FFGG}, \eqn{GFGF} with $F\rightarrow \FF$, $G\rightarrow \GG$. 
Equation \eqn{condscalars} holds because of \eqn{FFGG} and proves $Sp(2n,\rr)$
duality invariance of the theory with lagrangian $\LLL$. 
\sk
We end this subsection with few comments. We notice that \eqn{GFGF} is equivalent to 
the invariance of the lagrangian under the infinitesimal $SO(n)$ 
transformation $R\rightarrow \La R$. 

We also observe that under an $Sp(2n,\rr)$ duality transformation
$\big( {}^a_c{}^b_d\big)$, the dressed fields $\FF$ and $\GG$
transform via the field dependent rotation 
$\big( {}^{~0}_{-b'}{}^{b'}_0\big)=
\big( {}^{~~~0}_{-RbR^t}{}^{RbR^t}_{~~0}\big)$,
\eqa
\Delta \FF&=&\;R\,b\,R^t \,\GG~,\\
\Delta \GG&=&-R\,b\,R^t \,\FF~.
\ena

The geometry underlying the construction 
of $Sp(2n,\rr)$ duality invariant theories from $U(n)$ ones is that of coset 
spaces.
The scalar fields $\N$ parametrize the coset space 
$^{Sp(2n,\rr)}/ _{U(n)}$ (see proof in ...).
We also have $^{Sp(2n,\rr)}/ _{U(n)}\,=\,_{SO(n)}\backslash ^{GL^{+}(n)}\times \rr^{{n(n+1)\over 2}}$ where $GL^{+}(n)$ is the connected component of $GL(n)$ 
and the equivalence classes $[R]=\{R'\in GL^{+}(n);\;  R'R^{-1}=e^\La\in SO(n)\}$
parametrize the coset space $_{SO(n)}\backslash ^{GL^{+}(n)}$.
\sk
The proof of $Sp(2n,\rr)$ duality symmetry for the theory
described by the lagrangian $\LLL$ holds also if we add to $\LLL$
an $Sp(2n,\rr)$ invariant  lagrangian for the fields $\N$ like 
the lagrangian $\mathscr{L}$ in \eqn{kin}. 
Of course we can also  consider initial lagrangians in \eqn{inLUn} that 
depend on matter fields invariant under the $U(n)$ rotation, they will be
$Sp(2n,\rr)$ invariant in the corresponding lagrangian $\LLL$.
Moreover, by considering an extra scalar field $\Phi$, we can always extend 
an $Sp(2n,\rr)$ duality theory to an $\RR^{>0}\times Sp(2n,\rr)$ one.

\subsection{Nonlinear sigma models on $G/H$}
In this section we briefly consider the geometry of coset spaces $G/H$.
This is the geometry underlying the scalar fields and needed to
formulate their dynamics \cite{Wess1, Wess2}.

We study in particular the case $G=Sp(2n.\rr)$, $H=U(n)$
\cite{GZ} and give a kinetic term for the scalar fields 
$\N$.
\sk
The geometry of the coset space $G/H$ is
conveniently described in terms of coset representatives, 
local sections $L$ of the bundle $G\rightarrow G/H$.
A point $\phi$ in $G/H$ is an equivalence class 
$gH=\{\tilde g \,|\,g^{-1}\tilde{g}\in H\}$.
We denote by $\phi^i$ ($i=1,2\ldots m$) its coordinates 
(the scalar fields of the theory).
The left action of $G$ on $G/H$ is inherited from that of $G$ on $G$, it
is given by $gH\mapsto {g'}gH$, that we rewrite 
$\phi\mapsto {g'}\phi=\phi'$. 
Concerning the coset representatives we then have
\eq
g' L(\phi)=L(\phi')h~,\label{gLphi}
\en
because both the left and the right hand side are 
representatives of $\phi'$.
The geometry of $G/H$ and the corresponding physics can be 
constructed in terms of coset representatives.  
Of course the construction must be insensitive to the particular 
representative choice, we have a gauge symmetry with 
gauge group $H$.

When $H$ is compact 
the Lie algebra of $G$ splits in the direct sum 
${\mathbb G}={\mathbb H}+{\mathbb K}$, where 
\eq 
[{\mathbb H}, {\mathbb H}]\subset 
{\mathbb H}~~,~~~
[{\mathbb K}, {\mathbb K}]\subset 
{\mathbb H}+{\mathbb K}~~,~~~
[{\mathbb H}, {\mathbb K}]\subset 
{\mathbb K}~~.\label{HKK}
\en
The last expression defines the coset space 
representation of ${\mathbb H}$. The 
representations of the compact Lie algebra 
${\mathbb H}$ are equivalent to unitary ones, 
and therefore there exists a basis $(H_\al, K_a)$,
where $[H_\al ,K_a]=C_{\al a}^bK_b$ with
$C_\al=(C_{\al a}^b)_{a,b=1,...m=dimG/H}$
antihermitian matrices. Since  the coset 
representation is a real representation then 
these matrices $C_\al$ belong to the Lie 
algebra of $SO(m)$.

Given a coset representative $L(\phi)$, the 
pull back on $G/H$ of the ${\mathbb G}$ Lie algebra left
invariant 1-form $\Gamma=L^{-1}dL$  is decomposed as 
$$\Gamma=L^{-1}dL=P^a(\phi)K_a+
\upomega^\al(\phi)H_\al~.$$
$\Gamma$ and therefore $P=P^a(\phi)K_a$ and $\upomega=\upomega^\al(\phi)H_\al$ 
are invariant under 
diffeomorphisms generated by the left $G$ action. Under the local
right $H$ action of an element $h(\phi)$  
(or under change of coset representative
$L'(\phi)=L(\phi)h(\phi)$) we have
\eq
P\rightarrow h^{-1} P h ~~~,~~~~
\upomega\rightarrow h^{-1} \upomega h +h^{-1}dh~.
\en 
The 1-forms
$P^a(\phi)=P^a(\phi)_i d\phi^i$
are therefore vielbain on $G/H$ transforming in the
fundamental of $SO(m)$, while 
$\upomega=\upomega(\phi)_id\phi^i$ is an 
${\mathbb H}$-valued connection 
1-form on $G/H$.  We can then define the covariant derivative 
$\nabla P^a=[P,\upomega]^a=P^b\otimes -C^a_{\al b}\upomega^\al$. 

There is a natural metric on 
$G/H$,
\eq\label{metricGH}
g=\delta_{ab}P^a\otimes P^b~,
\en 
(this definition is well given because 
we have shown that the coset representation is via infiniesimal $SO(m)$ 
rotations). It is easy to see that the connection
$\nabla$ is metric compatible, $\nabla g=0$.

If the coset is furthermore a symmetric coset we have 
$$
[{\mathbb K}, {\mathbb K}]\subset 
{\mathbb H}~,$$
then the identity $d\Gamma+\Gamma\wedge \Gamma=0$, that is (the pull-back on $G/H$ of) the Maurer-Cartan equation, in terms 
of $P$ and $\upomega$ reads 
\eqa
R+P\wedge P=0~,~\label{RPP}\\
dP+P\wedge\upomega+\upomega\wedge P=0~.\label{RPPsec}
\ena
This last relation shows that $\upomega$
is torsionfree. Since it is metric compatible it is 
therefore the Riemannian connection on $G/H$. 
Equation \eqn{RPP} then relates the Riemannian 
curvature to the square of the vielbeins. 

\sk
By using the connection $\upomega$ and the vierbein $P$ we can 
construct couplings and actions invariant under the rigid $G$ and the local $H$ 
transformations, i.e. sigma models on the coset space $G/H$.

For example a kinetic term for the scalar fields, which are maps 
from spacetime to $G/H$, is given by pulling back to spacetime the invariant 
metric \eqn{metricGH} and then contracting it with the spacetime metric
\eq
\mathscr{L}_{\rm kin}(\phi)={1\over 2}P^a_\mu P_a^\mu={1\over 2}P^a_{~i}{\partial_\mu\phi^i
} P_{aj}{\partial^\mu\phi^j
}~.\label{TrPP}
\en
By construction the lagrangian  $\mathscr{L}_{\rm kin}(\phi)$ is invariant under $G$
and local $H$ transformations; it 
depends only on the coordinates of the coset space $G/H$.

\sk
\subsubsection{The case $G=Sp(2n,\rr)$, $H=U(n)$}
A kinetic term for the $Sp(2n,\rr)\over U(n)$ valued scalar fields is given 
by \eqn{TrPP}. This lagrangian is invariant under 
$Sp(2n,\rr)$  and therefore satisfies the duality condition \eqn{spdinvL} 
with $G=Sp(2n,\rr)$ and $\kappa=0$.
We can also write
\eq
\mathscr{L}_{\rm kin}(\phi)={1\over 2}P^a_\mu P_a^\mu={1\over 2}
{\rm Tr}(P_\mu P^\mu)~;\label{TrSPP}
\en
where in the last passage we have considered generators $K_a$ so that 
${\rm Tr}(K_a K_b)=\delta_{ab}$ (this is doable since $U(n)$ is 
the maximal compact subgroup of $Sp(2n,\rr)$). 
\sk
We now recall the representation 
of the group $Sp(2n.\rr)$ and of the associated coset 
$Sp(2n,\rr)\over U(n)$ in the complex basis discussed in the appendix 
(and frequently used in the later sections) and we give a more explicit 
expression for the lagrangian \eqn{TrSPP}.
\sk
Rather than using the symplectic matrix
$S=\big( {}^A_C {}^B_D\big)$ of the fundamental representation of 
$Sp(2n,\rr)$, we consider the conjugate matrix 
$\AA^{-1} S \AA$ where 
$
\AA={1\over \sqrt{2}}
\big( {}^{~\,1\!\!1}_{-i1\!\!1} {}^{~1\!\!1}_{\,\,i1\!\!1}\big)
$.
In this complex basis the subgroup $U(n)\subset Sp(2n,\rr)$ 
is simply given by the block diagonal matrices  $\big( {}^u_0 {}^0_{\bar u}\big)$.
We also define the $n\times 2n$ matrix
\eq
\left(\begin{array}{c}
f  \\
h
\end{array}\right)=
{1\over \sqrt{2}} \left(
\begin{array}{c}
A-iB\\
C-iD
\end{array}\right)~
\label{deffh} 
\en
and the matrix  
\eq
\VV=\left(\begin{array}{cc}
f  &\bar f\\
h  &\bar h
\end{array}\right)=\left(
\begin{array}{cc}
A & B\\
C & D
\end{array}\right){\cal A}~.
\en
Then (cf. \eqn{compsymp1}, \eqn{compsymp2}),
\eq
\VV^{-1}d \VV=
\left(
\begin{array}{cc}
i (f^\dagger dh-h^\dagger df) & i (f^\dagger 
d\bar h-h^\dagger d \bar f)   \\
-i (f^t dh-h^t df) & -i (f^t d\bar h - h^t d\bar f)
\end{array}\right)\equiv 
\left(
\begin{array}{cc}
\om & {\cal \bar P} \\
{\cal  P}  & \bar\om
\end{array}\right)~,
\label{GVdV}
\en
where in the last passage we have defined 
the $\nm\times \nm$ sub-blocks $\om$ and 
$\cal P$ corresponding to the $U(n)$ 
connection and the vielbein of $Sp(2n,\rr)/ U(n)$ in the complex basis,
(with slight abuse of notation we use the same letter $\om$ in this basis too).

We finally obtain the explicit expression
\eq \label{kin}
\mathscr{L}_{\rm kin}(\phi)={\rm{Tr}}({\cal \bar P}_\mu{\cal P}^\mu) ={1\over 4}
{\rm{Tr}}(\N_2^{-1}\partial_\mu\overline\N\;\N_2^{-1}\partial^\mu\N)
\en
where ${\cal P}={\cal P}_\mu dx^\mu={\cal P}_i\partial_\mu\phi^i dx^\mu$,
$\overline\N=\N_1-i\N_2$ and
$\N=\N_1+i\N_2=$ Re$\N+i$Im$N$. The matrix of scalars $\N$ parametrizes 
the coset space $Sp(2n,\rr)/U(n)$
(see Appendix 7.2); in terms of the $f$ and $h$ matrices it 
is given by (cf. \eqn{Nhf})
\eq
\N=fh^{-1}~,~~ \N^{-1}_2=-2ff^\dagger~.
\en
Under the symplectic rotation $\big( {}^A_C {}^B_D\big)\rightarrow 
\big( {}^{A'}_{C'} {}^{B'}_{D'}\big)\big( {}^A_C {}^B_D\big)$ the matrix $\N$ changes via
the fractional transformation $\N\rightarrow (C'+D'\N)\,(A'+B'\N)^{-1}$,
(cf. \eqn{sp2nrfract}).
\sk
Another proof of the invariance of the kinetic term \eqn{kin}
under the $Sp(2n,\rr)$
follows by observing that \eqn{kin} is obtained from the pullback to the spacetime manifold of
the  metric associated to the $Sp(2n,\rr)\over U(n)$ \K form
${\rm{Tr}}(\N_2^{-1} {\rm d}\,\overline\N\;\N_2^{-1} {\rm d}\,\N)$
(here ${\rm d}=\del+\overline\del$ is the exterior derivative). This metric is obtained from the \K potential 
\eq\label{kh}
\KK=-4\,{\rm Tr}\log\,i(\N-\overline \N)~.
\en 
Under the action of $Sp(2n,\rr)$, $\N$ and $\N-\overline \N$ change as in 
\eqn{sp2nrfract}, \eqn{Kahler}
and the \K potential changes by a \K transformation, 
thus  showing the invariance of the metric.
\sk
\sk
\noi{\bf{3.4.2~~~~{The case $G=\rr^{>0}\times Sp(2n,\rr)$, $H=U(n)$}}}
\sk

\noi In this case the duality rotation matrix $\big({}^a_c{}^b_d\big)$
belongs to the Lie algebra of $\rr^{>0}\times Sp(2n,\rr)$, as defined in \eqn{liedilsp}. In particular infinitesimal dilatations are given by the matrix
${\kappa\over 2}\big({}^{1\!\!1}_0{}^0_{1\!\!1}\big)$.
The coset space is
\eq\label{cs88}
{\rr^{>0}\times Sp(2n,\rr)\over U(n)}={\rr^{>0}\times}{Sp(2n,\rr)\over U(n)}~,
\en
there is no action of $U(n)$ on $\rr^{>0}$.
We consider a real positive scalar field $\Phi=e^\sigma$ 
invariant under $Sp(2n,\rr)$ transformations.
The fields $\Phi$ and $\N$ parametrize the coset space \eqn{cs88}.

Let's first consider the main example of Section 3.2. 
The duality symmetry conditions for the lagrangian \eqn{examplef2} 
are \eqn{DN1}-\eqn{spdinvL}.  
>From equations \eqn{DN1},\eqn{DN2} (that hold for 
$\big({}^a_c{}^b_d\big)$ in the Lie algebra of  $\rr^{>0}\times Sp(2n,\rr)$)
we see that the fields $\N$, and henceforth the lagrangian $\mathscr{L}_{\rm kin}(\phi)$, are invariant under the $\rr^{>0}$ 
action. 
It follows that  the scalar lagrangian
\eq\label{exception}
\Phi^{2}\mathscr{L}_{\rm kin}(\phi) +\partial_\mu\Phi\partial^\mu\Phi~
\en
satisfies the duality condition \eqn{spdinvL}.
This shows that the lagrangian \eqn{examplef2} with the scalar kinetic 
term given by \eqn{exception} has $\rr^{>0}\times Sp(2n,\rr)$ duality symmetry.
We see that in the lagrangian \eqn{examplef2} the scalar $\Phi$ does not couple to the field strenght $F$. The coupling of $\Phi$ to $F$ is however 
present in lagrangians where higher powers of $F$ are present.

More in general expression \eqn{exception} is a scalar kinetic term for lagrangians that satisfy the 
$\rr^{>0}\times Sp(2n,\rr)$ duality condition \eqn{thecondgen}.

\sk

\subsection{Invariance of energy momentum tensor}\label{enmomten}
Duality rotation symmetry is a symmetry of the equations of motion that does not leave invariant the lagrangian. The total change 
$\Delta\LL\equiv\LL(F',\ppi')-\LL(F,\ppi)$ of the lagrangian is  given in equation \eqn{thecondgen0}. Even if $\kappa=0$ this variation is not a total 
derivative because $F$ and $G$ are the curl of vector potentials $A_F$ and $A_G$ only on shell. 

  We show however that the variation of the action with respect to a duality rotation invariant parameter $\la$
is invariant under
$Sp(2n,\rr)$ rotations if the duality rotation \eqn{ppitrans} of the $\ppi$ fields is $\la$ independent.

Consider the $\la$-variation of $\Delta S[F,\ppi]\equiv S[F',\ppi']-S[F,\ppi]
=\int_y{\del \LL\over \del F}\Delta F+\Delta_\ppi S$,
\eqa
{\de\over\de\la}\Delta S &=&\int_y{\de\over\de\la}({\del \LL\over \del F})\;\Delta F
+\int_y{\del \LL\over \del F}{\de\over\de\la}(\Delta F) +
{\de \over\de\la}(\Delta_\ppi S)\nn \\[.3em]
& = &\int_y{\del\over\del F}({\de \LL\over \de \la})\;\Delta F
+{1\over 2}\int_y\tilde G {\de\over\de\la}(\Delta F) +
\Delta_\ppi({\de S\over\de\la} ) \nn\\[.3em]
& = &\Delta({\de S\over\de\la}) 
+{1\over 4} {\de\over\de\la} \int_y \tilde G\, b\,G 
\ena
where in the second line we used that ${\de \over\de\la}\Delta\ppi=0$. 
Thus $\Delta({\de S\over\de\la})= 
{\de\over\de\la}(\Delta S -{1\over 4} \int_y{\tilde G\, b\,G})$
and therefore from \eqn{thecondgen0} we have,
\eq
\Delta({\de S\over\de\la})= \kappa\, {\de S\over\de\la}
\en
thus showing invariance of ${\de S\over\de\la}$ under $Sp(2n,\rr)$ rotations ($\kappa=0$ rotations).
\sk
An important case is when $\la$ is the metric $g_{\mu\nu}$, this is invariant 
under duality rotations. This shows that the energy momentum tensor 
$\de S\over\de g_{\mu\nu}$ is invariant under $Sp(2n,\rr)$ duality rotations.
\sk
Another instance is when $\la$ is the dimensional parameter typically
present in a nonlinear theory. Provided the matter fields are properly 
rescaled $\ppi\rightarrow \hat\ppi=\la^s\ppi$, so that they become adimensional 
and therefore their transformation $\Delta\hat\ppi$, usually nonlinear,
does not explicitly involve $\la$,  then $\de S\over\de \la$ is invariant, where 
it is understood that ${\del \hat\ppi\over\del \la}=0$.

For the action of the Born-Infeld theory coupled to the  axion and 
dilaton fields,  $\LL={1\over \la}\big(1- \sqrt{1-{1\over 2}\la\N_2F^2-{1\over 16}\la^2\N_2(F\tilde F)^2}\,\big)$ we obtain the invariant 
${{\del \LL}\over{\del \la}}=-{1\over\la}(\LL-{1\over 4}F\tilde G)$;
we already found this invariant in \eqn{invLFG}.

\subsection{Generalized Born Infeld theory}

In this section we present the Born-Infeld theory 
with $n$ abelian gauge fields coupled to $n(n+1)/ 2$ scalar fields $\N$ 
and show that is has an $Sp(2n,\rr)$ duality symmetry. 
If we freeze the scalar fields $\N$ to the value $\N=-i1\!\!1$ 
then the lagrangian has $U(n)$ duality symmetry and reads
\eq
\LL=\Tr[1\!\!1-{\cal S}_{\al,\be}\sqrt{1\!\!1+2\al-\be^2}]~,\label{BITn}
\en
where as defined in \eqn{alphadef}, the componets of the $n\times n $ matrices $\al$ and $\be$ are
$
\alpha^{\La\Sigma} = \frac{1}{4} F^{\La} {F}^{\Sigma},
~\beta^{\La\Sigma} = 
\frac{1}{4}  \widetilde{F}^{\La} {F}^{\Sigma}.
$
The square root is to be understood in terms of its power series expansion,
and the operator ${\cal S}_{\al,\be}$ acts by symmetrizing each monomial 
in the $\al$ and $\be$ matrices. A world (monomial) in the letters $\al$ 
and $\be$ is symmetrized by averaging over all permutations of its letters.
The normalization of ${\cal S}_{\al,\be}$ is such that if $\al$ and $\be$
commute then ${\cal S}_{\al,\be}$ acts as the identity. Therefore in the case of just one abelian gauge field \eqn{BITn}
reduces to the usual Born-Infeld lagrangian.

The $Sp(2n,\rr)$
Born-Infeld lagrangian is obtained by coupling the lagrangian \eqn{BITn}
to the scalar fields $\N$
as described in Subsection \ref{csfndr} and explicitly considered in 
\eqn{BIrealfull}. 

Following \cite{ABMZ}  
we prove the duality symmetry of the Born-Infeld 
theory \eqn{BITn} by first showing that a  Born-Infeld  theory with $n$ 
{\sl complex} abelian gauge fields written in an auxiliary field formulation
has $U(n,n)$ duality symmetry. We then eliminate the auxiliary fields 
by proving a remarkable property of solutions of matrix equations \cite{ABMZ2}. Then we can consider real fields. 

\subsubsection{Duality rotations with complex field strengths}
>From the general study of duality rotations we know that a theory with 
$2n$ real fields $F_1^\La$ and $F_2^\La$ ($\La=1,\ldots n$) 
has at most $Sp(4n,\rr)$ duality if we consider duality rotations that leave 
invariant the energy-momentum tensor (and in particular the hamiltonian). 
We now consider the complex fields 
\eq
F^\La=F_1^\La+i F_2^\La~~,~~~~~\bar F^\La=F_1^\La-i F_2^\La~,
\en
the corresponding dual fields
\eq
G={1\over 2}(G_1+iG_2)~~,~~~~\bar G={1\over 2}(G_1-iG_2)\label{gcomp}~,
\en
and restrict the $Sp(4n,\rr)$ duality group to the subgroup of 
{\sl holomorphic} transformations,
\eqa
 \ddel \left( \matc F \\
G \emat \right)
&=&
\left( \matcc \asf & \bsf \\ \csf & \dsf \emat
\right) \left( \matc F \\ G \emat \right)\label{FGtransfhol}\\[.8em]
 \ddel \left( \matc \bar F \\
\bar G \emat \right)
&=&
\left( \matcc \bar\asf & \bar\bsf \\ \bar\csf & \bar\dsf \emat
\right) \left( \matc  \bar F \\ \bar G \emat \right)~.\label{FGtransahol}\\[-1.2em]\nn
\ena
This requirement singles out those matrices, acting on the vector 
{\footnotesize{${{\left( \matc F_1 \\ F_2\\ G_1\\
G_2 \emat \right)}}$}}},
that belong\\[-.9em] to the Lie algebra of $Sp(4n,\rr)$  and have the form

\eq
\left( \matcc \A {\footnotesize{\big( \matcc \asf & 0 \\ 0 & \bar\asf_{_{}} \emat
\big)}}\A^{-1}\; & {1\over 2}\A {\footnotesize{\big( \matcc \bsf & 0 \\ 0 & \bar\bsf \emat
\big)}}\A^{-1} \\[1.3em]
 2\A {\footnotesize{\big( \matcc \csf & 0^{{}} \\ 0 & \bar\csf \emat
\big)}}\A^{-1}\; &\; \A {\footnotesize{\big( \matcc \dsf & 0 \\ 0 & \bar\dsf \emat
\big)}}\A^{-1} \emat
\right)\label{unnsp4}
\en
where $
\AA={1\over \sqrt{2}}
\big( {}^{~\,1\!\!1}_{-i1\!\!1} {}^{~1\!\!1}_{\,\,i1\!\!1}\big)$.
The matrix \eqn{unnsp4} belongs to $Sp(4n,\rr)$ iff the $n\times n$ 
complex matrices $\asf, \bsf, \csf, \dsf$ satisfy
\eq
\asf^\dagger=-\asf~,~~\bsf^\dagger=\bsf~,~~\csf^\dagger=\csf~.\label{unn}
\en 
Matrices 
{\footnotesize{$\left( \matcc \asf & \bsf \\ \csf & \dsf \emat
\right)$}},
that satisfy \eqn{unn}, define the Lie algebra of the real form $U(n,n)$.
The group $U(n,n)$ is here the subgroup of $GL(2n,\cc)$ 
caracterized by the relations\footnote{In Appendix 7.1 we define $U(n,n)$ as the group of complex matrices that satisfy the condition 
$U^\dagger 
\big( {}^{1\!\!1}_{0} {}^{0}_{-1\!\!1}\big)
U=\big( {}^{1\!\!1}_{0} {}^{0}_{-1\!\!1}\big)$.
 The similarity transformation between these two definitions is 
$M=\A U\A^{-1}$.
} 
\eq
M^{\dagger}\,\left(\begin{array}{rr}
0 & -1\!\!1\\
1\!\!1 & 0
\emat
\right)  \,M = \left(\begin{array}{rr}
0 & -1\!\!1\\
1\!\!1 & 0
\emat
\right)~.
\label{MKMK}
\en
One can check that \eqn{MKMK} implies the following  relations for 
the block components of 
$
M={\left(\begin{array}{rr}
\Asf & \Bsf\\
\Csf & \Dsf
\emat
\right)}
$,
\eq
\Csf^{\dagger} \Asf = \Asf^{\dagger} \Csf~,~~ \Bsf^{\dagger} \Dsf = \Dsf^{\dagger} \Bsf~,
~~ \Dsf^{\dagger} \Asf-\Bsf^{\dagger} \Csf =1\!\!1~.
\label{sp2ncomp}
\en
The Lie algebra relations \eqn{unn} can be obtained from the
Lie group relations \eqn{sp2ncomp} by writing $\big({}^\Asf_\Bsf {}^\Csf_\Dsf\big)=\big({}^{1\!\!1}_0 {}^0_{1\!\!1}\big)+\epsilon\big({}^\asf_\csf {}^\bsf_\dsf\big)$ with $\epsilon$ infinitesimal. Equation  \eqn{unnsp4} gives the embedding of $U(n,n)$ in $Sp(4n,\rr)$.
\sk
The theory of holomorphic duality rotations can be seen as a special case
of that of real duality rotations, but 
(as complex geometry versus real geometry) it deserves also an independent 
formulation based on the holomorphic variables
{\footnotesize{$\left( \matcc F \\ G \emat
\right)$}} and maps {\footnotesize{$\left( \matcc \asf & \bsf \\ \csf & \dsf \emat
\right)$}}.

The dual fields in \eqn{gcomp}, 
or rather the Hodge dual of the dual field strength,
$\widetilde{G}^{~\mu \nu}_\La=
\frac{1}{2}\varepsilon_{\mu\nu\rho\sigma}G_\La^{~\rho\sigma}$,  is equivalently 
defined via

\beqn
\tilde{G}_{\La}^{~\mu\nu}\equiv 2\frac{\partial \Lag}
{\partial  \bar{F}^\La_{\,\mu\nu}}~,~~
\tilde{\bar{G}}_{\La}^{~\mu\nu}\equiv 2\frac{\partial \Lag}
{\partial  F^{\La}_{\,\mu\nu}}~.
\label{Gdef}
\eeqn
Repeating the passages of Section 3.1 we have that the 
Bianchi identities and equations of motion
${\pa}_{\mu}
{\tilde F}^{\La\,\mu\nu}  =0~,~~
{\pa}_{\mu}
\tilde G_{\!\La}^{~\mu\nu}=0~,~~
{\delta S[F,\bar F,\ppi]\over\delta \ppi^\al}=0~
$
transform  covariantly 
under the holomorphic infinitesimal transformations \eqn{FGtransfhol}
if the lagrangian satisfies the condition (cf. \eqn{thecondgen0})
\eq\label{thecondgen0comp}
\LL(F+\Delta F,\bar F+\Delta \bar F,\ppi+\Delta\ppi)-\LL(F,\bar F,\ppi)
-{1\over 2}  \tilde 
F\,\csf\, \bar F -{1\over 2} \tilde G\,\bsf\,\bar G ~=\; {\mbox{\it const}}_{\asf,\bsf,\csf,\dsf}
\en
Of course we can also consider dilatations $\kappa\not=0$, then 
in the left hand side of \eqn{thecondgen0comp} we have to add the 
term $-\kappa_{}\LL(F,\bar F,\ppi)$.

\sk
The maximal compact subgroup of $U(n,n)$  is $U(n)\times U(n)$
and is obtained by requiring \eqn{sp2ncomp} and
\[ 
\Asf= \Dsf~,~~ \Bsf= -\Csf~.
\]
The corresponding infinitesimal relations are \eqn{unn} and
$ 
\asf= \dsf\,,~ \bsf= -\csf~.
$
\sk
The coset space $U(n,n)\over {U(n)\times U(n)}$ is the space of all
negative definite hermitian matrices $\M$ of $U(n,n)$, see for 
example \cite{ABMZ} 
(the proof is similar to that for $Sp(2n,\rr)/U(n)$ in Appendix 7.2).  
All these matrices are for example of the form $\M=-{g^{\dagger}}^{-1} g^{-1}$
with $g\in U(n,n)$. 
These matrices can be factorized as
\eqa
\M&=&
\left(\begin{array}{cc}
1\!\!1 & -\N_1\\
0 & 1\!\!1
\end{array}\right)
\left(\begin{array}{cc}
\N_2 & 0  \\
0 & \N_2^{-1}
\end{array}\right)
\left(\begin{array}{cc}
1\!\!1 & 0\\
-\N_1^\dagger & 1\!\!1
\end{array}\right)
\nn\\
&=&
\left(\begin{array}{cc}
\N_2 +\N_1 \,\N_2^{-1}\, \N^\dagger_1 &~ -\N_1 \,\N_2^{-1}  \\
-\N_2^{-1}\,\N^\dagger_1 &~ \N_2^{-1}
\end{array}\right)\nn\\[.2em]
&=&
-i\left(\begin{array}{cc}
0 & -1\!\!1\\
1\!\!1& 0
\end{array}\right)
+\left(\begin{array}{cc}
\N \,\im \N^{-1}\, \N^\dagger &~ -\N \,\im \N^{-1}  \\
-\im \N^{-1}\,\N^\dagger &~ \im \N^{-1}
\end{array}\right)\nn\\[.2em]
&=&
-i\left(\begin{array}{cc}
0 & -1\!\!1\\
1\!\!1& 0
\end{array}\right)
+
\left(\begin{array}{cc}
\N & 0\\
-1\!\!1 & 0 
\end{array}\right)
\left(\begin{array}{cc}
\N_2 & 0  \\
0 & \N_2^{-1}
\end{array}\right)
\left(\begin{array}{cc}
\N^\dagger & -1\!\!1 \\
0& 0 
\end{array}\right)\label{MNcomp}
\ena
where $\N_1$ is hermitian, $\N_2$ is hermitian and negative definite,
and 
\eq
\N\equiv\N_1+i\N_2~.\label{decomp12}
\en
Since any complex matrix can always be decomposed into hermitian matrices as
in \eqn{decomp12}, the only requirement on $\N$ is that $\N_2$ is 
negative definite.

The left action of $U(n,n)$ on itself 
$g\rightarrow \tiny{\left(\begin{array}{rr}
\Asf & \Bsf\\
\Csf & \Dsf
\emat
\right)}g
$, induces the action on the coset space 
$\M\rightarrow \tiny{\left(\begin{array}{rr}
\Dsf & -\Csf\\
-\Bsf & \Asf
\emat
\right)}\,\M\,\tiny{\left(\begin{array}{rr}
\Dsf & -\Csf\\
-\Bsf & \Asf
\emat
\right)^\dagger}$ because
$\M=-{g^{\dagger}}^{-1} g^{-1}$. Expression  \eqn{MNcomp} 
then immediately gives the action of
$U(n,n)$ on the parametrization $\N$ of the coset space,
\eqa\label{Ncomptran}
\N\rightarrow \N'&=&(\Csf+\Dsf\N)\,(\Asf+\Bsf\N)^{-1}~,~~~\\[.3em]
\N_2\rightarrow \N_2'&=&(\Asf+\Bsf\N)^{-\dagger}\N_2(\Asf+\Bsf\N)^{-1}~.
\label{N2trancomp}
\ena
\sk
As in Section 3.4, given a theory depending on $n$ complex fields $F^\La$ and
invariant under the maximal compact duality group  $U(n)\times U(n)$ 
it is possible to extend it via the complex scalar fields $\N$, 
to a theory invariant under $U(n,n)$. 
The new lagrangian is
\eq
\LLL(F,R,\N_1) = \ll(RF)+{1\over 2}\tilde F\N_1 \bar F \label{LRNCOMP}
\en
where  $R=(R^\La_{~\Sigma})_{\La,\Sigma=1,...n}$ is now an arbitrary nondegenerate 
complex matrix. 
Because of the $U(n)$ maximal compact electric subgroup 
this new lagrangian depends only on the combination
\eq
\N_2=-R^\dagger R~,\label{N2RR}
\en
rather than on $R$. Thus $\LLL(F,R,\N_1)=\LLL(F,\N)$ where $\N=\N_1+i\N_2$.
A transformation for $R$ compatible with \eqn{Ncomptran} is
\eq
R' = R(\Asf +\Bsf \N )^{-1},
\en
whose infinitesimal transformation is $\Delta R = - R(\asf +\bsf \N)$\,.

Conversely, if we are given a Lagrangian $\Lag$ with equations of motion 
invariant under  $U(n,n)$ we can obtain a theory without the 
scalar field $\N$ by 
setting $\N=-i1\!\!1$. Then the duality group is broken 
to the stability group of $\N=-i1\!\!1$ which is $U(n)\times U(n)$, 
the maximal compact subgroup. 
\sk
Similarly to Section 3.4.1 we define 
the Lorentz invariant combinations
\beqn
\alpha_{{}_{}}^{ab} \equiv \frac{1}{2} F^{a} \bar{F}^{b},
~~\beta_{{}_{}}^{ab} \equiv 
\frac{1}{2}  \widetilde{F}^{a} \bar{F}^{b}.
\label{alphadefcomp}
\eeqn
If we consider lagrangians $\LL(F,\bar F) $ 
that depend only on gauge fields
and only through sum of traces (or of products of traces)
of monomials in $\alpha_{{}_{}}$ and $\beta_{{}_{}}$, then the necessary and sufficient 
condition for $U(n)\times U(n)$ holomorphic duality symmetry is still \eqn{bselfdual}, where  now $\al$ and $\be$ are as in \eqn{alphadefcomp}.

\subsubsection{Born-Infeld with auxiliary fields}
\label{BIaux}
A lagrangian that satisfies condition \eqn{thecondgen0comp} is
\beqn
\Lag={\rm Re}\,{\rm Tr}\,[\,
i(\N-\la)\chi -\frac{i}{2} \lambda \chi^\dagger \N_2 \chi
-i\lambda (\alpha_{} + i \beta_{})\,]~,\label{BIL}
\eeqn
The auxiliary fields $\chi$ and $\la$ and the scalar field $\N$ are $n$ 
dimensional complex matrices. We can also add to the lagrangian a duality 
invariant kinetic term for the scalar field $\N$, (cf \eqn{kin})
\eq \label{kincomp}
{\rm{Tr}}(\N_2^{-1}\partial_\mu\N^\dagger\;\N_2^{-1}\partial^\mu\N)~.
\en

In order to prove the duality of (\ref{BIL}) we first note 
that the last term in the Lagrangian can be written as
\[
-{\rm Re}\,{\rm Tr}\,[\,i\lambda (\alpha_{} + i \beta_{})\,]=
-{\rm Tr}(\lambda_2 \alpha_{} + \lambda_1\beta_{})~.
\]
If the field $\la$ transforms by fractional transformation and 
$\la_1$, $\la_2$ and the gauge fields are real this is the $U(1)^n$
Maxwell action \eqn{examplef2}, with the gauge fields interacting with the scalar
field $\la$. This term by itself has the correct
transformation properties under the duality group. 
Similarly for hermitian $\alpha$, $\beta$, $\lambda_1$ and $\lambda_2$ 
this term by itself satisfies equation~\rref{thecondgen0comp}.
It follows that the rest of the Lagrangian must be duality invariant.
The duality transformations of the scalar and auxiliary
fields are\footnote{In \cite{ABMZ} we use different notations: $\N\rightarrow S^\dagger, \la\rightarrow \la^\dagger, \chi\rightarrow \chi^\dagger,
\tiny{\left(\begin{array}{rr}
\Asf & \Bsf\\
\Csf & \Dsf
\emat
\right)\rightarrow \tiny{\left(\begin{array}{rr}
\Dsf & \Csf\\
\Bsf & \Asf
\emat
\right)  }}$.}
\eqa
\la'&=&(\Csf+\Dsf\la)\,(\Asf+\Bsf\la)^{-1}~,~~~\label{latr}\\[.3em]
\chi'&=&(\Asf+\Bsf\N)\chi(\Asf+\Bsf\la^\dagger)^\dagger \label{chitr}~,
\ena
and \eqn{Ncomptran}.
Invariance of ${\rm Tr}[ i(\N-\lambda)\chi]$
is easily proven by using \eqn{sp2ncomp} and by rewriting~\rref{latr} as
\eq
 \la'=(\Asf+\Bsf\la^\dagger)^{-\dagger}\,(\Csf+\Dsf\la^\dagger)^{\dagger}~.~~~\label{latrvar}
\en
Invariance of the remaining term which we write as
$
{\rm Re}\,{\rm Tr}\,[-\frac{i}{2} \lambda \chi^\dagger \N_2
  \chi]
={\rm Tr}\,[\frac{1}{2}\lambda_2 \chi^\dagger \N_2 \chi]~,
$
is straightforward by using \eqn{N2trancomp} and the following transformation
obtained from~\rref{latrvar},
\eq
\la_2'=(\Asf+\Bsf\la^\dagger)^{-\dagger}\la_2(\Asf+\Bsf\la^\dagger)^{-1}~.
\en

\subsubsection{Elimination of the Auxiliary Fields}
\label{NoAux}
The equation of motion obtained by varying $\la$ gives an equation for $\chi$,
\beqn
\chi+\frac{1}{2}\chi^\dagger\N_2  \chi +\alpha+i\beta=0~,
\label{EQ}
\eeqn
using this equation in the  Lagrangian (\ref{BIL}) we obtain 
\eqa
\Lag&=&{\rm Re}\,{\rm Tr}_{\,}\left(i \N\chi\right)~\\[.3em]
&=&{\rm Re}\,{\rm Tr}_{\,}\left(- \N_2\chi\right) + \Tr_{\,}(\N_1\be)~,
\label{BI2}
\ena
where $\chi$ is now a function of $\alpha$, $\beta$ and $\N_2$ that 
solves (\ref{EQ}). In the second line we observed that the anti-hermitian part of \eqn{EQ} implies $\chi_2=-\be$.

In this subsection we give the explicit expression of $\Lag$
in terms of $\alpha$, $\beta$ and $\N$.

First notice that (\ref{EQ}) can be simplified with
the following field redefinitions
\eqa
\widehat{\chi}&=&R\chi R^{\dagger}~, \nonumber \\
\widehat{\alpha}&=&R\alpha R^{\dagger}~,  \label{hatted}\\
\widehat{\beta}&=&R\beta R^{\dagger}~, \nonumber
\ena
where, as in~\rref{N2RR}, $R^{\dagger}R=-\N_2$.
The equation of motion for $\chi$ is then equivalent to
\beqn
\widehat{\chi}-\frac{1}{2}\widehat{\chi}^\dagger\widehat{\chi}
+\widehat\al-i\beh=0~.
\label{EQ'}
\eeqn
The anti-hermitian part of~\rref{EQ'} implies $\widehat\chi_2 = -\widehat\beta$\,, thus
$\widehat\chi^\dagger=\widehat\chi-2i\beta$. This can be used to eliminate $\widehat\chi^\dagger$ 
from~\rref{EQ'} and obtain a quadratic equation for $\widehat\chi$.
If we define $Q=\frac{1}{2}\widehat\chi$
this equation reads 
\beqn
Q=q+(p-q)Q+Q^2,
\label{Qeq}
\eeqn
where
\[
p \equiv -\frac{1}{2}(\alpha +i\beta) ~,~~
q \equiv -\frac{1}{2}(\alpha - i\beta)~. 
\]
The lagrangian is then
\beqn
\LL =  2\, {\rm Re}\,{\rm Tr}\,Q +\Tr_{\,}(\N_1\be)~.
\label{Lag}
\eeqn
If the degree of the matrices is one, we can solve for $Q$ in 
the quadratic equation~\rref{Qeq}. Apart from the fact that the gauge fields are complex,  
the result is the Born-Infeld Lagrangian coupled to the dilaton and axion 
fields $\N$,
\eq\label{BIusual}
\Lag=1-\sqrt{1-2 \N_2\alpha +{\N_2\!}^2 \beta^2} +\N_1\beta~.
\en
For matrices of higher degree, equation~\rref{Qeq} can be solved
perturbatively, 
\eq
Q_0=0~, ~~Q_{k+1}=q+(p-q)Q_k+Q_k^2~,\label{Qpertexp}
\en  
and by analyzing the first 
few terms in an expansion similar to \eqn{Qpertexp}  in~\cite{BMZ,ABMZ} it was conjectured 
that 
\beqn
{\rm Tr}\,Q =\frac{1}{2}\,{\rm Tr}\left[\,
1\!\!1+q-p -{\cal S}_{{p,q}}\sqrt{1\!\!1-2(p+q)+(p-q)^2}\,
\right]~,
\label{CONJ}
\eeqn
The right hand side formula is understood this way: first expand the square 
root as a power series in $p$ and $q$ assuming that $p$ and $q$ commute. Then
solve the ordering ambiguities arising from the noncommutativity of $p$ and $q$
by symmetrizing, with the operator  ${\cal S}_{p,q}$,
each monomial in the $p$ and $q$ matrices. A world 
(monomial) in the letters $p$ 
and $q$ is symmetrized by considering the sum of all the 
permutations of its letters, then normalize the sum by dividing 
by the number of permutations. This normalization of ${\cal S}_{p,q}$ is such 
that if $p$ and $q$
commute then ${\cal S}_{p,q}$ acts as the identity. Therefore in the case of just one abelian gauge field \eqn{BITn}
reduces to the usual Born-Infeld lagrangian.
An explicit formula for the coefficients of the expansion of
the trace of $Q$ is \cite{ABMZ2, ASVR}
\beqn
{\rm Tr}\,Q=
{\rm Tr}\left[\,
q+\sum_{r,s\geq 1}
\left(
\matc
r+s-2\\
r-1
\emat
\right)
\left(
\matc
r+s\\
r
\emat
\right)
{\cal S}(\,p^r q^s\,)
\,\right]~.
\label{pqLag}
\eeqn

In Appendix 8, following \cite{ABMZ2}, see also \cite{SchwarzA} 
and \cite{CerchiaiZumino}, we 
prove that the trace of $Q$ is completely symmetrized 
in the matrix coefficients $q$ and $p-q$.
Since this is equivalent to symmetrization in  $q$ and $p$ ~\rref{CONJ} follows.
Since symmetrization in $p$ and $q$ is equivalent to symmetrization in $\widehat\alpha$ and
$\beh$, the Born-Infeld lagrangian also reads
\eq
\LL=\Tr[1\!\!1-{\cal S}_{\al,\be}\sqrt{1\!\!1+2\alh-\beh^2}\,+\N_1\be]~.\label{BITnab}
\en

In \cite{ASVR} the convergence of perturbative matrix solutions of \eqn{EQ}, are studied. A sufficient condition for the convergence of the sequence 
\eqn{Qpertexp} to a solution of \eqn{Qeq} is that the norms of
 $p-q$ and $q$ have to satisfy  
$(1-||p-q||)^2>4||q||$. Here $||~||$ denotes any matrix norm with 
the Banach algebra property $||MM'||\leq ||M||\,||M'||$ (e.g. the usual norm). 
This condition is surely met if the field strengths $F_{\mu\nu}^\La$ are weak.

If equation \eqn{Qeq} is written as 
$(1\!\!1+q-p)Q=q+Q^2$, then 
the sequence given by 
$
Q_0=0~,~~Q_{k+1}=(1\!\!1+q-p)^{-1}q+(1\!\!1+q-p)^{-1}Q_{k}^2
$
converges 
and is a solution of equation \eqn{Qeq} if $||(1\!\!1+q-p)^{-1}||\,|| (1\!\!1+q-p)^{-1}q||<1/4$. 
Notice that the matrix $1\!\!1+q-p$ is always invertible, use
${1\over 2}(1\!\!1+q-p) +{1\over 2}(1\!\!1+q-p)^\dagger= 1\!\!1$,
and the same argument as in \eqn{fhpsi}. Notice also that if $p$ and $q$
commute then $\sqrt{1\!\!1-2(p+q)+(p-q)^2}=
(1\!\!1+q-p)\sqrt{1\!\!1-4(1\!\!1+q-p)^{-2}q}$
and convergence of the power series expansion of 
this latter square root  holds if 
 $||(1\!\!1+q-p)^{-2}q||<1/4$. 

\subsubsection{Real field Strengths }
\label{RealF}

We here construct a Born-Infeld 
theory with $n$ real field strengths which is duality invariant under the
duality group $Sp(2n,\rr)$. 

We first study the 
case without scalar fields, i.e.
$\N_1=0$ and $-\N_2=R=1\!\!1$.
Consider a Lagrangian $\LL=\LL(\al,\be)\,$ with $n$ complex gauge fields 
which describes a theory symmetric under the maximal compact group
 $U(N)\times U(N)$ of holomorphic duality rotations.
Assume that the Lagrangian is a sum of
traces (or of products of traces) of monomials in $\alpha$\, and $\beta$\,.
It follows that this Lagrangian satisfies the self-duality equations~\rref{bselfdual} with $\al$ and $\be$ complex (recall  
end of Section 3.7.1). This equation remains true in the special case that $\al$ 
and $\be$ assume real values. That is $\LL=\LL(\al,\be)$
satisfies the self-duality equation (\ref{bselfdual}) 
with $\al=\al^T=\bar{\al}$
and $\be=\be^T=\bar{\be}$. 
We now recall that equation (\ref{bselfdual}) is also 
the self-duality condition
for Lagrangians with real gauge fields provided that $\al$ and $\be$
are defined as in \eqn{alphadef} as functions of field strengths $F^\La$ that are real (cf. the different complex case definition \eqn{alphadefcomp}).
%
This implies that the theory described by the lagrangian 
$\LL(\al,\be)$ that is now function of $n$ real field strengths
is self-dual with duality group $U(n)$, the 
maximal compact subgroup of  
$Sp(2n,\rr)$.
The duality group can be extended to the full noncompact $Sp(2n,\rr)$, 
by introducing the symmetric matrix of scalar fields $\N$ 
via the prescription (\ref{extension}).

As a straighforward application we obtain the Born-Infeld Lagrangian with 
$n$ real gauge fields describing an  $Sp(2n,\rr)$ duality invariant theory
\eq
\Lag_{}={\rm Tr}\; [\, 1\!\!1 - 
{\cal{S}}_{_{\!\widehat{\al},\widehat{\be}}}\sqrt{1+2 \widehat{\al} - 
\widehat{\be}^2} +
\N_1\beta\, ]~, ~\label{BIrealfull}
\en
where $\alh=R\al R^t$, $\beh=R\be R^t$, $\N_2=-R^t R$, and 
$\al^{\La\Sigma}={1\over 4}F^\La F^\Sigma$, $\be^{\La\Sigma}={1\over 4}\tilde F^{\La} F^\Sigma$ as in \eqn{alphadef}.

\subsubsection{Supersymmetric Theory }
\label{SUSY}

In this section we briefly discuss supersymmetric versions of some of
the Lagrangians introduced. First we discuss the supersymmetric form
of the Lagrangian~\rref{BIL}.
Consider the superfields $V^\La=\frac{1}{\sqrt{2}}( V^\La_1+iV^\La_2)$ 
and  $\check{V}^\La=\frac{1}{\sqrt{2}}(V^\La_1-iV^\La_2)$ 
where $V^\La_1$ and $V^\La_2$ are real vector superfields, and define
\[
W^\La_{\alpha}=-\frac{1}{4}\bar{D}^2 D_{\alpha} V^\La~,~~
 \check{W}^\La_{\alpha}=-\frac{1}{4}\bar{D}^2 D_{\alpha} \check{V}^\La~.
\]
Both $W^\La$ and $\check{W}^\La$ are chiral superfields
and can be used to construct a matrix of chiral superfields
\[
{M}^{\La\Sigma}\equiv W^\La \check{W}^\Sigma~.
\]
The supersymmetric version of the Lagrangian~\rref{BIL} is then given by
\eq
\Lag={\rm Re}
\int \, d^2 \theta
\left[{\rm Tr}\,(
i(\N-\lambda)\chi -\frac{i}{2} \lambda \bar{D}^2(\chi^\dagger \N_2 \chi)
+ i  \lambda{ M}
)\right]~,\nonumber\label{BILSUSY}
\en
where $\N$, $\lambda$ and $\chi$ denote chiral superfields with the same
symmetry properties as their corresponding bosonic fields.
While the bosonic fields $\N$ and $\lambda$ appearing in~\rref{BIL} are 
the lowest component of the superfields denoted by the same letter, 
the field $\chi$  in the action~\rref{BIL} is the
highest component of the superfield $\chi$.
A supersymmetric kinetic term for the scalar field $\N$ can be written
using the K\"ahler potential~\rref{kh} as described 
in~\cite{Zumino1}. 

Just as in the bosonic Born-Infeld theory, one would like to eliminate the
auxiliary fields. This is an open problem if $n\not=1$.
For $n=1$ just as in the bosonic case the theory with auxiliary fields 
also admits both a real and a complex version, i.e. one can also consider a
Lagrangian with a single real superfield. Then by integrating 
out the auxiliary superfields  the supersymmetric
version of the Born-Infeld lagrangian~\rref{BIusual} is obtained  
\beqn
\Lag=\int \, d^4 \theta\frac{\N_2^2 W^2\bar{W}^2}{1+A+\sqrt{1+2A+B^2}}
+
{\rm Re}\left[
\int \, d^2 \theta (\frac{i}{2} \N W^2)\right]~,
\label{sbi}
\eeqn
where
\[
A=\frac{1}{4}(D^2(\N_2W^2)+\bar{D}^2(\N_2\bar{W}^2))~,~~
B=\frac{1}{4}(D^2(\N_2W^2)-\bar{D}^2(\N_2\bar{W}^2))~.
\]
If we only want a $U(1)$ duality invariance we can set $\N=-i$ 
and then the lagrangian~\rref{sbi} reduces to the supersymmetric Born-Infeld lagrangian described in~\cite{DP,CF,BG}. 

In the case of weak fields the
first term of~\rref{sbi} can be neglected and the Lagrangian is
quadratic in the field strengths. Under these conditions 
the combined requirements of
supersymmetry and self duality can be used~\cite{BG2} 
to constrain
the form of the weak coupling limit of the effective Lagrangian from
string theory. Self-duality of  Born-infeld theories with $N=2$ supersymmetries is discussed in \cite{Kuzenko:2000uh}.

\section{Dualities in $N>2$ extended Supergravities}

In this section we consider $N>2$ supergravity theories in $D = 4$; 
in these theories the graviton is also coupled to gauge fields and scalars. 
We study the corresponding duality groups, that are subgroups of the symplectic group. It is via the geometry of these subgroups of the symplectic group that we can obtain the scalars kinetic terms, the supersymmetry transformation rules and the structure of the central and matter charges of the theory with their differential equations and their duality invariant combinations $\VBH$ and $\SBH$ 
(that for extremal black holes are the effective potential and the entropy).

Four dimensional $N$-extended supergravities contain in the bosonic sector, besides the metric, a number $\nm$ of vectors and $m$ of (real) scalar  fields. The relevant bosonic action is known to have the following general form:
\begin{eqnarray}
 {\cal S}&=&{1\over 4}\int\sqrt{-g}\,
d^4x\left(-\frac{1}{2}\,R+\im{\cal N}_{\Lambda
\Gamma}F_{\mu\nu}^{\Lambda } F^{\Gamma \,\mu\nu}+
\frac{1}{2\,\sqrt{-g}}\,\rea{\cal N}_{\Lambda \Gamma  }
\epsilon^{\mu\nu\rho\sigma}\, F_{\mu\nu}^{\Lambda } F^{\Gamma
}_{\rho\sigma}+\right.\nonumber\\
& &~~~~~~~~~+\left.\frac{1}{2} \,g_{ij}(\phi) \partial_{\mu}
\phi^{i}\partial^{\mu}\phi^{j}\right)\,,\label{bosonicL}
\end{eqnarray}
where $g_{ij}(\phi)$ ($i,j,\cdots =1,\cdots ,m$) is the scalar
metric on the $\sigma$-model described by the scalar manifold ${
M}_{scalar}$ of real dimension $m$ and the vectors kinetic matrix
${\cal N}_{\Lambda\Sigma}(\phi) $ is a complex, symmetric, ${\nm} \,
\times \, {\nm}$ matrix depending on the scalar fields. The
 number  of vectors and scalars, namely 
$\nm$ and $m$, and the geometric properties of the scalar manifold
${ M}_{scalar}$ depend on the number $N$ of supersymmetries and
are summarized in Table \ref{topotable}.

The duality group of these theories is in general not the maximal one
$Sp(2n,\rr)$ because the requirement of supersymmetry constraints 
the number and the geometry of the scalar fields in the theory.
In this section we study the case where the scalar fields manifold is a coset 
space $G/H$, and we see that the duality group in this case is $G$.

\sk

In  Section 5 we then study the general $N=2$ case where the target space 
is a special K\"ahler manifold $M$ and thus in general we do not have 
a coset space.
There the  $Sp(2n,\rr)$ transformations are needed in order to globally define the supergravity theory. We do not have a duality 
symmetry of the theory; $Sp(2n,\rr)$ is rather a gauge symmetry 
of the theory, in the sense that only $Sp(2n,\rr)$ invariant 
expressions are physical ones.
\sk
The case of duality rotations in $N=1$ supergravity is  considered in 
\cite{cdfv}, \cite{Andrianopoli:2007rm}, see also \cite{Kuzenko:2002vk}. 
In this case 
there is no vector potential in the graviton multiplet hence no scalar 
central charge in the supersymmetry algebra. Duality symmetry is due to
the number of matter vector multiplets in the theory, the coupling to 
eventual chiral multiplets must be via a kinetic matrix $\N$ holomorphic 
in the chiral fields.  
We see that the structure of duality rotations is similar to that 
of $N=1$ rigid supersymmetry. 
For duality rotations in $N=1$ and $N=2$ rigid
supersymmetry using superfields see the review \cite{Kuzenko:2000uh}.

\subsection{Extended supergravities with target space $G/H$}
In $N\geq 2$ supergravity theories where the scalars target space is a coset 
$G/H$, the scalar sector has a Lagrangian invariant under the global $G$ 
rotations. Since the scalars appear in supersymmetry multiplets the symmetry $G$
should be a symmetry of the whole theory. This is indeed the case
and the symmetry on the vector potentials is duality symmetry.

Let's examine the gauge sector of the theory.
We recall from Section 3.1 that we have an $Sp(2n,\rr)$ duality group
if the vector $(^F_G)$ transforms in the fundamental of $Sp(2n,\rr)$, 
and the gauge kinetic term $\N$ transforms via fractional 
transformations,  if $\big(  _{C~D} ^{A~B} \big)\in Sp(2n,\rr)$,
\eq
\label{fracN}
\N\rightarrow \N'= (C+D\N)\, (A+B\N)^{-1}~.
\en
Thus in order to have $G$ duality symmetry, $G$ needs to act on the 
vector $(^F_G)$ via symplectic transformations, i.e. via matrices 
$\big(  _{C~D} ^{A~B} \big)$ in the fundamental of $Sp(2n,\rr)$.  
This requires a homomorphism 
\eq\label{Liesubgroup}
S : G\rightarrow Sp(2n,\rr)~.
\en
Different infinitesimal $G$  transformations should correspond to 
different infinitesimal symplectic rotations so that the induced map 
Lie$(G)\rightarrow$ Lie$(Sp(2n,\rr))$ is injective, and equivalently 
the homomorphism $S$ is a local embedding (in general $S$ it is not 
globally injective, the kernel of $S$ may contain 
some discrete subgroups of $G$).
 
Since $U(n)$ is the maximal compact subgroup of $Sp(2n,\rr)$
and since $H$ is compact, we have that the image of $H$ under 
this local embedding is in $U(n)$. It follows that we have a 
$G$-equivariant map
\eq\label{equivariant}
\N: G/H\rightarrow Sp(2n,\rr)/U(n)~,
\en 
explicitly,
for all $g\in G$,
\eq
\N(g\phi)=(C+D\N(\phi))\, (A+B\N(\phi))^{-1}~,
\en
where with $g\phi$ we denote the action of $G$ on $G/H$,
while the action of $G$ on $Sp(2n,\rr)/U(n)$ 
is given by fractional transformations. 
Notice that we have identified $Sp(2n,\rr)/U(n)$ with
the space of complex symmetric matrices $\N$ that have
imaginary part $\im \N=-i(\N-\overline{\N})$ negative definite 
 (see Appendix 7.2). 
\sk
The $D=4$ supergravity theories with $N>2$ have all target space $G/H$, they 
are characterized by the number $\nm$ of total vectors, the number $N$ of supersymmetries, and the coset space $G/H$, see Table 1$\,$\footnote{In Table 1 the group $S(U(p)\times U(q))$ is the 
group of block diagonal matrices 
$\big({}^P_0 {^{\,0}_Q}\big)$ with $P\in U(p)$, $Q\in U(q)$ and det$P\,$det$Q=1$.
There is a local isomorphism between 
$S(U(p)\times U(q))$ and the direct product group $U(1)\times SU(p)\times SU(q)$, 
in particular the corresponding Lie algebras coincide. Globally these groups
are not the same, for example $S(U(5)\times U(1))=U(5)=U(1)\times PSU(5)\not=U(1)\times SU(5)$. }.

{\footnotesize
\begin{table}[h]
\begin{center}
\caption{\sl Scalar Manifolds of $N>2$ Extended Supergravities}
\label{topotable}
\begin{tabular}{|c||c|c|c|c|c|}
\hline N &  Duality group $G$ & isotropy $H$ & ${M}_{scalar}$ & $\nm $&$ m$   \\[.6em]
\hline \hline
$3$  &  $SU(3,\pn)$ & $S(U(3) \times U(\pn))$
 & ${\frac{SU(3,\pn)}
{S(U(3)\times U(\pn))}}$ & $3+\pn$& $6\pn$ \\[.6em]
\hline $4$  &   $SU(1,1)\times SO(6,\pn)$ & $U(1)\times S(O(6) \times O(\pn))$ &
$\frac{SU(1,1)}{U(1)} \times \frac{SO(6,\pn)}{S(O(6)\times O(\pn))}$ & $6+ \pn$& $6\pn +2$ \\[.6em]
\hline $5$  &  $SU(5,1)$ &$S(U(5)\times U(1))$  & $\frac{SU(5,1)^{}}
{S(U(5)\times U(1))}$ & 10& 10 \\[.6em]
\hline $6$  &  $SO^\star(12)$ &$U(6)$ &
$\frac{SO^\star(12)}{U(6)}$ & 16& 30 \\[.6em]
\hline $7,8$&  $E_{7(7)}$ & $SU(8)/\mathbb{Z}_2$ &
$\frac{ E_{7(7)} }{SU(8)/\mathbb{Z}_2}$ & 28& 70 \\[.6em]
\hline
\end{tabular}
\end{center}
{\sl In the table, $n$ stands for the number of vectors and 
$m=\,$dim${\, M}_{scalar}$ for
the number of real scalar fields. In all the cases the duality
group $G$ is (locally) embedded in ${Sp}(2\,n,\mathbb{R})$. The number $n$ 
of vector potentials of the theory is given by $n= n_g + \pn $ where $\pn$ 
is the number of vectors potentials in the matter multiplet while $n_g$ 
is the number of graviphotons (i.e. of vector potentials that belong to the graviton multiplet). We recall that $n_g = \frac{N(N-1)}{2}$ if $N\not= 6$ ; 
and $n_g = \frac{N(N-1)}{2}+1=16$ if $N = 6$ ; we also have $\pn = 0$ if 
$N > 4$.
The scalar manifold of the $N=4$ case is usually written as 
$SO_o(6,n')/SO(6)\times SO(n')$ where $SO_o(6,n')$ is the component of 
$SO(6,n')$ connected to the indentity. 
The duality group of the $N=6$ 
theory is more precisely the double cover of $SO^*(12)$.
Spinors fields transform according to $H$ or its double cover.}
\end{table}
}
In general the isotropy group  $H$ is the product
\eq
H=H_{\rm Aut}\times H_{\rm matter}
\en
where $H_{\rm Aut}$ is the authomorphism group of the 
supersymmetry algebra, while $H_{\rm matter}$ depends on the 
matter vector multiplets, that are not present 
in $N>4$ supergravities.
\sk
In Section 3.5 we have described the geometry of the coset space $G/H$
in terms of coset representatives, 
local sections $L$ of the bundle $G\rightarrow G/H$.
Under a left action of $G$ they transform as
$g L(\phi)=L(\phi')h\,$, where the $g$ action on $\phi\in G/H$ 
gives the point $\phi'\in G/H$. 

We now recall that duality symmetry is implemented by the
symplectic  embeddings (\ref{Liesubgroup})  and (\ref{equivariant})
and conclude that the embeddings of the coset representatives 
$L$ in $Sp(2n,\rr)$ will play a central role. Recalling 
(\ref{deffh}) these embeddings are determined by defining
\eq
L\rightarrow f(L) ~~\mbox{ and } ~~L\rightarrow h(L)~.
\en

In the following we see that the matrices $f(L)$ and $h(L)$ 
determine the scalar kinetic term $\N$, the supersymmetry 
transformation rules and the structure of the central and 
matter charges of the theory. We also derive the differential equations
that these  charges satisfy  and  consider their 
 positive definite and duality invariant quadratic expression $\VBH$.
These relations are similar to the Special Geometry ones of $N=2$ supergravity.
\sk
>From the equation of motion 
\eqa
dF^\La  &=& 4\pi j^\La_m\\  
dG^\la  &=& 4\pi j_{e\La} 
\ena 
we associate with a field strength 2-form $F$  a magnetic charge $p^\La$  
and an electric charge $q_\La$  given respectively by: 
\eq
p^\La  = \frac{1}{4\pi} \int_{S^2} F^\La~~,~~~~
q_\La  = \frac{1}{4\pi} \int_{S^2} G_\La~~~~\label{pq}
\en
where $S^2$ is a spatial two-sphere containing these electric and magnetic charges. These are not the only charges of the theory, in particular we are interested in the central charges of the supersymmetry algebra and other charges related to the vector multiplets. These latter charges result to be the electric and 
magnentic charges $p^\La$  and $q_\La$  dressed with the scalar  fields of the theory. In particular these dressed charges are invariant under the duality group $G$ 
and transform under the isotropy subgroup $H = H_{Aut} \times  H_{matter}$.

While the index $\La$  is used for the fundamental representation of $Sp(2\nm ;R)$ 
the index $M$ is used for that of $U(\nm )$. According to the local embedding
\eq
H = H_{Aut} \times  H_{matter}\rightarrow U(\nm)
\en
the index $M$ is further divided as $M = (AB, \bar I)$ where  $\bar I$ refers to 
$H_{matter}$ and $AB=-BA~$ $(A=1,\ldots , N)$ labels the two-times antisymmetric representation of the $R$-symmetry group $H_{Aut}$.
We can understand the appearence of 
this representation of $H_{Aut}$ because this is a typical representation acting on the 
central charges. 
The index  $\bar I$ rather than $I$ is used 
because the image of $H_{matter}$ in $U(\nm)$
will be the complex conjugate of the fundamental of $H_{matter}$, 
this agrees with the property that under K\"ahler transformations of the 
$U(1)$ bundle $Sp(2\nm ,\rr)/SU(\nm ) \rightarrow Sp(2\nm,\rr)/U(\nm )$
the coset representatives of the scalar  fields in the gravitational 
and 
matter multiplets transform with opposite K\"ahler weights. 
This is also what happens in the generic $N = 2$ case (cf. \eqn{VdelbarVi}).\sk

The dressed graviphotons field strength 2-forms $T_{AB}$ may be identified from 
the supersymmetry transformation law of the gravitino 
field in the interacting theory, namely:   
\eq
\delta\psi_A = \nabla \epsi_A +  \al T_{AB\,\mu\nu}\gamma^a\gamma^{\mu\nu}    \epsi^B V_a +\ldots \label{tragra}
\en
Here $\nabla$ is the covariant derivative in terms of the space-time spin connection and the composite connection of the automorphism group $H_{Aut}$, $\al$  
is a coefficient  fixed by supersymmetry, $V^a$ is the space-time vielbein. 
Here and in the following the dots denote trilinear fermion terms which are characteristic of any supersymmetric theory but do not play any role in the following discussion. The $2$-form  field strength $T_{AB}$ is constructed by dressing the bare  field strengths $F^\La$  
with the image $f(L(\phi ))$, $h(L(\phi ))$ 
in $Sp(2\nm ;R)$ of the coset representative $L(\phi )$ of $G/H$. 
Note that the same  field strengths $T_{AB}$ which appear in the gravitino 
transformation law are also present in the dilatino transformation law in the following way:   
\eq
\delta\chi_{ABC} = {\cal P}_{ABCD\,\ell}\partial_\mu\phi^\ell\gamma^\mu\epsi^D  
+ \beta T_{[AB\,\mu\nu}\gamma^{\mu\nu}\epsi_{C]} \label{tradil}
\en 
Analogously, when vector multiplets are present, the matter vector  field strengths $T_I$ appearing in the transformation laws of the gaugino  fields, are linear combinations of the  field strengths dressed with a different combination of the scalars:
\eq
 \delta  \la_{IA} = i {\cal P}_{IAB\,r}\partial_\mu\phi^r \gamma^\mu\epsi^B  
+  \gamma T_{I\,\mu\nu}\gamma^{\mu\nu}\epsi_A +\ldots  \label{tragau}
\en 
Here ${\cal P}_{ABCD} = {\cal P}_{ABCD\,\ell}\,d \phi^\ell$ and 
${\cal P}^I_{ AB} = {\cal P}^I_{ AB\,r}\,d\phi^r$ 
are the vielbein of the scalar manifolds spanned by the scalar fields 
$\phi^i=(\phi^\ell,\phi^r)$  of the gravitational and vector multiplets respectively (more precise definitions are given below), and   $\be$ and $\ga$   are constants  fixed by supersymmetry. 
\sk
According to the transformation of the coset representative 
$g L(\phi)=L(\phi')h\,$, 
under the action of 
$g\in G$ on $G/H$ we have 
\eq
S(\phi )\A \rightarrow S(\phi')\A = S(g)S(\phi)S(h^{-1})\A
=S(g)S(\phi)\A U^{-1}\label{SphiA}
\en
where $
\AA={1\over \sqrt{2}}
\big( {}^{~\,1\!\!1}_{-i1\!\!1} {}^{~1\!\!1}_{\,\,i1\!\!1}\big)$ is unitary and symplectic (cf. (\ref{defAAm1})),
$S(g) = \big({}^A_C {}^B_D\big)$ and $S(h)$ are the 
embeddings of $g$ and $h$ in the fundamental of $Sp(2n,\rr)$, 
while $U=\A^{-1}S(h)\A$ is the embedding of $h$ in the complex basis of $Sp(2n,\rr)$. Explicitly $U=\big({}^u_0 {}^0_{\bar u})$, where $u$ is in the fundamental of $U(n)$
(cf. \eqn{USfh} and \eqn{UinUspSp}). Therefore the symplectic matrix 
\eq\label{Vphi}
\VV=S\A=
\left(
\begin{array}{cc}
f & \bar f  \\
h & \bar h
\end{array}\right)
\en
transforms according to
\eq\label{SonVphi}
\VV(\phi)\rightarrow \VV(\phi')=S(g)\VV(\phi)\left(
\begin{array}{cc}
u^{-1} & 0  \\
0 & \bar u^{-1}
\end{array}\right)~.
\en
The dressed field strengths transform only under a unitary representation of $H$ and, in accordance with (\ref{SonVphi}), are given by \cite{adf96}
\eq
\left(\begin{array}{c}
T \\
-\bar T 
\end{array}\right)=-i_{\,}\overline {\VV(\phi)}^{-1}
\left(\begin{array}{c}
F \\
G 
\end{array}\right)~;
\en
\eq
T\rightarrow\bar u T~.
\label{TshouldbecalledbarT}
\en
Explicitly, since 
\eq\label{vbarm}
-i_{\,}\bar \VV^{-1}=
\left(
\begin{array}{cc}
h^t & -f^t  \\
-h^\dagger & f^\dagger
\end{array}\right)
\en 
we have
\eqa\label{gravi}
T_{AB}&=& h_{\La\,AB}F^\La-f^\La_{~AB}G_\La\nn\\[.2em]
\bar T_{\bar I}&=&\bar h_{\La \bar I}F^\La-\bar f^\La_{\bar I} G_\La
\ena
where we used the notation 
$T = (T^{\bar M})=(T_M) = (T_{AB}, \bar T_{\bar I})$,  
\eqa\label{enphasizefh}
f &=& (f^\La_{~M}) = (f^\La_{~AB},   \bar f^\La_{~\bar I})~,\nn\\[.2em]
h &=& (h_{\La M}) = (h_{\La AB},   \bar h_{\La\bar I})~,
\ena
that enphasizes that 
(for every value of $\La$) the sections 
$\left(^{\bar f^\La_{~\bar I}}
_{\bar h_{\La\bar I}}\right)$ 
have \K weight opposite to the 
$\left(^{f^\La_{~AB}}_{h_{\La AB}}\right)$ ones.
This may be seen from the supersymmetry transformation rules of the supergravity fields, in virtue of the fact that gravitinos and fotinos with the same chirality have opposite K\"ahler weight.
Notice that  this notation (as in \cite{Andrianopoli:2006ub}) differs  from the one in \cite {adf96}, where $(f^\La_{~M}) = (f^\La_{~AB},   f^\La_{~ I})~,~~
(h_{\La M}) = (h_{\La AB}, h_{\La I})~.$
\sk

Consequently the central charges are
\eqa
Z_{AB}&=&-\frac{1}{4\pi}\int_{S^2_\infty} T_{AB}
=f^\La_{~AB}q_\La-h_{\La\,AB}p^\La\label{centralcharges1}\\
\bar Z_{\bar I}&=&-\frac{1}{4\pi}\int_{S^2_\infty} \bar{T}_{\bar I}=\bar f^\La_{\bar I}q_\La-\bar h_{\La\bar I\,}p^\La
\label{centralcharges2}
\ena
where the integral is considered at spatial infinity and, for spherically symmetric configurations, $f$ and $h$ in \eqn{centralcharges1}, \eqn{centralcharges2} are $f(\phi_\infty)$ 
and $h(\phi_\infty)$ 
with $\phi_\infty$ the constant value assumed 
by the scalar  fields at spatial infinity.

The integral of the graviphotons $T_{AB\,\mu\nu}$   gives the value of the central charges $Z_{AB}$ of the supersymmetry algebra, while by integrating the matter  field strengths $T_{I\,\mu\nu}$ one obtains the so called matter charges $Z_I$ . The charges of these dressed  field strength that appear in the supersymmetry transformations of the fermions have a profound meaning and play a key role in the physics of extremal black holes. 
In particular, recalling \eqn{SonVphi} 
the quadratic combination (black hole potential)
\eq
\mathscr{V}_{\!BH}:={1\over 2}\bar{Z}^{AB} {Z}_{AB}+\bar Z^IZ_I
\en
(the factor $1/2$ is due to our summation convention that treats the $AB$ indices as independent)
is invariant under the symmetry group $G$.
In terms of the charge vector
\eq
Q=\left(
\begin{array}{c}
p^\La \\
q_\La
\end{array}\right)~,
\en
we have the formula for the potential (also called charges sum rule)
\eq\label{mothersumrule}
\VBH={1\over 2}\bar{Z}^{AB} {Z}_{AB}+\bar Z^IZ_I=
-{1\over 2}Q^t{\cal M}({\cal N})Q
\en
where 
\eq
{\cal M}({\cal N})=-(i\bar \VV^{-1})^\dagger i\bar \VV^{-1}=-(S^{-1})^t S^{-1}~
\en is a negative definite 
matrix, here depending on $\phi_\infty$. 
In Appendix 7.2 we show that the set of matrices of the kind $SS^{t}$
with $S\in Sp(2n,\rr)$ are the coset space $Sp(2\nm,\rr)/U(\nm)$, hence the matrices ${\cal M}(\N)$ parametrize $Sp(2\nm,\rr)/U(\nm)$.
Also the matrices $\N$ parametrize $Sp(2\nm,\rr)/U(\nm)$. The relation between ${\cal M}(\N)$ and $\N$ is 
\eq
\M(\N)=
\left(\begin{array}{cc}
1\!\!1 & -\rea \N\\
0 & 1\!\!1
\end{array}\right)
\left(\begin{array}{cc}
\im \N & 0  \\
0 & \im \N^{-1}
\end{array}\right)
\left(\begin{array}{cc}
1\!\!1 & 0\\
-\rea \N & 1\!\!1
\end{array}\right)~.\label{Mlastbis}
\en
This and further properties of the ${\cal M}({\cal N})$ matrix are derived in 
 Appendix 7.2.
\sk
For each of the supergravities with target space $G/H$ there is another 
$G$ invariant expression $\mathscr{S}$ quadratic in the charges \cite{uinvar}; 
the invariant $\mathscr{S}$ is independent from the
scalar fields of the theory and thus depends only on the electric and 
magnetic charges $p^\La$ and $q_\La$. In extremal black hole configurations 
$\pi\mathscr{S}$ is the entropy of the black hole. In the $N=3$ 
supergravity theory $\mathscr{S}$ is the absolute value of a
quadratic combination of the charges, 
while for $N\geq 4$ it is the square root of the absolute value of a
quartic combination of the charges. The 
positive or negative value of this quadratic combination is related to the
 different BPS properties of the black hole. It turns out that $\mathscr{S}$
 coincides with the potential $\mathscr{V}_{\!BH}$ computed at its critical
 point (attractor point) \cite{fegika, FM, uinvar}.
In the next section we give the explicit expressions of the invariants $\mathscr{S}$.
They are obtained by considering among the $H$ invariant combination of the 
charges those that are also $G$ invariant, i.e. those that do not depend on the scalar fields. This is equivalent to require invariance of  $\mathscr{S}$
under the coset space covariant derivative $\nabla$ defined in Section 3.5, see also  
\eqn{diffrelwithoutindices}.

\sk\sk
We now derive some differential 
relations among the central and matter 
charges. We recall the symmetric 
coset space geometry $G/H$ studied in Section 3.5,
and in particular relations \eqn{RPP}, \eqn{RPPsec} that express 
the Maurer-Cartan equation $d\Gamma+\Gamma\wedge \Gamma=0$ in terms 
of the vielbein $P$ and of the Riemannian connection $\upomega$.
Using the (local) embedding of $G$ in $Sp(2n,\rr)$ we consider the pull 
back on $G/H$ of the $Sp(2\nm ,\rr)$ Lie algebra left 
invariant one form $V^{-1}dV$ given in \eqn{GVdV}, we have 
\eq
\VV^{-1}d \VV=
\left(
\begin{array}{cc}
i (f^\dagger dh-h^\dagger df) & i (f^\dagger 
d\bar h-h^\dagger d \bar f)   \\
-i (f^t dh-h^t df) & -i (f^t d\bar h - h^t d\bar f)
\end{array}\right)=
\left(
\begin{array}{cc}
\om & {\cal \bar P} \\
{\cal  P}  & \bar\om
\end{array}\right)~,
\label{GVdV1}
\en
where with slight abuse of notation 
we use the same letters $V$, ${\cal P}$ and $\om$ 
for the pulled back forms (we also recall that  ${\cal P}$
denotes $P$ in the complex basis).
Relation \eqn{GVdV1} equivalently reads 
\eq\label{GVdV2}
d\VV=\VV\left(
\begin{array}{cc}
\om & {\cal \bar P} \\
{\cal  P}  & \bar\om
\end{array}\right)~,
\en
that 
is equivalent to the  $\nm\times   \nm $ 
matrix equations:
\eqa
\nabla f &=&\bar f \,{\cal P}~,\label{CMuno}\\
\nabla h &=&\bar h \,{\cal P}~,\label{CMdue}
\ena
where
\eq\label{diffrelwithoutindices}
\nabla f=df-f\om~~,~~~~\nabla h=dh-h\om~.
\en
Recalling that ${\cal P}$ is symmetric 
(cf. \eqn{LieUsp}) we equivalently have  
$
\nabla f = {\cal P} \bar f\,,~
\nabla h ={\cal P} \bar h\, .
$ 
In these equations we can now see $\om$ and ${\cal P}$ as our 
data (vielbein and Riemannian connection) on a manifold $M$, 
while $f$ and $h$ are the unknowns. 
By construction these equations are 
automatically satisfied if $M=G/H$ and
$G$ is a Lie subgroup of $Sp(2n,\rr)$.
More in general equations \eqn{CMuno},\eqn{CMdue} hold (with $f$ and $h$ invertible) iff
the integrability condition, i.e. the Cartan-Maurer equation,
$d\big({}^\om_{\bar{\cal P}}{}^{\cal P}_\om\big)+\big({}^\om_{\bar{\cal P}}{}^{\cal P}_\om\big)\wedge \big({}^\om_{\bar{\cal P}}{}^{\cal P}_\om\big)=0$
holds. With abuse of terminology we sometimes call \eqn{CMuno}, \eqn{CMdue}
the Maurer-Cartan equations.

The differential relations among the 
charges $Z_{AB}$ and $\bar Z_{\bar I}$
follow after rewriting 
\eqn{CMuno}, \eqn{CMdue}
with $AB$ and $\bar I$ indices. 
The embedded connection $\om$ and vielbein ${\cal P}$ are decomposed as follows:
\eq
\om=(\om^N_{~M})=
\left(
\begin{array}{cc}
\om^{AB}_{~CD}& 0\\
0 & \om^{\bar I}_{\bar J}
\end{array}\right)~,
\en
\eq
{\cal P}=({\cal P}^{\bar N}_{~M})=({\cal P}_{NM})=
\left(
\begin{array}{cc}
{\cal P}^{\bar A \bar B}_{~CD}& 
{\cal P}^{\bar A \bar B}_{~\,\bar J} \\[.2em]
{\cal P}^{\bar I}_{~CD} & 
{\cal P}^{\bar I}_{\bar J}
\end{array}\right)=
\left(
\begin{array}{cc}
{\cal P}_{ABCD}& {\cal P}_{AB\bar J} \\[.2em]
{\cal P}_{I CD}  & {\cal P}_{I \bar J}
\end{array}\right)~,\label{subblokP}
\en
the subblocks being related to the vielbein of $G/H$, written in terms of the indices of $H_{Aut}\times   H_{matter}$. 
We used the following indices conventions:
\eq
f=\big(f^\La_{~M}\big)~~,~~~f^{-1}=\big(f^M_{~\,\La}\big)=(f_{\bar M\La}\big)~~~~
{\rm etc.}~
\en
where in the last passage, since we are in $U(\nm)$, we have lowered the index $M$ with the
$U(\nm)$ hermitian form 
$\eta=\big(\eta_{M\bar N}\big)_{M,N=1,...\nm}=diag({1,1,....1})$.
 Similar conventions hold for the $AB$ and $I$ indices, for example $\overline{f^\La_{~I}}=\bar f^\La_{~\bar I}=\bar f^{\La I}$.
\sk
Using further the index decomposition $M = (AB, \bar I)$, 
relations \eqn{CMuno}, \eqn{CMdue} read (the factor $1/2$ is due to our summation convention that treats the $AB$ indices as independent):
\eqa
\nabla f^\La_{AB}&=&
\frac{1}{2}\bar f^{\La CD}{\cal P}_{CDAB}
+f^\La_{~I}{\cal P}^I_{\;AB}~,\\[.2em]
\nabla h^\La_{AB}&=&
\frac{1}{2}\bar h^{\La CD}{\cal P}_{CDAB}
+h^\La_{~I}{\cal P}^I_{\;AB}~,\\[.2em]
\nabla f^\La_{~\bar I}&=
&\frac{1}{2} \bar f^{\La CD}{\cal P}_{CD \bar I}+ 
f^{\La \bar J}{\cal P}_{\bar J\bar I}~,\\[.2em]
\nabla h^\La_{~\bar I}&=
&\frac{1}{2} \bar h^{\La CD}{\cal P}_{CD \bar I}+ 
h^{\La \bar J}{\cal P}_{\bar J\bar I}~.
\ena
As we will see, depending on the coset manifold, some of the sub-blocks of \eqn{subblokP} can be actually zero. For $N > 4$ (no matter indices) we have that ${\cal P}$ 
coincides with the vielbein ${\cal P}_{ABCD}$ of the relevant $G/H$. Using the 
definition of the charges (21) we then get the differential 
relations among charges: 
$\nabla Z_M={\bar Z}_{\bar N}{\cal P}^{\bar N}_{~M}$, where 
$\nabla Z_{M}=\frac{\partial Z_{M}}{\partial \phi^i_\infty}\, d\phi^i_\infty -Z_N\om^N_{~M}$, with $\phi^i_\infty$ the value of the $i$-th
coordinate of $\phi_\infty\in G/H$ and $\phi_\infty=\phi(r=\infty)$. 
Explicitly, using the $AB$ and $I$ indices,
\eqa 
\nabla Z_{AB}&=&Z_{I}{\cal P}^I_{\;AB}+\frac{1}{2}\bar Z^{CD}{\cal P}_{CDAB}~,\\
\nabla \bar Z_{\;\bar I}&=
&\frac{1}{2} \bar Z^{AB}{\cal P}_{AB \bar I}+ 
Z^{\bar J}{\cal P}_{\bar J\bar I}~.
\ena
\sk
The geometry underlying the differential 
equation \eqn{GVdV2} is that of a flat 
symplectic vector bundle of rank $2\nm$, 
a structure that appears
also in the special \K manifolds of scalars 
of $N=2$ supergravities. 
Indeed if we are able to find $2\nm$ linearly independent row vectors $V^\xi=(V^\xi_{\,~\zeta})_{\zeta=1,...2\nm}$ then the matrix $\VV$ in \eqn{GVdV2}  is invertible and therefore the connection
$\big({}^\om_{\cal P}{\,}^{\bar{\cal P}}_{\bar\om}\big)$ is flat. If these vectors are mutually symplectic then we have a symplectic frame, the transition functions are constant symplectic matrices, the connection is symplectic.

 In the present case 
we naturally have a flat symplectic bundle, 
$$G\times_H \rr^{2\nm}\rightarrow G/H\,;$$
this bundle is the space of all equivalence classes $[g,v]=\{(gh,S(h)^{-1}v)\,,~g\in G, 
v\in \rr^{2\nm} , h\in H\}$. 
The symplectic structure on
$\rr^{2\nm}$ immediately extends to a 
well defined symplectic structure on the 
fibers of the bundle.
Using the local sections of
$G/H$ and the usual basis $\{e_\xi\}=
\{e_M,e^M\}$ of  
$\rr^{2\nm}$ 
($e_1$ is the column vector with with 1 as first and only nonvanishing entry, etc.)
we obtain immediately the local sections 
$s_\xi=[L(\phi),e_\xi]$ of $G\times_H \rr^{2\nm}\rightarrow G/H$.
Since the action of $H$ on $\rr^{2\nm}$ extends to the action  of $G$
on $\rr^{2\nm}$, we can consider the new sections 
$
\mathsf{e}_\xi=s_\zeta S^{-1}(L(\phi))^\zeta_{~\xi}
=[L(\phi),S^{-1}(L(\phi))e_\xi]~,
$ that are
determined by the column vectors 
$S^{-1}(L(\phi))_{~\xi}=
(S^{-1}(L(\phi))^\zeta_{~\xi})_{\zeta=1,...2\nm}$. These sections
are globally defined and linearly independent. Therefore this bundle is not only flat, 
it is trivial.
If we use the complex local frame
$\VVV_\xi=\{s_\zeta\A^\zeta_{~\xi}\}$ 
rather than the $\{s_\xi\}$ one 
(we recall that $\AA={1\over \sqrt{2}}
\big( {}^{~\,1\!\!1}_{-i1\!\!1} {}^{~1\!\!1}_{\,\,i1\!\!1}\big)$, 
 cf. \eqn{defAAm1}), then the global sections $\mathsf{e}_\xi$ are determined 
by the column vectors
$V^{-1}(L(\phi))_{~\xi}=(V^{-1}(L(\phi))^\zeta_{~\xi})_{\zeta=1,...2\nm}$, 
\eq
\mathsf{e}_\xi=\VVV_\eta\, {V^{-1}}^\eta_{~\xi}~.
\en
The sections $\VVV_\xi$ too form 
a symplectic frame (a symplectonormal basis, 
indeed $V^\rho_{~\xi}{\sf\Omega}_{\rho\sigma}
V^\sigma_{~\zeta}={\sf\Omega}_{\xi\zeta}$, 
where ${\sf\Omega}=
\big({}^0_{1\!\!1} {}^{-1\!\!1}_{~0}\big)$), and the last $\nm$ sections are the complex conjugate of the first $\nm$ ones, $\{\VVV_\xi\}=\{\VVV_M,\bar{\VVV}_{\bar M}\}$.
Of course the column vectors 
$V_{~\eta}=(V^\xi_{~\eta})_{\xi=1,...2\nm}$,
are the coefficients of the sections 
$\VVV_\eta$ with respect to the flat basis 
$\{\mathsf{e}_\xi\}$.

Also the rows of the $\VV$ matrix define
global flat sections. Let's consider the dual bundle 
of the vector bundle 
$G\times_H \rr^{2\nm}\rightarrow G/H$,
i.e. the bundle with fiber the dual 
vector space. If $\{s_\zeta\}$
is a frame of local sections of 
$G\times_H \rr^{2\nm}\rightarrow G/H$,
 then 
$\{s^\zeta\}$,  with $\le s^\zeta,s_\xi\re=\delta^\zeta_\xi$, is the dual frame of local 
sections of the dual bundle. Concerning the transition functions, if $s'_\zeta=s_\eta S^\eta_{~\zeta}$ then $s'^\xi={S^{-1}}^\xi_{~\lambda} s^\lambda$.
This dual bundle is also a trivial bundle  
and a trivialization is 
given by the global symplectic sections 
$\mathsf{e}^\xi=V^\xi_{~\eta}\VVV^\eta$,
whose coefficients are the row vectors 
$V^\xi=(V^\xi_{~\,\zeta})_{\zeta=1,...2\nm}$
i.e.,  the rows of the 
symplectic matrix $\VV$ defined in 
\eqn{Vphi},
\eqa
\left(V^\La_{\,~\zeta}\right)_{\zeta=1,...2\nm}&=&\left(f^\La_{~\, M},\bar f^\La_{~\,\bar M}\right)_{M=1,...\nm}~,\nn\\[.2em]
\left(V^{}_{\La\zeta}\right)_{\zeta=1,...2\nm}&=&\left(h_{\La M},\bar h_{\La\bar M}\right)_{M=1,...\nm}~.
\ena
\sk

\subsection{Specific cases}
We now  describe in more detail  
the supergravities of  Table \ref{topotable}.
The aim is to write down the group theoretical structure of each theory,
 their symplectic  (local) embedding 
$S\,:\, G\rightarrow Sp(2n,\rr)$ and 
$\N: G/H\rightarrow Sp(2n,\rr)/U(n)$, the vector kinetic
matrix $\N$, the supersymmetric transformation laws, the structure
 of the central and matter charges, their
differential relations
originating from the Maurer-Cartan equations \eqn{RPP},\eqn{RPPsec}, 
and the invariants $\VBH$ and $\SBH$.
As far as the boson transformation rules are concerned we prefer to write
 down the supercovariant definition of the
field strengths (denoted by a superscript hat), from which the supersymmetry transformation
laws are retrieved. 
 As it has been mentioned in previous section it is here that
the symplectic sections 
$( f^\Lambda _{~AB},
\bar f^\Lambda_{~\bar I},
\bar f^\Lambda_{~AB}, f^\Lambda_{~I} )$ 
appear as coefficients of the bilinear 
fermions in the supercovariant 
field strengths while the analogous 
symplectic section $(h_{\Lambda AB},
\bar h_{\Lambda \bar I},
\bar h_{\Lambda AB},
 h_{\Lambda I })$ would appear in 
the dual magnetic theory.
We include in the supercovariant field strengths also the supercovariant
vielbein of the $G/H$ manifolds.
Again this is equivalent to giving the susy transformation laws of the scalar fields.
The dressed field strengths from which the central and matter charges
 are constructed appear instead in the susy transformation laws of the
 fermions for which we give the expression up to trilinear fermion terms.
We stress that the numerical coefficients in the aforementioned
 susy transformations and supercovariant field strengths
are fixed by supersymmetry (or, equivalently, by Bianchi identities in superspace),
but we have not worked out the relevant computations
being interested in the general structure rather that
 in the precise numerical expressions.
 These 
numerical factors could
also be retrieved by comparing our formulae with those written
 in the standard literature on supergravity and performing the necessary
redefinitions.
The same kind of considerations apply to the central and matter
 charges whose precise normalization has not been fixed.
\sk\sk

Throughout this section we denote by $A,B,\ldots$ indices of $SU(N)$,
 $SU(N)\times U(1)$, being $H_{aut}$ the automorphism group of the $N$--extended
supersymmetry algebra. Lower and upper $SU(N)$ indices on the fermion fields are related to
 their left or right chirality respectively.
If some fermion is a $SU(N)$ singlet, chirality is denoted by the usual (L) or (R) suffixes.

Furthermore for any boson field $v$ carrying $SU(N)$ indices
 we have that lower and upper indices are related by complex conjugation,
namely:
$\overline{(v_{AB\cdots})} = \bar v^{AB\cdots}
$.

\sk
\subsubsection{The $N=4$ theory}
The field content is given by the
\sk
$- ~$ Gravitational multiplet (vierbein for the 
graviton, gravitino, graviphoton, dilatino, dilaton):
\begin{equation}
(V^a_\mu,\psi_{A\mu},A_\mu^{AB},\chi_{ABC},\ns) \quad\quad (A,B=1 ,\cdots,4)
\end{equation}
frequently the upper half plane parametrization $S=\bar \ns$ 
is used for the axion-dilaton field. 
\sk
$- ~$ Vector multiplets:
\begin{equation}
(A_\mu,\lambda^{A}, 6\,\phi)^I 
\quad \quad (I=1,\cdots,n)
\end{equation}

The coset space is the product
\begin{equation}
G/H= {SU(1,1)\over U(1)}\times {SO(6,n)\over {S(O(6) \times O(n))}}
\end{equation}
We have to embed 
\eq
Sp(2,\rr) \times SO(6,n)\rightarrow Sp(2(6+n),\rr)~.
\en
We first consider the embedding of $SO(6,n)$, 
\eqa
S : SO(6,n)&\rightarrow & Sp(2(6+n),\rr)\nn\\
L&\mapsto &S(L)=
\left(
\begin{array}{cc}
{L^t}^{-1} & 0  \\
0 & L
\end{array}\right)
\ena
we see that under this embedding $SO(6,n)$ is a symmetry 
of the action (not only of the equation of motions) that rotates 
electric fields into electric fields and magentic fields into 
magnetic fields. 
The natural embedding of 
$SU(1,1)\simeq SL(2,\rr)\simeq Sp(2,\rr)$ 
into $Sp(2(6+n),\rr)$ is the $S$-duality that rotates 
each electric field in its corresponding magnetic field,
we also want the image of $Sp(2,\rr)$ in $Sp(2(6+n),\rr)$
to commute with 
that of $SO(6,n)$ 
(since we are looking for a symplectic embedding 
of all $Sp(2,\rr)\times SO(6,n)$) and therefore 
we have
\eqa
S : Sp(2,\rr)&\rightarrow & Sp(2(6+n),\rr)\nn\\
\big(\begin{array}{cc}
_A & _B  \\
^C & ^D
\end{array}\big)
&\mapsto &S\big(\begin{array}{cc}
_A & _B  \\
^C & ^D
\end{array}\big)=
\left(\begin{array}{cc}
{{{\mbox{\footnotesize{$A$}}}}} 1\!\!1 & {{{\mbox{\footnotesize{$B$}}}}}\eta  \\
{{{\mbox{\footnotesize{$C$}}}}}\eta & {{{\mbox{\footnotesize{$D$}}}}}1\!\!1
\end{array}\right)
\ena
where $\eta=diag(1,1,...,-1,-1,...)$ 
is the $SO(6,n)$ metric.

Concerning the coset representatives, on 
one hand  we denote by $L(t)$ the 
representative in $SO(6,n)$ 
of the point $t\in SO(6,n)/S(O(6)\times O(n))$. 
On the other hand we have that 
$SU(1,1)/U(1)\simeq Sp(2,\rr)/U(1)$ is the lower half 
plane (see appendix) and is spanned by the complex number 
$\ns$ with $\im \,\ns<0$, (frequently the upper half plane parametrization $S=\bar \ns$ 
is used). A coset representative of $SU(1,1)/U(1)$ is
\eq
U(\ns)={1\over n(\ns)}\left(\begin{array}{cc}
1 & {{i-\bar\nt}\over{i+\bar\nt}} \\
{{i+\nt }\over{i-\nt}}   & 1
\end{array}\right)
~~,~~~n(\ns)=\sqrt{{-4\im \ns}\over{1+|\ns|^2-2\im\ns}}
\en
(In order to show that the $SU(1,1)$ matrix $U(\ns)$ 
projects to $\ns$ use \eqn{USfh} and \eqn{Nhf}, 
that reads $\ns=hf^{-1}$ with $h$ and $f$ complex numbers).
The coset representative $U(\ns)$ is defined for any 
$\ns$ in the lower complex plane and therefore $U(\ns)$ 
is a global section of the bundle $SU(1,1)\rightarrow 
SU(1,1)/U(1)$. 
(The projection $SU(1,1)\rightarrow SU(1,1)/U(1)$ can be 
also obtained by extracting $\ns$ from $\M(\ns)=
(^0_{1} {}^{\,-1}_{~0}) {\cal A} UU^\dagger{\cal A}^{-1}
( {}^{\,0}_{-1} {}^{\,1}_{\,0} )\,$, cf. \eqn{MversusS}).

\sk
With the given coset parametrizations the symplectic embedded
 section $\left(^{f^\Lambda_{~\Sigma}} _{h_{\Lambda \Sigma}}\right)$ is
\eqa
f^\Lambda_{\ \Sigma}&=&
(f^\Lambda_{~AB},\bar f^\La_{~\bar I})=
{1\over n(\ns)}\Big({2\over 1+i\ns} {{L^t}^{-1}\,}^\La_{~AB},
{{2\over 1-i\bar\ns}}
{{L^t}^{-1}\,}^\La_{~\bar I}\Big)
\nn\\
h_{\Lambda \Sigma}&=&
(h_{\Lambda AB},\bar h_{\La \bar I})=
{1\over n(\ns)}\Big({2\ns\over i\ns+1} {L}^\La_{~AB},
{{2\bar \ns\over i\bar\ns-1}}
{L}^\La_{~\bar I}\Big)
\label{n44}
\ena

We now have all ingredients to compute the matrix $\N$ in terms of $\ns$ and 
$L$. 
The coset representative in $Sp(2(6+m),\rr)$ of $(\ns,L)$
is 
$
S({\cal A} U(\ns){\cal A}^{-1})S(L)
$, and recalling that $\N=hf^{-1}$ and \eqn{deffh}, 
we obtain after elementary algebra the kinetic matrix
\eq
\N=\rea \N +i\im \N=\rea \ns\;\eta +i\im \ns\, LL^t~.
\en


\begin{table}[ht]
\caption{Group assignments of the fields in 
$D=4$, $N=4$}
\label{tab4,4}
\begin{center}
\begin{tabular}{|c||c|c|c|c|c|c|c|c|}
\hline
& $V^a_\mu$  & $\psi_{A \vert \mu}$ & $A^\Lambda_{\mu}$ & $\chi_{ABC}$ &
$\lambda_{IA} $ & $ U(\ns)L^\Lambda_{AB} $ & $U(\ns) L^\Lambda_I$ & $R_H$ \\
\hline
\hline
$SU(1,1)$ & 1 & 1 & - & 1 & 1 & $2\times 1$ &  $2\times 1$ & - \\
\hline
$SO(6,\pn)$ & 1 & 1 & $6+\pn$ & 1 & 1 & $1 \times (6+\pn)$ & $1 \times (6+\pn)$ & - \\
\hline
$SO(6)$ & 1 & 4 & 1 & $\bar 4 $ & $ \bar 4$ & $ 1\times 6$ & 1 & 6 \\
\hline
$SO(\pn)$ & 1 & 1 & 1 & 1 & $\pn$ & 1 & $\pn$ & $\pn$ \\
\hline
$U(1)$& 0 & ${1\over 2}$ & 0 & ${3\over 2}$ & $ -{1\over 2}$ & 1 & 1 & 0 \\
 \hline
\end{tabular}
\end{center}
{\sl In this and in the following tables, $R_H$  is the representation  under which 
the scalar fields of the linearized theory, 
or the vielbein $\cal P$ of $G/H$ of the full 
theory transform (recall text after \eqn{HKK} and that ${\cal P}$
is $P$ in the complex basis).
Only the left--handed fermions are quoted, 
right handed fermions transform in the complex conjugate representation of $H$.  Care must be taken in the transformation properties under the $H$ subgroups; indeed according to 
\eqn{SonVphi} the inverse right rep. of the one listed should really appear, i.e. since we are dealing with unitary rep., the complex conjugate}
\end{table}
The supercovariant field strengths and 
the vielbein of the coset manifold are:
\begin{eqnarray}
\hat F^\Lambda &=& dA^\Lambda +\bigl [
f^\Lambda_{AB}(c_1 \bar \psi^A \psi ^B + c_2 \bar \psi_C \gamma_a
\chi^{ABC}V^a)\nonumber\\
& & + f ^\Lambda_I (c_3 \bar\psi^A
\gamma_a \lambda^I_A V^a + c_4 \bar \chi^{ABC} \gamma_{ab} \lambda^{ID}\epsilon_{ABCD}V^a V^b) + h.c. \bigr ]\\
\hat {\Ps} &=&  {\Ps} - \bar \psi_A \chi_{BCD} \epsilon^{ABCD}\\
\hat {\cal P}^I_{AB}&=& {\cal P}^I_{AB} - (\bar\psi_A \lambda^I_B +\epsilon_{ABCD} \bar\psi^C\lambda^{ID})\\
\end{eqnarray}
where ${\Ps}= {\Ps}_{\!\ns\,} d\ns$ and ${\cal P}^I_{AB} ={\cal P}^I_{AB\, i\,}d\phi^i $ are the vielbein of ${SU(1,1)\over U(1)}$ and ${SO(6,\pn) \over S(O(6)
     \times O(\pn))}$
respectively.
The fermion transformation laws are:
\begin{eqnarray}
\delta\psi_A &=& D \epsilon_A +a_1 T_{AB \, \mu\nu}\ga^a\ga^{\mu\nu} \epsilon^B V_a
 + \cdots\\
\delta\chi_{ABC} &=&a_2 \Ps_{_{\!\!\ns\,}}\partial_\mu \ns\, \gamma^\mu \epsilon^D \epsilon_{ABCD} +
a_3 T_{[AB \, \mu\nu}\gamma^{\mu\nu} \epsilon_{C]} + \cdots\\
\delta\lambda^I_A &=&a_4 {\cal P}^I_{AB\,i\,}\partial_a \phi^i \gamma^a \epsilon^B + a_5 T^{- I}_{\mu\nu} \gamma^{\mu\nu}\epsilon_A + \cdots
\end{eqnarray}
where the 2--forms $T_{AB}$ and $T_{I}$ are defined in eq.(\ref{gravi}).
By integration of these two-forms
we find the  central and matter dyonic charges
given in equations (\ref{centralcharges1}), (\ref{centralcharges2}).
>From the equations \eqn{CMuno},\eqn{CMdue} for $f,h$ and the
definitions of the charges one easily finds:
\begin{eqnarray}
  \label{charge4,4}
  \nabla^{ SU(4)\times U(1)} Z_{AB} &=& \bar  Z^I {\cal P}_{I AB} + {1 \over 2} \epsilon_{ABCD}\bar Z^{CD}\Ps  \\
  \nabla^{ SO(\pn)} Z_{I} &=&{1 \over 2} \bar  Z^{AB} {\cal P}_{I AB} + Z_I
  \bar \Ps
\end{eqnarray}
where ${1\over 2}\epsi_{ABCD}\bar Z^{CD}=\bar Z_{AB}$.
In terms of the kinetic matrix (\ref{n44}) the invariant $\VBH$  
for the charges is given by, cf. (\ref{mothersumrule}),
 \begin{equation}
\VBH={1 \over 2} Z_{AB} \bar Z^{AB} + Z_I \bar Z^I = -{1\over 2} Q^t \cM (\cN) Q~.
\end{equation}

The  unique   $SU(1,1)  \times SO(6,\pn) $  invariant 
combination of the charges that is independent from the scalar fields is 
$I_1^2 - I_2\bar{I_2}$, so that 
\begin{equation}
\mathscr{S}= \sqrt{|I_1^2 - I_2\bar{I_2}|}~.
\label{invar4}
\end{equation}
Here,
 $I_1$, $I_2$ and $\bar I_2$ are the three $SO(6,\pn)$ invariants given by
\eq
  I_1 = {1 \over 2}  Z_{AB} \bar Z^{AB} - Z_I \bar Z^I~~,~~~~
I_2 = {1\over 4} \epsilon^{ABCD}  Z_{AB}   Z_{CD} - \bar Z_I \bar Z^I~.
\label{invar42}
\en

\subsubsection{The $N=3$ theory}
In the $N=3$ case \cite{Castellani:1985ka}
the coset space is:
\begin{equation}
G/H= {SU(3,\pn)\over {S(U(3)\times U(\pn))}}
\end{equation}
and the field content is given by:
\begin{eqnarray}
&(V^a_\mu, \psi_{A \mu}, A^{AB}_\mu , \chi_{(L)}) \quad\quad  A=1,2,3 \quad\quad
\hbox{(gravitational multiplet)}& \\
&(A_\mu , \lambda_A, \lambda_{(R)} , 3\,z)^I  \quad \quad I=1,\ldots,\pn \quad \quad
\hbox{(vector multiplets)}&
\end{eqnarray}
The transformation properties of the fields are given in Table \ref{4,3}. 
\begin{table}[ht]
\caption{Transformation properties of fields in $D=4$, $N=3$}
\label{4,3}
\begin{center}
\begin{tabular}{|c||c|c|c|c|c|c|c|c|c|}
\hline
& $V^a_\mu$ & $\psi_{A\mu}$ & $A^\Lambda_\mu$ & $\chi_{(L)}$
& $\lambda^{I}_A$ &
$\lambda^I_{(L)}$ &$ L^\Lambda_{AB} $&$ L^\Lambda_I $& $R_H$ \\
\hline
\hline
$SU(3,\pn)$ & 1  & 1 & $3+\pn$ & 1 & 1 & 1 & $3+\pn$& $3+\pn$ & -  \\
\hline
$SU(3)$ & 1 & 3 & 1 & 1 & 3 & 1 & ${\bar 3}$ & 1 & $3$  \\
\hline
$SU(\pn)$ & 1  & 1 & 1 & 1 & $\pn$ & $\pn$ & 1 & $\pn$ & $\pn$ \\
\hline
  $U(1)$ & 0  & ${\pn\over 2}$ & 0 & 3${\pn\over 2}$ & 3+${\pn\over 2}$
  &$-3(1  + {\pn\over 2})$ & $\pn$ & $-3$ &$3+\pn$ \\
\hline
\end{tabular}
\end{center}
\end{table}
We consider the (local) embedding of $SU(3,\pn)$ 
in $Sp(3+\pn,\rr)$
defined  by the following dependence of  
the matrices $f$ and $h$ in terms of the $G/H$ coset representative $L$,
\begin{eqnarray}
f^\Lambda_{\ \Sigma}&=& \frac{1}{\sqrt{2}} (L^\Lambda_{\ AB},\bar L^\Lambda_{\ I})
\label{n3f}  \\
h_{\Lambda \Sigma}&=&-{ i}(\eta f\eta)_{\Lambda\Sigma} \quad \quad \quad \quad
\eta= \left(\begin{array}{cc}
1\!\!1_{3\times 3}&0\\ 0 &-
1\!\!1_{\pn\times \pn}\end{array} \right) 
\label{n3h}
\end{eqnarray}
where $AB$ are antisymmetric $SU(3) $ indices, $I$ is an index of
$SU(\pn)$ and $\bar L^\Lambda_{\ I}$ denotes the complex conjugate
of the coset representative.
We have:
\begin{equation}
 \cN_{\Lambda\Sigma} = (h f^{-1})_{\Lambda \Sigma}
 = -{i}(\eta f\eta f^{-1})_{\Lambda\Sigma} \label{kin2}
\end{equation}
The supercovariant field strengths and the supercovariant scalar vielbein
are:
\begin{eqnarray}
\hat F^\Lambda &=& dA^\Lambda +\big[\, {{i}\over 2} f^{\Lambda }_I
\bar \lambda^I_A \gamma_a \psi^A V^a - {1\over 2} f^{\Lambda }_{AB}
 \bar \psi ^A \psi^B + { i}f^\Lambda_{AB}\bar
\chi_{(R)}\gamma_a \psi_C \epsilon^{ABC} V^a \nonumber +h.c.\big]\\
\hat  {\cal P}^{\ A}_I &=& {\cal P}^{\ A}_{I} - \bar\lambda^I_B \psi_C
 \epsilon^{ABC} - \bar \lambda_{I(R)}\psi^A
\end{eqnarray}
where the only nonvanishing entries of the vierbein ${\cal P}$ are 
\begin{eqnarray}
  {\cal P}^{ A}_I & = & {1 \over 2}  \epsilon^{ABC}{\cal P}_{I BC}   
= {\cal P}^{A}_{I\,i\,}dz^i 
\end{eqnarray}
$z^i$ being the (complex) coordinates of $G/H$. 
The chiral fermions transformation laws are given by:
\begin{eqnarray}
\delta \psi_A &=& D \epsilon_A + 2{ i} T^{}_{AB \,
\mu\nu} \ga^a\ga^{\mu\nu}V_a \epsilon^B + \cdots \\
\delta \chi_{(L)} &=& 1/2\, T^{} _{AB \, \mu\nu} \gamma^{\mu\nu} \epsilon_C
\epsilon^{ABC} + \cdots \\
\delta\lambda_{IA} &=& -{ i} {\cal P}_{I\ \,i}^{\ B}\partial_\mu z^i\gamma^\mu \epsilon^C \epsilon_{ABC}
+ T_{I \, \mu\nu} \gamma^{\mu\nu} \epsilon_A + \cdots
\\
 \delta\lambda^I_{(L)} &=& { i} {\cal P}_{I\ \,i}^{\ A} \partial_\mu z^i\gamma^\mu \epsilon_A
+ \cdots
\end{eqnarray}
 where $T_{AB} $ and $T_{I}$ have the general form given
 in equation (\ref{gravi}).
>From the general form of the equations \eqn{CMuno}, \eqn{CMdue}
for $f$ and $h$ we find:
\eqa
\nabla f^\La_{AB}&=&
f^\La_{~I}{\cal P}^I_{\;AB}~,\\[.2em]
\nabla h^\La_{AB}&=&
h^\La_{~I}{\cal P}^I_{\;AB}~,\\[.2em]
\nabla f^\La_{~\bar I}&=
&\frac{1}{2} \bar f^{\La CD}{\cal P}_{CD \bar I}~,\\[.2em]
\nabla h^\La_{~\bar I}&=
&\frac{1}{2} \bar h^{\La CD}{\cal P}_{CD \bar I}~.
\ena
According to the general study of  Section 4.1,
using (\ref{centralcharges1}), (\ref{centralcharges2}) one finds
 \begin{eqnarray}
\nabla^{(H)} Z_{AB} &=& \bar Z^I
{\cal P}_I^{\ C} \epsilon_{ABC}   \\
\nabla^{(H)} Z_{I} &=&{1\over 2} \bar Z^{AB}
{\cal P}_I^{\ C} \epsilon_{ABC}
\end{eqnarray}
and the formula for the potential, cf. \eqn{mothersumrule},
\begin{equation}
\VBH=      {1  \over 2}Z ^{  AB}\bar Z_{AB} + Z^{ I} \bar Z_{I} = -{1 \over 2}Q^t\cM(\cN)Q
\end{equation}
where the matrix $\cM(\cN)$ has the same form as in equation (\ref{Mlastbis}) in terms of the kinetic
matrix $ \cN$ of equation (\ref{kin2}), and $Q$ is the charge vector $Q=\left(^g _e \right)$.

\sk
The $G=SU(3,\pn)$ invariant  is $Z^A \bar Z_A - Z_I \bar Z^I$
(one can check that $ {\partial_i} (Z^A \bar Z_A - Z_I \bar Z^I)=
\nabla^{(H)}_i (Z^A \bar Z_A - Z_I \bar Z^I) =0 $)
so that 
\begin{equation}
\mathscr{S}=   |Z^A \bar Z_A - Z_I \bar Z^I|~.
\label{invar3}
\end{equation}

\sk
\subsubsection{The $N=5$ theory} 
For $N>4$ the only available supermultiplet is the gravitational one, so that $H_{\rm matter}=1$. The coset manifold of the scalars of the 
$N=5$ theory 
\cite{dwni} is:
\begin{equation}
G/H = {SU(5,1) \over U(5) }
\end{equation}
The field content and the group assignments are displayed in Table
\ref{tab4,5}.
\begin{table}[ht]
\caption{Transformation properties of fields in $D=4$, $N=5$}
\label{tab4,5}
\begin{center}
\begin{tabular}{|c||c|c|c|c|c|c|}
\hline
& $V^a$ &$ \psi _A; $ &$\chi_{ABC},\chi_L $
&$A^{\Lambda\Sigma}$ &$L^{x}_{A  }$ & $R_H$  \\
\hline
\hline
$SU(5,1)$& 1 & 1 & 1 & -          &6   & -     \\
\hline
$SU(5)$ & 1 & 5 & $(10,1)$ & 1 & 5 &${\bar 5}$ \\
\hline
$U(1)$ & 0 & ${1\over 2}$ & $({3\over 2}, - {5\over 2}
)$ & 0 & 1 & 2 \\
\hline
\end{tabular}
\end{center}
\end{table}
\noindent

In Table \ref{tab4,5} the incides $x,y,\ldots = 1,\ldots,6$ and  $A,B,C,\ldots
=1,\ldots,5$ are  indices of the fundamental representations of $SU(5,1)$ and
 $ SU(5)$, respectively.
$L^x_A$ denotes as usual the coset representative in the
fundamental representation of $SU(5,1)$. The antisymmetric couple $\Lambda
\Sigma$, $\Lambda,
\Sigma = 1,\ldots,5$, enumerates the ten vector potentials.
 The local embedding
of $SU(5,1)$ into  the Gaillard-Zumino  group $Usp(10,10)$ is given in terms
of the three-times antisymmetric representation of $SU(5,1)$, this is a 20 dimensional 
complex representation,  we denote by 
$t^{xyz}$ a generic element. This representation is reducible to a complex 10 dimensional 
one by imposing the self-duality condition
\begin{equation}
  \label{txyz}
  \bar t^{\bar x\bar y\bar z}={1 \over 3!} \epsilon^{\bar x\bar y\bar z}_{~~~~uvw}t^{uvw}
\end{equation}
here indices are raised with the $SU(5,1)$
hermitian structure $\eta=diag(1,1,1,1,1,-1)$.
The self duality condition \eqn{txyz} is compatible with the $SU(5,1)$ action (on 
$\bar t^{\bar x\bar y\bar z}$ acts the complex conjugate of the three-times antisymmetric of $SU(5,1)$).
Due to the self-duality condition 
we can decompose $t^{xyz}$ as follows:
\begin{equation}
 \label{txyz1}
  t^{xyz}=
\left(
\begin{array}{c}
t^{\Lambda\Sigma 6}\\
\bar t^{\bar\Lambda\bar\Sigma\bar 6}
\end{array}\right)
\end{equation}
where $(\Lambda,\Sigma,\cdots= 1,\cdots,5)$.
In the following we set
$t^{\Lambda\Sigma}\equiv t^{\Lambda\Sigma 6}$,
$\,\bar t^{\bar\Lambda\bar\Sigma}\equiv 
\bar t^{\bar\Lambda\bar\Sigma\bar 6}$, 
$\,\bar t_{\La\Sigma}\equiv 
\bar t_{\La\Sigma 6}=-\bar t_{\La\Sigma}^{~~~\bar 6}$.
The symplectic structure in this complex 
basis is given by the matrix  $\left(
\begin{array}{cc}
0 & -1\!\!1\\
1\!\!1 & 0
\end{array}\right)$,
\eqa
\le t,\ell\re &:=& 
{1\over 2}\big(t^{\La\Sigma}, 
\bar t^{\bar\La\bar\Sigma}\big)
\left(
\begin{array}{cc}
0 & -\delta_{\La\Sigma\,\bar\Ga\bar\Pi}\\
\delta_{\bar\La\bar\Sigma\,\Ga\Pi}& 0
\end{array}\right)
\left(
\begin{array}{c}
\ell^{\Ga\Pi}\\
\ell^{\bar\Ga\bar\Pi} 
\end{array}\right)\\
&=&
{1\over 2}t^{\La\Sigma}\bar\ell_{\La\Sigma}
-{1\over 2}\bar t_{\La\Sigma}
\ell^{\La\Sigma}\nn\\
&=&
{1\over 3! 3!}t^{xyz}\epsi_{xyzuvw}\ell^{uvw}
\ena
this last equality implies that
the $SU(5,1)$ action preserves the 
symplectic structure. We have thus
 embedded\footnote{Strictly speaking we have immersed 
$SU(5,1)$ into $Sp(20,\rr)$, in fact this map 
is a local embedding but fails to be injective, indeed the three $SU(5,1)$ 
elements ${\sqrt[3]{1}}\,1\!\!1$ are all mapped into the identity element of 
$Sp(20,\rr)$.}
$SU(5,1)$ into $Sp(20,\rr)$ 
(in the complex basis).

The 20 dimensional real vector $(F^{\Lambda\Sigma}, G_{\Lambda\Sigma})$
transforms under the $20$ of $SU(5,1)$, as well as, for fixed $AB$, each of the 20 dimensional vectors
$\left( ^{f^{\Lambda\Sigma}_{~~AB}} _{h_{\Lambda\Sigma AB}}\right)$ of the
embedding matrix:
\begin{equation}\label{defu}
U = {1\over \sqrt{2}}\left(\begin{array}{cc}
f + {i } h & \bar f + { i }\bar h \\
f - {i } h & \bar f - { i }\bar h
\end{array}\right)~.
\end{equation}
The supercovariant field strengths and  vielbein are:
\begin{eqnarray}
\hat F^{\Lambda\Sigma} &=& d A ^{\Lambda\Sigma} +
\bigl [f^{\Lambda\Sigma}_{\ \ \ AB} (a_1 \bar\psi^A \psi^B + a_2
\bar \psi_C \gamma_a \chi^{ABC}V^a ) + h. c.\bigr ] \\
\hat {\cal P}_{ABCD} &=& {\cal P}_{ABCD}- \bar \chi_{[ABC}\psi_{D]}
- \epsilon_{ABCDE} \bar\chi^{(R)} \psi^E
\end{eqnarray}
where ${\cal P}_{ABCD}= \epsilon_{ABCDF} {\cal P}^F$ is the complex vielbein, completely antisymmetric
in $SU(5)$
indices and $\overline{{\cal P}_{ABCD}} = \bar {\cal P}^{ABCD}$.\\
The fermion transformation laws are:
\begin{eqnarray}
\delta\psi_A &=& D \epsilon _A  + a_3 T_{
 AB \, \mu\nu} \gamma^a \gamma^{\mu\nu}\epsilon^B V_a
+ \cdots\\
\delta\chi_{ABC} &=&a_4 {\cal P}_{ABCD \, i}\partial_\mu \phi^i \gamma^\mu \epsilon^D +
a_5 T_{
 [AB \, \mu\nu}  \gamma^{\mu\nu}
\epsilon_{C]} + \cdots \\
\delta \chi_{(L)} &=& a_6 \bar {\cal P}^{ABCD}_{\,~~~~~~~\bar \imath}\,\partial_\mu \bar\phi^{\bar\imath} \gamma^\mu \epsilon^E
\epsilon_{ABCDE} + \cdots
\end{eqnarray}
where:
\begin{eqnarray}
  T_{AB}
&=&{1 \over 2} (h_{\Lambda\Sigma AB}F^{\Lambda\Sigma} -
f^{\Lambda\Sigma}_{\ \ AB} G_{\Lambda\Sigma})\\\cN_{\Lambda\Sigma\; \Delta\Pi}& =& {1 \over 2}
h_{\Lambda\Sigma\, AB} (f^{-1})^{AB}_{\ \ \Delta\Pi}~.\label{NLASIDEPI}
\end{eqnarray}
With a by now familiar procedure one finds the following (complex)
central charges:
\begin{equation}
Z_{AB} = i\overline {\VV(\phi_\infty)}^{-1}Q
\en
where the charge vector is
\eq
Q=\left(\begin{array}{c}
p^{\Lambda\Sigma} \\[.3em] 
 q_{\Lambda\Sigma}
\end{array}\right)=
\left(\begin{array}{c}
{1\over 4\pi}\int_{S^2} F^{\Lambda\Sigma} \\[.3em] 
{1\over 4\pi} \int_{S^2} G_{\Lambda\Sigma}
\end{array}\right)
\en
and $\phi_\infty$ is the constant value 
assumed by the scalar fields at spatial infinity.
>From the equations (Maurer-Cartan equations)
\begin{equation}
\nabla^{(U(5))} f^{\Lambda\Sigma}_{~~AB} = {1 \over 2}\bar
f^{\Lambda\Sigma \, CD}{\cal P}_{ABCD}
\end{equation}
and the analogous one for $h$ we find:
\begin{equation}
\nabla^{(U(5))} Z_{AB} = {1 \over 2}\bar
Z^{CD}{\cal P}_{ABCD}
~.\end{equation}
Finally, the formula for the potential is, cf. \eqn{mothersumrule},
\begin{equation}
\VBH=    {1\over 2}  \bar Z ^{  AB} Z_{AB} 
 = -{1\over 2}Q^t\cM(\cN)Q
\end{equation}
where the matrix $\cM (\cN)$ has exactly the same form as in 
equation (\ref{Mlastbis}), and $\N$ is given in \eqn{NLASIDEPI}.

\sk

For $SU(5,1)$ there are only two  $U(5)$ quartic invariants.
In terms of the matrix $A_A^{\ B} = Z_{AC} \bar Z^{CB}$ they are:
\eq
 \Tr A  =  Z_{AB} \bar Z^{BA} ~~,~~~~~
 \Tr ( A^2) =  Z_{AB} \bar Z^{BC} Z_{CD} \bar Z^{DA}~.
\en
The $SU(5,1)$ invariant expression is
\begin{equation}
\mathscr{S}= {1\over 2} \sqrt{  |4 \Tr(A^2) - (\Tr A)^2 |}~.
\label{invar5}
\end{equation}

\subsubsection{The $N=6$ theory} The scalar manifold of the $N=6$ theory has the coset structure \cite{cj}:
\begin{equation}
G/H = {SO^\star (12) \over U(6)}
\end{equation}
We recall that $SO^\star (2n)$ is the real form of $O(2n,\cc)$ defined by the relation:
\begin{equation}
L^\dagger C L = C~,~~~~ C= \left(\begin{array}{cc}0 & -1\!\!1 \\ 1\!\!1 & 0 \\
\end{array}\right)
\end{equation}
The field content and transformation properties are given in
Table \ref{tab4,6},
\begin{table}[ht]
\caption{Transformation properties of fields in $D=4$, $N=6$}
\label{tab4,6}
\begin{center}
\begin{tabular}{|c||c|c|c|c|c|c|}
\hline
& $V^a$ &$ \psi _A$ &$\chi_{ABC},\chi_A$
&$A^\Lambda$ &$S^\alpha_r $  & $R_H$ \\
\hline
\hline
$SO^\star(12)$& 1 & 1 & 1 &- & 32 & - \\
\hline
$SU(6)$ & 1 & $6$ & $(20 + 6)$ & 1 &$( 15, 1)+(\bar{15},\bar 1)$ & $ {\bar {15}}$  \\
\hline
$U(1)$ & 0 & ${1\over 2}$ & $({3\over 2}, -{5\over 2})$ & 0 &$(1,-3) + (-1,3)$ & 2 \\
\hline
\end{tabular}
\end{center}
\end{table}
where $A, B, C = 1,\cdots,6$ are $SU(6)$ indices in the fundamental
representation and $ \Lambda = 1,\cdots, 16 $.
 The ${32}$ spinor
representation of $SO^\star (12)$  can be given in terms of
a  $Sp(32,\rr)$ matrix, which in the complex basis we denote by 
$S^\alpha_r$ ($\alpha,r = 1,\cdots,32 $). It is the double cover of
$SO^\star(12)$ that embeds in $Sp(32,\rr)$
and  therefore the duality group is this spin group. 
Employing the usual notation we may set:
 \begin{equation}
S^\alpha_r = {1\over \sqrt{2}}\left(\begin{array}{cc}f^\Lambda _{~M} + {\rmi} h_{\Lambda M} &
 \bar f^\Lambda _{~M} + {\rmi} \bar h_{\Lambda M}    \\
 f^\Lambda _{~M} - {\rmi} h_{\Lambda M} &
 \bar f^\Lambda _{~M} - {\rmi} \bar h_{\Lambda M}    \\
\end{array}\right)
\end{equation}
where $\Lambda,M=1,\cdots,16$.
With respect to  $SU(6)$, the sixteen symplectic vectors
$(f^\Lambda_{~M},h_{\Lambda M})$,
($M = 1,\cdots,16$) are  reducible into the antisymmetric 15
dimensional representation plus a singlet of $SU(6)$:
\begin{equation}
(f^\Lambda_M,h_{\Lambda M})\to (f^\Lambda_{~AB},h_{\Lambda AB})+
(\bar f^\Lambda,\bar h_{\Lambda})~.
\end{equation}
It is precisely the existence of a $SU(6)$ singlet which allows
for the Special Geometry structure of ${SO^*(12) \over U(6)}$
(cf. \eqn{5-70}, \eqn{5-71})\footnote{Due to its Special Geometry
structure the coset space ${SO^*(12) \over U(6)}$ is also the scalar manifold of an $N=2$ supergravity. The two supergravity theories
 have the same bosonic fields however the 
fermion sector is different.}.
Note that the element $S^\alpha_r$ has no definite $U(1)$
weight since the submatrices
$f^\Lambda_{~AB},\bar f^\Lambda$ have the weights 1 and $-3$ respectively.
The vielbein matrix is
\eq
{\cal P}=\left(
\begin{array}{cc}
{\cal P}_{ABCD}& {\cal P}_{AB} \\
{\cal P}_{CD}  & 0
\end{array}\right)~,
\en
where 
\begin{equation}
  {\cal P}_{AB} = {1 \over 4!}\epsilon_{ABCDEF} {\cal P}^{CDEF} ; \quad \bar {\cal P}^{AB} = \overline{{\cal P}_{AB}}~.
\label{pab}
\end{equation}
The supercovariant field strengths and the coset manifold vielbein have the
following expression:
\begin{eqnarray}
\hat F^{\Lambda} &=& d A ^{\Lambda} +\bigl [
f^{\Lambda}_{\  AB} (a_1 \bar\psi^A \psi^B + a_2
\bar \psi_C \gamma_a \chi^{ABC}V^a ) \nonumber\\
& & + a_3
f^\Lambda \bar \psi_C \gamma_a \chi^{C}V^a + h. c. \bigr ] \\
\hat {\cal P}_{ABCD} &=& {\cal P}_{ABCD} -  \bar \chi_{[ABC}\psi_{D]}
- \epsilon_{ABCDEF} \bar\chi^E \psi^F
\end{eqnarray}
The fermion transformation laws are:
\begin{eqnarray}
\delta\psi_A &=& D \epsilon _A  + b_1 T_{AB
\, \mu\nu}
\gamma^a \gamma^{\mu\nu} \epsilon^B V_a
+ \cdots\\
\delta\chi_{ABC} &=&  b_2 {\cal P}_{ABCD\, i}\partial_a z^i \gamma^a \epsilon^D +
b_3T_{[AB\, ab} \gamma^{ab}
\epsilon_{C]} + \cdots\\
\delta \chi_A &=& b_4 {\cal P}^{BCDE}_{~~~~~~~i} \partial_a z^i \gamma^a \epsilon^F
\epsilon_{ABCDEF} +b_5T_{ab}\gamma^{ab} \epsilon_A + \cdots
\end{eqnarray}
where according to the general definition \eqn{gravi}:
\eqa
T_{AB}&=& h_{\La\,AB}F^\La-f^\La_{~AB}G_\La\nn\\[.2em]
\bar T&=&\bar h_{\La }F^\La-\bar f^\La G_\La
\ena
With the usual procedure we have the following complex dyonic central
charges:
\begin{eqnarray}
Z_{AB} &=&  h_{\Lambda AB} p^\Lambda - f^\Lambda_{AB} q_\Lambda \\
 \bar Z &=&   \bar h_{\Lambda} p^\Lambda - \bar f^\Lambda q_\Lambda
\end{eqnarray}
 in the $\overline{15}$ (recall \eqn{TshouldbecalledbarT}) and singlet representation of $SU(6)$
 respectively.
Notice that although we have 16 graviphotons, only 15 central charges
are present in the supersymmetry algebra.
The singlet charge plays a role analogous to a ``matter'' charge (hence our notation $\bar Z$, $\bar f^\La$, $\bar h_\La$).
The charges differential relations are
\begin{eqnarray}
\nabla ^{(U(6))} Z_{AB} &=& {1 \over 2} \bar Z^{CD} {\cal P}_{ABCD} + {1 \over 4!} Z
\epsilon_{ABCDEF}{\cal P}^{CDEF} \\
\nabla^{(U(1))} \bar Z &=& {1 \over 2! 4!}\bar Z^{AB} \epsilon_{ABCDEF}{\cal P}^{CDEF}
\end{eqnarray}
and the formula for the potential reads, cf. (\ref{mothersumrule}), 
\eq
\VBH={1\over 2}\bar{Z}^{AB} {Z}_{AB}+\bar ZZ=
-{1\over 2}Q^t{\cal M}({\cal N})Q~.
\en
\sk

The quartic $U(6)$ invariants are
\begin{eqnarray}
I_1&=& (Tr A)^2 \label{invar61}\\
I_2&=& Tr(A^2)\label{invar62} \\
I_3 &=& 
{1\over 2^3 3!}
\rea( \epsilon^{ABCDEF}Z_{AB}Z_{CD}Z_{EF} Z)\label{invar63}\\
I_4 &=& (Tr A) Z \bar Z\label{invar64}\\
I_5&=& Z^2 \bar Z^2\label{invar65}
\end{eqnarray}
where $A_A^{\ B} = Z_{AC} \bar Z^{CB}$. The unique $SO^*(12)$ invariant is
\eq
\mathscr{S}={1\over 2} \sqrt{|4 I_2 - I_1 + 32 I_3 +4I_4 + 4 I_5| }
\label{invar6} ~. 
\en
%

\subsubsection{The $N=8$ theory}
In  the $N=8$ case \cite{crju}  the coset manifold is:
\begin{equation}
G/H={E_{7(7)}\over SU(8)/\mathbb{Z}_2}.
\end{equation}
The field content and group assignments are 
given in  Table \ref{tab4,8}.
\begin{table}[ht]
\begin{center}
  \caption{ Field content and group assignments in $D=4$, $N=8$ supergravity}
  \label{tab4,8}
  \begin{tabular}{|c||c|c|c|c|c|c|}
\hline
&$V^a $ &$ \psi_A $ & $A^{\Lambda\Sigma}$ &
 $ \chi_{ABC}$ & $S^\alpha_r$ & $R_H$ \\
\hline
\hline
$E_{7(7)}$ & 1 & 1 & - & 1 & 56 & - \\
\hline
$SU(8)$ & 1 & 8 & 1 & 56 & $ 28 + \bar{28}$ & 70 \\
\hline
 \end{tabular}
\end{center}
\end{table}

The embedding in $Sp(56,\rr)$ is automatically realized because
the ${56}$ defining representation of $E_{7(7)}$ is a real symplectic 
representation. The components of the $f$ and $h$ matrices and their 
complex conjugates are
\eq
f^{\Lambda\Sigma}_{\ \ AB} ~,~~
 h_{\Lambda\Sigma AB}~,~~
\bar f_{\Lambda\Sigma}^{~~ AB} ~,~~ \bar h^{\Lambda\Sigma\, AB}~,
\en
here $\Lambda\Sigma,AB$ are couples of antisymmetric  indices, with $\Lambda,\Sigma,A,B$ 
running from 1 to 8. The $70$ under which 
the vielbein of $G/H$ transform is obtained 
from the four times antisymmetric of $SU(8)$ 
by imposing the self duality condition
\eq\label{70rep}
\bar t^{\bar A\bar B\bar C\bar D}={1\over 4!}\epsilon^{\bar A\bar B\bar C\bar D}_{~~~~~~A'B'C'D'}
t^{A'B'C'D'}
\en
The supercovariant field strengths and coset 
manifold vielbein are:
\begin{eqnarray}
  \hat F^{\Lambda\Sigma} &=& dA^{\Lambda\Sigma} +[ f^{\Lambda\Sigma}_{\ \ AB}(
a_1\bar\psi^A \psi^B + a_2 \bar\chi^{ABC} \gamma_a \psi_C V^a) + h.c.]\\
\hat {\cal P}_{ABCD} &=&  {\cal P}_{ABCD} -   \bar\chi_{[ABC}\psi_{D]} + h.c.
\end{eqnarray}
where $ {\cal P}_{ABCD}= {1 \over 4!} \epsilon_{ABCDEFGH}\bar {\cal P}^{EFGH}\equiv (L^{-1} \nabla^{SU(8)} L)_{AB\, CD}=
{\cal P}_{ABCD\,i}d\phi^i$ 
($\phi^i$ coordinates of $G/H$).
In the complex basis the vielbein ${\cal P}_{ABCD}$ 
of $G/H$  are $28\times 28$ matrices 
completely antisymmetric and self dual as 
in \eqn{70rep}. 
 The fermion transformation laws are given by:
\begin{eqnarray}
  \delta\psi_A &=& D \epsilon_A + a_3 T_{AB \, \mu\nu}  \gamma^a\gamma^{\mu\nu}
\epsilon^B V_a + \cdots\\
\delta\chi_{ABC} &=& a_4 {\cal P}_{ABCD\,i} \partial_a \phi^i \gamma^a\epsilon^D + a_5T_{[AB\, \mu\nu}
\gamma^{\mu\nu} \epsilon_{C]}+ \cdots
\end{eqnarray}
where:
\eq
T_{AB}= {1 \over 2}(h_{\Lambda\Sigma AB} F^{\Lambda\Sigma}- f^{\Lambda\Sigma}_{\ \ AB}
G_{\Lambda\Sigma} )
\en
with:
\eq
\cN_{\Lambda\Sigma\, \Gamma\Delta}={1 \over 2}h_{\Lambda\Sigma AB}
 (f^{-1})^{ AB}_{\ \ \Gamma\Delta}  \label{NLASIGMAGAD}~.
\en

With the usual manipulations we obtain the central charges:
\begin{equation}
  Z_{AB}={1 \over 2}( h_{\Lambda\Sigma AB}
p^{\Lambda\Sigma} - f^{\Lambda\Sigma}_{\ \ AB} q_{\Lambda\Sigma}),
\end{equation}
the differential relations:
\begin{equation}
  \nabla^{SU(8)}Z_{\ AB}= {1 \over 2} \bar Z^{\ CD} {\cal P}_{ABCD}
\end{equation}
and the formula for the potential, cf. \eqn{mothersumrule},
\begin{equation}
\VBH=   {1\over 2}   \bar Z ^{  AB} Z_{AB} 
 = -{1\over 2}Q^t\cM(\cN)Q
\end{equation}
where the matrix $\cM (\cN)$ is given 
 in equation (\ref{Mlastbis}), and $\N$ in \eqn{NLASIGMAGAD}.
\sk

For $N=8$ the $SU(8)$ invariants are
\begin{eqnarray}
I_1 &=& (Tr A) ^2 \\
I_2 &=& Tr (A^2) \\
I_3 &=& P\!f \, Z 
 ={1\over 2^4 4!} \epsilon^{ABCDEFGH} Z_{AB} Z_{CD} Z_{EF} Z_{GH}
\end{eqnarray}
where $P\!f Z$ denotes the Pfaffian of the antisymmetric matrix $(Z_{AB})_{A,B=1,...8}$, and where $A_A^{\ B} = Z_{AC} \bar Z^{CB}$. One finds the following 
$E_{7(7)}$ invariant \cite{kako}:
\begin{equation}
\mathscr{S}= {1\over 2} \sqrt{|4 \Tr (A^2) - ( Tr A)^2 + 32 \rea (P\!f \,
Z) |}
\end{equation}

   For a very recent 
study of $E_{7(7)}$ duality rotations and 
of the corresponding conserved charges see 
\cite{Kallosh:2008ic}.

\sk
\noi
\subsubsection{Electric subgroups and the  $D=4$ and $N=8$ 
theory. }
A duality rotation is really a strong-weak 
duality if there is a rotation between electric and magnetic fields, more precisely if some of the rotated field strengths $F'^\La$ 
depend on the initial dual fields $G^\Sigma$, 
i.e. if the submatrix $B\not=0$ 
in the symplectic matrix $\left({}^A_C {}^B_D\right)$. 
Only in this case the gauge kinetic term 
may transform nonlinearly, via a fractional 
transformation. On the other hand, under
infinitesimal duality rotations 
$\left({}^{1\!\!1}_0 {}^0_{1\!\!1}\right)+
\left({}^a_c {}^0_d\right)$, with $b=0$,
the lagrangian changes by a total derivative
so that (in the absence of instantons) these 
transformations are symmetries 
of the action, not just of the equation of 
motion. Furthermore if $c=0$ the lagrangian itself is invariant.
 
We call electric any subgroup $G_e$ of the 
duality group $G$  with the property that it (locally) embeds in the symplectic group 
via matrices $\left({}^A_C {}^B_D\right)$ with $B=0$. The parameter space of 
true strong-weak duality rotations is  $G/G_e$.
\sk
The electric subgroup of $Sp(2n,\rr)$
is the subgroup of all matrices of the kind
\eq\label{ACAt}
\left(
\begin{array}{cc}
A&  \,0\\ 
C  & {A^t}^{-1}
\end{array}\right)~;
\en
we denote it by $Sp_e(2n,\rr)$. It is  {\sl the} electric subgroup because
any other electric subgroup is included in $Sp_e(2n,\rr)$.
This subgroup is maximal in $Sp(2n,\rr)$ (see for example the appendices in\cite{LledoV, Craps}). 
In particular if an action
is invariant under infinitesimal
$Sp_e(2n,\rr)$ transformations, and if the equations of motion 
admit also a $\pi/2$ duality rotation symmetry 
$F^\La\rightarrow G^\La$, $G^\La\rightarrow -F^\La$ for one or more indices $\La$ (no transformation on the other indices)
then the theory has $Sp(2n,\rr)$ duality.

It is easy to generalize the results of Section 2.2 and prove that 
duality symmetry under these $\pi/2$
rotations is equivalent to the following invariance property of the 
lagrangian under the Legendre transformation associated to $F^\La$,
\eq
\LL_D(F,\N')=\LL(F,\N)~,\label{Legelec}
\en
where $\N'= (C+D\N)(A+B\N)^{-1}$ are the transformed scalar fields,
the matrix $\left({}^A_C {}^B_D\right)$
implementing the $\pi/2$ rotation
 $F^\La\rightarrow G^\La$, $G^\La\rightarrow -F^\La$.
We conclude that $Sp(2n,\rr)$ duality symmetry holds if
there is $Sp_e(2n,\rr)$ symmetry and if the lagrangian
satisfies \eqn{Legelec}.
\sk
When the duality group $G$ is not $Sp(2n,\rr)$ then there may exist  
different maximal electric subgroups of $G$, say $G_e$ and $G'_e$.
Consider now a theory with $G$ duality symmetry, the electric subgroup 
$G_e$ hints at the existence of an action $S=\int \LL$ 
invariant under the Lie algebra Lie$(G_e)$ 
and under Legendre transformation that are $\pi/2$ duality rotation in $G$.
Similarly $G'_e$ leads to a different action $S'=\int\LL'$ that is
invariant under Lie$(G_e')$
and under Legendre transformations that are $\pi/2$ duality rotation in $G$.
The equations of motion of both actions have $G$ duality symmetry.
They are equivalent if
$\LL$ and $\LL'$ are related by a Legendre transformation. Since 
$\LL'(F,\N')\not=\LL(F,\N)$, this Legendre 
transformation cannot be a duality symmetry, it is a 
$\pi/2$ rotation  $F^\La\rightarrow G^\La$, $G^\La\rightarrow -F^\La$
that is not in $G$, this is possible since $G\not=Sp(2n,\rr)$.

As an example consider the $G_e=SL(8,\rr)$ symmetry
of the $N=8$, $D=4$ supergravity lagrangian  whose duality group 
is $G=E_{7,(7)}$ this is the formulation of Cremmer-Julia.
An alternative formulation, obtained from dimensional reduction of the $D=5$ 
supergravity, exhibits an electric group $G'_e=[E_{6,(6)}\times SO(1,1)]\ltimes
 T_{27}$ where the nonsemisimple  group $G'_e$ is realized as a lower 
triangular subgroup of $E_{7,(7)}$ in its  fundamental (symplectic) $56$ 
dimensional representation. $G_e$ and $G'_e$ are both 
maximal subgroups of $E_{7(7)}$.
The corrseponding lagrangians can be related only after a proper 
duality rotation of electric and magnetic fields which involves 
a suitable Legendre transformation.
\sk

A way to construct new supergravity theories 
is to promote a compact rigid electric subgroup 
symmetry to a local symmetry, thus constructing gauged supergravity 
models (see for a recent  review  \cite{Trigiante07}, and references 
therein).  Inequivalent choices of electric subgroups give different 
gauged supergravities. Consider again $D=4$, $N=8$ supergravity.
The maximal compact subgroups of $G_e=SL(8,\rr)$ and of
$G'_e=[E_{6,(6)}\times SO(1,1)]\ltimes T_{27}$ are $SO(8)$ and 
$Sp(8)=U(16)\cap Sp(16,\cc)$  respectively.
The gauging of $SO(8)$ corresponds to the gauged $N=8$ supergravity of 
De Witt and  Nicolai \cite{dwni}. As shown in \cite{Andrianopoli:2002mf} the 
gauging of the nonsemisimple group $U(1)\ltimes T_{27}\subset G'_e$ 
corresponds to the gauging 
of a flat group in the sense of 
Scherk and Schwarz dimensional reduction \cite{Scherk:1979zr},
and gives the massive deformation of the $N=8$ supergravity as obtained by 
Cremmer, Scherk and Schwarz \cite{Cremmer:1979uq}.


\section{Special Geometry and $N=2$ Supergravity}

In the case of $N=2$ supergravity the 
requirements imposed by
supersymmetry on the scalar manifold ${M}_{scalar}$ of the
theory dictate that it should be the following direct product: $
{M}_{scalar}={ M}\, \times \, { M}^Q$ where
${ M}$ is a special K\"ahler manifold of complex
dimension $n$ and ${M}^Q$ a qua\-ter\-nio\-nic manifold
of real dimension $4n_H$,  here $n$ and $n_H$ are respectively
the number of vector multiplets and hypermultiplets
contained in the theory. 
The direct product structure imposed by
supersymmetry precisely reflects the fact that the quaternionic and
special K\"ahler scalars belong to different supermultiplets. We do
not discuss the hypermultiplets any further and refer to \cite{n=2} 
for the full structure of N=2 supergravity. 
Since we are concerned with duality rotations we here concentrate our 
attention to an $N=2$ supergravity where 
the graviton multiplet, containing besides the graviton $g_{\mu \nu}$ 
also a graviphoton $A^0_{\mu}$, is coupled to $n'$  vector multiplets. Such a
theory has a bosonic action of type \eqn{bosonicL} where the number of
(real) gauge fields is ${\nm}=1+\pn$ and the number of (real) scalar fields
is $2\pn$.  Compatibiliy of their couplings  with local $N=2$ 
supersymmetry lead to the formulation of special \K geometry \cite{spegeo},\cite{str}. 
\sk
The formalism we have developed so
far for the $D=4$, $N>2$ theories is 
completely determined by the (local) embedding 
of the coset representative of the scalar manifold  $M={G}/H$ in
$Sp(2n,\mathbb{R})$. It leads to a flat 
-actually a trivial- symplectic bundle with 
local symplectic sections $\VVV_\eta$, 
determined by the symplectic 
matrix $\VV$, or equivalently by the 
matrices $f$ and $h$. 
We want now to show that these matrices,
the differential relations among charges 
and their quadratic invariant $\VBH$ (\ref{mothersumrule}) are 
also central for the description of $N=2$ 
matter-coupled supergravity.
This follows essentially
from the fact that, 
though the scalar 
manifold $M$ of the
$N=2$ theory is not in general a coset 
manifold, nevertheless, as for the $N>2$ 
theories, we have a flat symplectic 
bundle associated to $M$, with symplectic sections $\VVV_\eta$. 
While the formalism is very similar there is a difference,
the bundle is not a trivial bundle 
anymore, and it is in virtue of duality rotations that the theory can be
globally defined  on $M$.
\sk
In the next section we study the geometry of the scalar 
manifold $M$ and in detail its associated flat symplectic bundle. 
Then in Section 5.2 we see how, in analogy with $N>2$ supergravities,
the flat symplectic bundle geometry of $M$ enters  
the supersymmetry transformations laws of $N=2$ 
supergravity 
and the differential relations among the matter and 
central charges. 

  \subsection{Special Geometry}
There are two kinds of special geometries: rigid and local. 
While rigid special \K manifolds are the 
target space of the scalar 
fields present in the vector multiplets of $N = 2$ Yang Mills theories, 
the (local) special \K manifolds, in the mathematical literature called 
projective special \K manifolds, describe the target space of the 
scalar fields in the vector multiplets of $N = 2$ supergravity (that has 
local supersymmetry).
In order to describe the structure of a (local or projective) special 
\K manifold it 
is instructive to recall that of rigid \K manifold.  

\subsubsection{Rigid Special Geometry}
In short a rigid special \K manifold  is a \K
manifold $M$ that has a flat connection on its tangent bundle. This connection
must then be compatible with the symplectic and complex structure of $M$. 

More precisely, following \cite{Freed}, see also \cite{LledoV}, 
a {\bf rigid                  special \K 
structure} on a \K manifold $M$ with \K form 
$K$ is a 
connection $\nabla$ that is real, flat, torsionfree,
compatible with the symplectic structure $\om$:
\eq
\label{simplecticcomp}
\nabla\om=0~
\en
and compatible with the almost complex structure $J$ of $M$:
\eq
\label{complexcomp}
d_\nabla J=0~
\en
where 
$d_\nabla: \Om^1(TM)\rightarrow \Om^2(TM)$ is the covariant 
exterior derivative on vector-valued forms. Explicitly,
if $J=J^\xi\,\partial_\xi$ where $J^\xi$ are $1$-forms, and 
$\nabla \partial_\xi= A_{~\xi}^{\zeta}\,\partial_\zeta$,
with $A_{~\xi}^{\zeta}$ $1$-forms, then 
$d_\nabla J=dJ^\xi\,\partial_\xi-J^\xi\wedge A_{~\xi}^\zeta\, \partial_\zeta=(dJ^\xi+A^\xi_{~\zeta}\wedge J^\zeta)\,\partial_\xi$.
Notice that the torsionfree condition can be 
similarly written $d_\nabla I=0$, where $I$ is the identity map in $TM$, locally 
$I=dx^\xi\otimes\partial_\xi$. The two conditions $d_\nabla J=0$, $d_\nabla I=0$ for the real connection $\nabla$ can be written in the complexified tangent bundle simply as 
\eq\label{realitytorsionfreeJcompatibility}
d_\nabla \pi^{1,0}=0~,
\en
where $\pi^{1,0}$ is the projection onto the $(1,0)$ part of the complexified tangent bundle; locally 
$\pi^{1,0}=dz^i\otimes {\del\over\del z^i}$.

The flatness condition is equivalent 
to require the existence of
a covering of $M$ with local frames 
$\{\mathsf{e}_\xi\}$ that are covariantly constant, $\nabla \mathsf{e}_\xi=0$.
The corresponding transition functions 
of the real tangent bundle  
$TM$ are therefore constant invertible 
matrices; compatibility with the symplectic structure, equation \eqn{simplecticcomp}, further implies that these matrices belong to
the fundamental of $Sp(2n,\rr)$, where $2n$ is the real dimension of $M$ (each frame
$\{\mathsf{e}_\xi\}$ can be chosen to have mutually symplectic vectors $\mathsf{e}_\xi$).

Flatness of $\nabla$ (i.e., the vanishing of the  curvature $R_\nabla$ or equivalently $d_\nabla^2=0$) implies that
\eqn{realitytorsionfreeJcompatibility}
is equivalent to the existence of a local 
complex vector field $\xi$ that satisfies
\eq 
\nabla\xi=\pi^{1,0}
\label{localcomaptJ}
\en
[hint: in a flat reference frame $d_\nabla=d$, and Poincar\'e lemma for $d$ implies that
any $d_\nabla$-closed section is also 
$d_\nabla$-exact]. 
Studying the components of this vector field (with respect to a flat Darboux coordinate system) we obtain the existence of local holomorphic
coordinates on $M$, called special coordinates, their transition 
functions are constant $Sp(2n,\rr)$ matrices, so that the holomorphic tangent 
bundle $TM$ is a flat symplectic holomorphic one.
Corresponding to these special coordinates
we have a holomorhic function $\F$, the
holomorphic prepotential.
In terms of this data the \K potential and the \K form read
\eq\label{Kpotspec}
{\cal K}={1\over 2} {\rm Im}\big({{\partial\F}\over{\partial z^i}}\bar z^i\big)
dz^i\wedge d\overline{z}^j~,
\en
\eq\label{Kformspec}
K=i\del\bar\del \KK={i\over 2} {\rm Im}\big({{\partial^2\F}\over{\partial z^i\partial z^j}}\big)dz^i\wedge d\overline{z}^j
={i\over 2} {\rm Im}({\tau_{ij}})dz^i\wedge d\overline{z}^j~,
\en
where $z^i$ are special coordinates, 
and $\tau_{ij}={{\partial^2\F}\over{\partial z^i\partial z^j}}$.

\sk
An equivalent way of characterizing 
rigid special \K manifolds is via a 
holomorphic symmetric 3-tensor $C$. This 
tensor measures the difference between the
symplectic connection $\nabla$ and the 
Levi-Civita connection $D$, whose connection 
coefficients we here denote 
$\gamma_{ij}^k$ and 
$\bar\gamma_{\bar \imath \bar \jmath}^{\bar k}$.
 
Define $${\cal P}_\rr=\nabla-D~.$$
The nonvanishing components of ${\cal P}_\rr$ are
\eq\label{AGAAG}
A^k_{ij}-\gamma^k_{ij}~,~~
A^{\bar{k}}_{ij}~,~
A^{\bar k}_{\bar i\bar j}-\gamma^{\bar k}_{\bar i\bar j}
~,~A^k_{\bar i\bar j}~,~
\en
this is so because the components $A$ of the
connection $\nabla$ are constrained 
by condition
\eqn{realitytorsionfreeJcompatibility}.
Since $D$ and $\nabla$ are real and 
torsionfree we further have that the lower 
indices in \eqn{AGAAG} are symmetric, and
the reality conditions $\overline{A^k_{ij}-\gamma^k_{ij}}=
A^{\bar k}_{\bar i\bar j}-\gamma^{\bar k}_{\bar i\bar j}$, 
$\overline{A^{\bar k}_{\bar i\bar j}}=A^k_{\bar i\bar j}
$.
Since both $D$ and $\nabla$ are symplectic
we have that for any vector $u\in T_mM$, 
$({\cal P}_\rr)_u: T_mM\rightarrow T_mM$ is a generator of a symlectic transformation,
\eqa\label{KBvw}
u(K(v,w))=
D_u(K(v,w))=
K(D_u v,w)+K( v,D_u w)\nn\\
u(K(v,w))=
\nabla_u(K(v,w))=
K(\nabla_u v,w)+K( v,\nabla_u w)\nn\\
~~~~~~0=K(({\cal P}_\rr)_uv,w)+K(v, ({\cal P}_\rr)_u w)~.
\ena
If we set $u=\partial_k$, $v=\partial_i$, 
$w=\bar\partial_{\bar\jmath}$, and use 
that $K$ is a $(1,1)$-form,
we obtain
\eq
A^k_{ij}-\gamma^k_{ij}=0~.
\en
Then the components of $${\cal P}_\rr={\cal P}+\overline{\cal P}$$ 
are just 
$A^{\bar{k}}_{ij}$ and $A^k_{\bar i\bar j}$.
This leads to define the tensor 
\eq
C_{ijk}=-i g_{i\bar \ell}A^{\bar \ell}_{jk}~.
\en
Setting $u=\partial_k$, $v=\partial_i$, 
$w=\partial_{j}$ in \eqn{KBvw} we obtain
that $C_{ijk}$  is totally symmetric in 
its indices. 
Since $D_j\pi^{(1,0)}=0$ we easily compute, recalling \eqn{localcomaptJ},
$C_{ijk}=-\le\nabla_i\xi,\nabla_j\nabla_k\xi\re$, hence we obtain the coordinate independent expression for
$C=C_{ijk}dz^i\otimes dz^j\otimes dz^k$,
\eq\label{Cnablaxi}
C=-\le\nabla\xi,\nabla\nabla\xi\re~.
\en

Flatness of $\nabla=D+{\cal P}_\rr$, i.e. $d_\nabla^2=0$, is equivalent to 
\eq\label{RdB}
R+d_D{\cal P}+d_D{\overline {\cal P}}+{\cal P}\wedge \overline {\cal P}
+\overline {\cal P}\wedge {\cal P}=0
\en 
where $R=d^2_D$ is the Levi-Civita curvature
and $d_\nabla {\cal P}$ is the exterior covariant derivative action on the 1-form ${\cal P}$ with values in $T_\cc M\otimes T^*_\cc M$
(where $T^*_\cc M$ is the 
complexified cotangent bundle).
Now in \eqn{RdB}, 
the term $R+{\cal P}\wedge \overline 
{\cal P}+\overline {\cal P}\wedge {\cal P}\in \Om^{(1,1)}(M,End(T_\cc M,T_\cc M))$,  i.e., this term maps 
$T^{(1,0)}M$ (or $T^{(0,1)}M$) vectors into 
(1,1)-forms valued in $T^{(1,0)}M$ (or $T^{(0,1)}M$). On the other hand
${\cal P}\in \Om(End(T_\cc M,\overline {T_\cc M}))$,
in particular it maps 
$T^{(1,0)}M$ vectors into 
forms valued in $T^{(0,1)}M$, and annihilates 
$T^{(0,1)}M$ vectors (hence ${\cal P}\wedge {\cal P}=0$).
Similar properties hold for the complex conjugate $\overline {\cal P}$, 
with $T^{(1,0)}M$ replaced by $T^{(0,1)}M$,
and for $d_D{\cal P}$ and $d_D\overline {\cal P}$. 
It follows that equation \eqn{RdB} is 
equivalent to two independent equations, 
\eq\label{RdB1}
R+{\cal P}\wedge \overline {\cal P}
+\overline {\cal P}\wedge {\cal P}=0
\en 
\eq\label{RdB2}
d_D{\cal P}=0~.
\en
Since the covariant derivative of the metric 
vanishes, this last equation is equivalent to
$d_DC=0$. 
In local coordinates we have
\eq
d C_{\ell j}
-\gamma^k_{~\ell}\wedge C_{kj}-
\gamma^k_{~j}\wedge C_{\ell k}=0~.
\en
where $C_{ij}=C_{ikj}dz^k$. This equation splits in the condition
\eq
\bar\del C=0~,
\en
so that $C$ is holomorphic,
and the condition $\del_DC=0$, that can be 
equivalently written 
\eq
D_iC_j=D_jC_i\label{DCDC}
\en 
where $C_i$ is the matrix 
$C_i=(C_{ki\ell})_{k,\ell=1,...n}$,
i.e., $C_i\in\Om^0(M, 
{T^*}^{(1,0)}M\otimes {T^*}^{(1,0)\!}M)$, 
so that  $D_i$ is the covariant derivative on
functions valued in 
${T^*}^{(1,0)}M\otimes {T^*}^{(1,0)\!}M$. 

The local coordinates expression 
of
\eqn{RdB1}
is
\eq
R_{\bar\imath j \bar k \ell}=-\overline{C}_{\bar\imath\bar k\bar s}g^{\bar s p}C_{pj\ell}~.
\en
\sk
In conclusion a rigid special \K structure 
on $M$ implies the existence of a 
holomorphic symmetric 3-tensor (cubic form) 
$C$ that satisfies \eqn{RdB1} and \eqn{DCDC}.

Viceversa if a \K manifold $M$ admits a 
symmetric holomorphic 3-tensor $C$ that satisfies \eqn{RdB1} and \eqn{DCDC}, then $M$ 
is a special \K manifold. 
Indeed the contraction of $C$ 
with the metric gives ${\cal P}$, so that
we can define $\nabla=D-{\cal P}_\rr$. The symmetry 
of $C$ implies that $d_\nabla\pi^{1,0}=0$ so 
that $\nabla$ is torsionfree and compatible 
with the complex structure, $d_\nabla J=0$. 
The symmetry of $C$ also implies \eqn{KBvw} 
so that $\nabla$ is symplectic. Finally
\eqn{RdB1} and \eqn{DCDC} imply that $\nabla$
is flat.

In special coordinates the
holomorphic 3-tensor $C$ is simply
given by $C_{ijk}={1\over 4}{{\partial^3\F}
\over{\partial z^i\partial z^j\partial 
z^k}}$.
\sk
\subsubsection{Local Special Geometry} 
We have recalled that to a rigid special \K manifold of dimension $n$ there 
is canonically associated a holomorphic $n$ dimensional flat 
symplectic vector bundle.  On the other hand, to a projective (or local) 
special \K manifold $M$, of dimension $\pn$ 
there is canonically associated a holomorphic 
$\nm=\pn+1$ dimensional flat symplectic 
vector bundle. The increase by one unit of the rank of the vector bundle with respect to the dimension of the manifold
is due to the graviton multiplet. 
The mathematical 
description involves the $\nm=\pn+1$ dimensional manifold $L$, total 
space of a line bundle over $M$. 
%
\sk
\noi\textbf{{\K$\!\!$-Hodge manifolds and their associated principal bundles $\tilde M\rightarrow M$}}  \\
Consider a \K$\!$-Hodge manifold, i.e. a triple $(M,L,K)$, where $M$ is 
\K with integral \K form $K$, so that it defines 
a class 
$[K]\in H^2(M,\zz)$, and 
$$L\stackrel{\pi}{\rightarrow} M$$ 
is a holomorphic hermitian line 
bundle with first Chern class equal to $[K]$, and 
with curvature equal 
to $-2\pi i  K$ (recall that on a hermitian holomorphic 
vector bundle there is a unique connection compatible with the hermitian holomorphic structure).

Consider the complex manifold $\tilde M$, that is 
$L$ without the zero 
section of  $L\stackrel{\pi}{\rightarrow} M$.
The manifold $\tilde M$ is a principal bundle over $M$, with structure group
$\cc^\times$ (complex numbers minus the zero); the action of $\cc^\times $ on $\tilde M$ is holomorphic.  The hermitian connection canonically associated 
to $L\rightarrow M$ induces a connection on $\tilde M$ so that in
$T\tilde M$ we have the subspaces of horizontal and vertical tangent vectors.  

Another property of the manifold $\tilde M$ is that it 
has a canonical hermitian line bundle $\pi^*L \rightarrow \tilde M$;
it is the pullback to $\tilde M$ of  $L\rightarrow M$, so that the 
fiber on the point $\tilde m\in \tilde M$ 
is just the fiber of $L$ on the point $m=\pi(\tilde m)\in M$,
\eq
\begin{CD}
{{\pi^* L}}                   @> >>  {L}\\
 @V  {} VV                                      @V \pi VV\\
{\tilde M}@> \pi >>{M}
\end{CD}
\en
\sk
\noi
Explicitly $\pi^\st L=\{(\tilde m, \ell) \,;~ \pi(\ell)=\pi(\tilde m)\}$. 
The line bundle $\pi^*L$ 
is trivial indeed we have the globally defined nonzero holomorphic section 
\eqa \Omm\,:~ \tilde M~~~&\rightarrow& \pi^*L\nn\\
        ~~~~~~\tilde m ~~~&\mapsto &(\tilde m,\tilde m)~\nn\\
        ~~~~~~(m,\la)&\mapsto &(m,\la,\la)~.
\ena
In the last line we used a local trivialization of $\tilde M\rightarrow M$ 
(and henceforth of $L\rightarrow M$) given by a local section $s$, say
$\tilde m=\la s(m)\sim (m,\la)$. This induces a local trivialization 
$\tilde s =\pi^*s$ of the line bundle $\pi^*L\rightarrow \tilde M$.
Explicitly 
$\tilde s$ associates to $\tilde m$ the point $s(m)$ of $L$, 
so that a generic element  $\tilde\ell=\sigma \tilde s(\tilde m)\in
\tilde L$ is described by the triple $(m,\la,\sigma)$, and
in particular 
\eq
\Omm(\tilde m)=\Omm(\la s(m) )=\la\tilde s(\tilde m)\sim (m,\la, \la)~. 
\label{OMmlas}
\en
\sk
It can be shown that $\tilde M$ is a pseudo-\K manifold (i.e. a \K 
manifold where the metric has pseudo-Riemannian signature).
The \K form is
\eq\label{pseudoKform}
\tilde K={i\over 2\pi}\bar\del\del|\Omm|^2~,
\en
where $|\Om|^2$ is the evaluation on $\Om$ of the 
hermitian structure of $\pi^\st(L)$ (this latter is trivially inherited from the hermitian structure of $L$). With respect to the corresponding \K metric, horizontal 
and vertical vectors
are orthogonal, moreover the \K metric is negative definite along 
vertical vectors,
and  positive definite along  horizontal vectors, where $\tilde K|_{_{hor}}= |\Om|^2\pi^*K$.\footnote{Hint: in the coordinates $(z^i,\la)$, associated to the local trivialization $\tilde m=\la s(m)\sim (m,\la)$ induced by a section $s$ of $L$, we have 
$|\Om|^2=\la\bar\la|s|^2$. Moreover horizontal vectors read $u=u^i\partial_i-u^ia_i\la{\del\over\del\la}$ where the local connection 1-form on 
$M$ is $a=a_idz^i={|s|}^{-2}\del|s|^2$. The pseudo-\K form reads
$-2\pi i \tilde K=\la\bar\la\del_i\del_{\bar\jmath}|s|^2dz^i\wedge d\bar z^{\bar \jmath} +|s|^2d\la\wedge d\bar\la +\la\partial_i|s|^2dz^i\wedge d\bar\la+
\bar\la\partial_{\bar\jmath}|s|^2d\la\wedge d\bar z^{\bar\jmath}$.}
Thus $(\tilde M, \tilde K)$ has Lorentzian signature.

Concerning the pullback $\pi^*K$ on $\tilde M$ 
of the \K form 
$K$ on $M$; while $K$ is in general only closed, $\pi^*K$ 
is exact, 
\eq\label{pullom}
\pi^*K={i\over {2\pi}}\bar\del\del \lg |\Omm|^2~.
\en
This last formula easily follows by pulling back the usual 
local curvature formula for the hermitian connection 
$K={i\over 2\pi}\bar\del\del \log  |s|^2$ and by 
observing that $\pi^*\lg  |s|^2 =\lg  |\tilde s|^2
=\lg |\Omm|^2 - \lg\la-\lg\bar\la$.
\sk
In conclusion, one can canonically associate to a  
\K$\!$-Hodge manifold $(M,L,K)$ a pseudo-\K manifold 
$(\tilde M,\tilde K)$ that carries a free and holomorphic $\cc^\times $ 
action, and a line bundle $\pi^*L\rightarrow \tilde M$ 
that has a canonical global holomorphic section $\Omm$. 

The bundle $\tilde L$ can be naturally identified as the
holomorphic subbundle of $T\tilde M$ given by the vertical 
vectors of $\tilde M$ with respect to the holomorphic $\cc^\times $ 
action.  
The global holomorphic section $\Om$ corresponds
to the vertical vector field that gives the infinitesimal 
$\cc^\times $ action. Under this identification we have
\eq \label{identifyOm}
\tilde K(\Om,\Om)=-{i\over 2\pi}|\Om|^2~.
\en
This equation shows that under the identification $T\tilde M|_{vert}\simeq 
L$ the corresponding hermitian structures are mapped one into minus the other.
 
%
%

%
\sk
\noi{\textbf{{Special \K manifolds}}}  \\
Following \cite{Freed}, $(M,L,K)$ is special \K
if $(\tilde M,\tilde K)$ is rigid special \K and if
$\Omm$ is compatible with the symplectic connection 
$\tilde\nabla$.

\sk

A
 (projective or local) {\bf special \K manifold} 
is a \K$\!$-Hodge manifold
$(M,L, K)$ such that the associated pseudo-\K manifold 
$(\tilde M,\tilde  K)$ has a 
rigid special pseudo-\K structure $\tilde \nabla$ which satisfies 
\eq\label{comaptJ}
\tilde\nabla \Omm=\pi^{(1,0)}~.
\en

Notice that \eqn{comaptJ} is equivalent to the condition
$\tilde\nabla_u \Omm=u$ for any $u\in T^{(1,0)}\tilde M$.
As shown in \cite{LledoV}, since $\tilde\nabla$ is torsionfree and flat, then condition \eqn{comaptJ} implies the $\cc^\times$ invariance of $\tilde\nabla$, i.e. $dR_b(\tilde\nabla_uv)=\tilde\nabla_{dR_b u}dR_b v$ where $R_b$ denotes
the action of $b\in\cc^\times$.
Notice also that equation \eqn{comaptJ} is the global version of eq. \eqn{localcomaptJ}.

For ease of notation in the following we denote the flat 
torsionfree symplectic connection $\tilde\nabla$ on $\tilde M$
simply by $\nabla$.

 \sk\sk
We now construct a flat symplectic $2n=2\pn+2$ dimensional bundle ${\cal H}$ on $M$
that is frequently used in the literature in order to characterize projective 
special \K manifolds.  We introduce a new $\cc^\times$ action on $T\tilde M$. 
On $\tilde M$ it is the usual one $R_b\tilde m=\tilde m b=b\tilde m$, where 
$b\in \cc^\times$, while on vectors we have 
\eq\label{newC}
v_{\tilde m}\mapsto b^{-1}dR_b\, v_{\tilde m}~.
\en
>From now on by $\cc^\times$ action we understand the new above defined one. 
Thus for example since $b^{-1}dR_b\Om_{\tilde m}=b^{-1}\Om_{\tilde m b}$, then $\Om$ is not invariant under \eqn{newC}.
On the other hand  the local section (vertical vector field)
$\tilde s$,  obtained from a local section $s$ of $L$, 
satisfies $b^{-1}dR_b\tilde s_{\tilde m}=\tilde s_{b\tilde m}$ (or $b^{-1}{R_b}_*\tilde s=\tilde s$)
and is therefore $\cc^\times$ invariant.
A $\cc^\times$ invariant frame
associated with local coordinates $z^i$ of $M$ and with 
the local section $s$ of $L$ 
is $(\la^{-1}{\del\over\del{z^i}},{\del\over{\del\la}})$; it is given by the 
coordinates $(X^i,X^0)=(\la z^i,\la)$, they are $\cc^\times$ invariant ($b^{-1}{R_b}^* X=X$) and therefore are homogeneous (projective) coordinates of $M$.

\sk
 
We define the $2n=2\pn+2$ dimensional real vector bundle on $M$ (dim$_\rr M=2\pn$),
\eq
{\cal H}\rightarrow M
\en
by identifying its local sections with the 
$\cc^\times$ invariant sections of $T\tilde M$. In other words ${\cal H}$ is the quotient of $T\tilde M$ via the $\cc^\times$ action \eqn{newC}.   
A point $(m,h)\in{\cal H}$ is the equivalence class $[(\tilde m,v_{\tilde m})]$
where $(\tilde m,v_{\tilde m})\sim (\tilde m', u_{\tilde m'})$ if $m'=mb$ and 
$b^{-1}dR_bv_{\tilde m}=u_{\tilde m'}$. Under this quotient $\pi^\st L\subset T\tilde M$ becomes $L$, while the subbundle $T\tilde M|_{hor}$ of 
horizontal vectors becomes $L\otimes TM$.\footnote{Hint: denote by 
$\widehat{v_m}|_{\tilde m}$ the horizontal lift in $T_{\tilde m}\tilde M$ of the vector 
$v_m\in T_m M$. Then the map $L\otimes TM \rightarrow (T\tilde M|_{hor}) / _{\cc^\times \rm{action}}$
 defined by $(\ell_m\otimes v_m) \mapsto
[(\ell_m,\widehat{v_m}|_{\ell_m})]$ if $\ell_m\not= 0$, and by $0\mapsto 0$
is well defined, linear and injective. }
Therefore we have two natural inclusions 
\eq\label{natincl}
L\subset {\cal H}~~~~\mbox{and}~~~~L\otimes TM\subset {\cal H} ~.
\en
Since the $\cc^\times$ action is holomorphic, then ${\cal H}$ is a holomorphic 
vector bundle on $M$ of rank $\pn+1$. Since $\tilde K$ is a $\cc^\times$ invariant 
2-form the symplectic structure of $T\tilde M$ goes to 
the quotient ${\cal H}$: indeed $\tilde K(u,v)$ is a homogeneous function on $M$ if
$u$ and $v$ are $\cc^\times$ invariant vector fields of $T\tilde M$. Similarly also the flat symplectic connection $\nabla$ induces a flat symplectic connection on ${\cal H}$ (see for example \cite{LledoV}). The inclusion $L\subset {\cal H}$
implies that \eq
L^{-1}\otimes {\cal H}\rightarrow M
\en
 has a nonvanishing global holomorphic section.
\sk
In the following we work in $T\tilde M$, but we choose $\cc^\times$ invariant 
tensors and therefore our results immediately apply to the bundle ${\cal H}$.
Let's consider a $\cc^\times $ invariant flat local symplectic 
framing of $T\tilde M$, that we denote by 
$\{\mathsf{e}_\xi\}=\{\mathsf{e}_\La,\mathsf{f}^\La\}$, 
$\xi=1,\ldots 2\nm$, $\La=1,\ldots \nm$. The framing is flat because 
$\nabla \mathsf{e}_\La=0, \nabla \mathsf{f}^\La=0$, and it is 
symplectic because in this basis the symplectic 
matrix is in canonical form: the components 
$\tilde K(\mathsf{e}_\La, \mathsf{e}_\Sigma)$, $\tilde K(\mathsf{e}_\La, \mathsf{f}^\Sigma)$,
$\tilde K(\mathsf{f}^\La, \mathsf{e}_\Sigma)$, $\tilde K(\mathsf{f}^\La, \mathsf{f}^\Sigma)$
read
\eq
\left(
\begin{array}{cc}
0 & -1\!\!1\\
1\!\!1 & 0
\end{array}\right)
\en
With respect to the $\{ \mathsf{e}_\La, \mathsf{f}^\La\}$ frame, 
the global section $\Om$ has local components
$\Om=\Om^\xi\mathsf{e}_\xi=X^\La\mathsf{e}_\La +F_\La\mathsf{f}^\La$. 
We also denote by $\Om$ this column vector 
of coefficients,
\eq
\Om=(\Om^\xi)=\left(
\begin{array}{c}
X^\La  \\
F_\La
\end{array}\right)~.
\en
The local functions $X^\La$, $F_\La$ on $\tilde M$ are holomorphic, indeed 
\eqn{comaptJ} implies that $\nabla\Om$ is a
$(1,0)$-form valued in $T\tilde M$, since $\nabla(\Om^\xi\mathsf{e}_\xi)=
d\Om^\xi\,\mathsf{e}_\xi= \del\Om^\xi\,\mathsf{e}_\xi
+\bar\del\Om^\xi\,\mathsf{e}_\xi$, we obtain $\bar\del\Om^\xi=0$.
In conclusion $(X^\La, F_\La)$ are 
local components of the global symplectic section $\Om$ of the 
tangent bundle $T\tilde M$. 

Each entry $X^\La$, $F_\La$ is also a local 
holomorphic section of the line bundle $L^{-1}\rightarrow M$. 
Indeed from the transformation properties of $\Om$ under the $\cc^\times$ action 
$\tilde m\mapsto R_{e^{-f(m)}}(\tilde m)=e^{-f(m)}\tilde m $ (or under a change of local trivialization $s'(m)=e^{f(m)}s(m)$) we have
\eq
\left(
\begin{array}{c}
X^\La  \\
F_\La
\end{array}\right)'
=
e^{-f(m)}\left(
\begin{array}{c}
X^\La  \\
F_\La
\end{array}\right)~,
\en
therefore for each invertible $\Om^\xi$ we have that 
${\Om^\xi}^{-1}(s) s$ is a section of $L\rightarrow M$ 
or equivalently each $X^\La$ and each $F_\La$ 
are the coefficients of sections of $L^{-1}\rightarrow M$.
\sk
In conclusion $(X^\La, F_\La)$ are 
local components of the global symplectic section $\Om$ of the 
tangent bundle $T\tilde M$. Each entry is also a local 
holomorphic section of the line bundle $L^{-1}\rightarrow M$. 
Under change of local trivialization of $T\tilde M$ we have
\eq
\left(
\begin{array}{c}
X^\La  \\
F_\La
\end{array}\right)'
=S\left(
\begin{array}{c}
X^\La  \\
F_\La
\end{array}\right)
=
\left(
\begin{array}{cc}
A & B  \\
C & D
\end{array}\right)
\left(
\begin{array}{c}
X^\La  \\
F_\La
\end{array}\right)~,\label{uglt}
\en
where $S=\left({}^A_B{}^C_D\right)$ is a 
constant symplectic matrix.
We can also consider a change of coordinates on $M$, say $z\rightarrow z'$. Provided we keep fixed the frame of $T\tilde M$ and 
the trivalization of $L$ we then have that $X^\La$ and $F_\La$ behave like local functions on $M$,
$X^\La(z)=X'^\La(z')$, $F_\La(z)=F'_\La(z')$ (here $X^A(z)=X^A(s(z))$ etc.).
\sk
It can be shown \cite{LledoV} that from the set of
$2n$ elements $\{X^\La, F_\La\}$ one can always choose 
a subset of $n$ elements that form a local coordinate 
system on $\tilde M$.  Contrary to the \K case (where the metric 
is Riemanninan) in this pseudo-\K case in general neither 
$\{X^\La\}$ nor $\{F_\La\}$ are coordinates systems on $\tilde M$. 
The frame $\{\mathsf{e}_\La, \mathsf{f}^\La\}$ is determined 
up to a symplectic transformation, if using this freedom
we have that the $\{X^\La\}$ are coordinates functions then
the $\{X^\La\}$ are named special 
coordinates. The sections $F_\La$ can then 
be seen as functions of the 
$X^\La$ and are obtained via a prepotential $\F$,
\eq
F_\La={\del\, \F\over{\del X^\La}}~.
\en 
Recalling \eqn{pullom} and \eqn{identifyOm} we have
\eq
\pi^* K
={i\over 2\pi}\bar\del\del\lg\, i \le\Omega,\overline \Omega\re
\en
and for the corresponding
``\K$\!$'' potential $\cal K$ 
we have\footnote{As usual when $K$ is integral  $K={i\over 2\pi}g_{i\bar\jmath}dz^i\wedge d\bar z^{\bar\jmath}=
{i\over 2\pi}\del_i\del_{\bar\jmath}\KK dz^i\wedge d\bar z^{\bar\jmath}={i\over 2\pi}\del\bar\del\KK$.}
\eq
{\cal K}=-\lg \,i \le\Omega,\overline \Omega\re~;\label{ortOm}
\en
in these formulae we used the standard notation $$\le\Omega,\overline \Omega\re=\tilde K(\Omega,\overline \Omega)~.$$ 
Using the components $(X^\La, F_\La)$   
expression \eqn{ortOm} reads
\eq
{\cal K}=-{\lg}\big[ i (X, F)
\left(
\begin{array}{cc}
0 & -1\!\!1\\
1\!\!1 & 0
\end{array}\right)
\left(
\begin{array}{c}
\bar X  \\
\bar F
\end{array}\right)\big]=-\lg  [i( F_\La\bar X^\La- X^\La\bar F_\La)]~.
\en
By considering local sections of the bundle 
$\tilde M\rightarrow M$, we can then 
pull back the potential 
$\cal K$ to local \K potentials on $M$.

Under the action of $e^{-f(m)}\in\cc^\times$ 
on $\tilde M$ (or equivalently under change of 
trivialization of $\tilde M\rightarrow M$) we have
\eq
{\cal K}'={\cal K}+f+\bar f
\en
thus showing that $e^{-{\cal K}}$ defines 
a global nonvanishing section of the bundle 
$L\otimes \overline L\rightarrow M$, 
in particular this bundle is trivial. 
Explicitly this global section is 
$e^{{\cal K}(s)}[s,\bar s]$ where 
$s$ is any local section of
$\tilde M\rightarrow M$ and $[s,\bar s]
=\{(s\la,\la^{-1}\bar s), 
\la\in \cc^\times\}$
is the corresponding local section of 
$L\otimes \bar L$.
\sk

\sk
\noi{\textbf{{Symplectic Sections and Matrices from local coordinates frames on $M$}}}\\
Let's examine few more properties of special \K manifolds and 
introduce those symplectic vectors 
that we have seen characterizing the 
geometry of the 
supergravity scalar fields.
Consider a vector $u\in T_m^{(1,0)}M$, this can be lifted 
to a horizontal vector $\hat u\in T^{(1,0)}_{\tilde m}\tilde M$.
Because of \eqn{comaptJ} the covariant derivative 
$\nabla_{\hat{u}}\Om$ 
is again a vector in 
$T^{(1,0)}_{\tilde m}\tilde M$, then 
\eq
\le\Om,\nabla_{\hat{u}}\Om\re=0~~,
~\le\bar\Om,\nabla_{\hat{u}}\Om\re=0~~;
\label{Omnabu}
\en
the first relation holds because 
$\tilde  K=\le~,~\re$ is a $(1,1)$-form,
the second relation holds because horizontal
and vertical vectors are orthogonal under 
$\tilde K$ (recall paragraph after 
\eqn{pseudoKform}).

Subordinate to a holomorphic coordinate 
system $\{z^i\}$ of $M$, and a 
local section $s$ of $L\rightarrow M$
we have the local coordinates 
$(z^i, \la)$ on $\tilde M$.
The corresponding vector fields are 
$(\del_i,{\del\over\del\la})$.
A more natural frame on $\tilde M$ is given 
by considering
the vertical vector field associated to the  action of $\cc^\times$ on $\tilde M$,
\eq
\hat\del_0 \equiv \Om=\la{\del\over{\del\la}}~,
\en
and the horizontal lift 
$\hat\del_i$ of the vector fields $\del_i$ 
on $M$
\eq\label{lifti}
\hat\del_i =
\del_i-|s|^{-2}\partial_i|s|^2
\la{\del\over{\del\la}}=\del_i+\del_i{\cal K}
\la{\del\over{\del\la}}~.
\en  
In \eqn{lifti}, $|s|^2=h(s,s)$ is the hermitian form of  $L\rightarrow M$. All these vector fields have degree 1 and are independent from the 
section $s$ of $L\rightarrow M$.

We define 
\eq
\nabla_i=\nabla_{\hat\partial_i}~.
\en 
The new sections $\nabla_i\Om$ are exactly 
the horizontal vector fields $\hat\partial_i$,
indeed from
\eqn{comaptJ} we obtain
\eq
\nabla_i\Om=\hat\del_i~~,~~
\nabla_0\Om=\hat\del_0=\Om~~.
\en
Similarly
\eq
\bar\nabla_{\bar\imath}\bar\Om=0~~,~~ 
\bar\nabla_{\bar 0}\Om=0~~.\label{holomorphicOm}
\en
Recalling 
\eqn{Omnabu} we obtain
\eqa
\le\Om,\nabla_i\Om\re&=&0 \label{ort1}~\\
\le \nabla_i\Om,\nabla_j\Om\re&=&0 \label{ort2}~\\
\le\bar\Om,\nabla_i\Om\re&=&0 \label{ort3}~.
\ena
Notice also that $\le\Om,\bar\Om\re$ is invariant
under horizontal vector fields,
\eq\label{horonOm}
\hat\del_i\le\Om,\bar\Om\re=
\nabla_i\le\Om,\bar\Om\re=
\le\nabla_i\Om,\bar\Om\re+
\le\Om,\nabla_i\bar\Om\re=0
\en
where in the last passage we used 
\eqn{Omnabu} and \eqn{holomorphicOm}. 
Similarly 
$\bar\nabla_{\bar\imath}\le\Om,\bar\Om\re=0$.

The metric associated to the \K form 
\eqn{pseudoKform} on $\tilde M$
is block diagonal in the $\hat\del_0,\hat\del_i$ basis, (see paragraph following 
\eqn{pseudoKform}), 
\eq\label{lorentianmetric}
\left(
\begin{array}{cc}
\tilde g_{0\bar 0} & 0\\
 0 & \tilde g_{i\bar\jmath}
\end{array}\right)=
\left(
\begin{array}{cc}
-\la\bar \la |s|^2& 0\\
 0 &\la\bar\la |s|^2 g_{i\bar\jmath}\!\circ\!\pi
\end{array}\right)=
\left(
\begin{array}{cc}
-|\Om|^2& 0\\
 0 &|\Om|^2\,g_{i\bar\jmath}\!\circ\!\pi
\end{array}\right)~.
\en
Because of \eqn{horonOm} the associated Levi-Civita 
connection coefficients of $\tilde M$ in the $\hat\del_i$ basis of horizontal 
vectors coincide with those of $M$ in the  $\partial\over{\partial z^i}$ basis,
\eq
\tilde\Ga^\ell_{ij}=\tilde g^{\bar k\ell}\hat\del_i\tilde g_{j\bar k}=
g^{\bar k\ell}\del_i g_{j\bar k}=
\Ga^\ell_{ij}~.
\en

In terms of  the symplectic 
frame $\{\mathsf{e_\xi}\}=\{\mathsf{e}_\La, 
\mathsf{f}^\La\}$, that is flat,  
we have $\nabla \Om=
\nabla(\Om^\xi\mathsf{e_\xi})=
d(\Om^\xi)\mathsf{e_\xi}$, and
$
\nabla_i\Om= \hat\del_i(\Om^\xi)
\mathsf{e_\xi}=\del_i\Om^\xi
+\del_i\KK \Om^\xi$, i.e.,
\eq\label{nablaXF}
\nabla_i \left(
\begin{array}{c}
X^\La  \\
F_\La
\end{array}\right)=\del_i \left(
\begin{array}{c}
X^\La  \\
F_\La
\end{array}\right)+\del_i\KK \left(
\begin{array}{c}
X^\La  \\
F_\La
\end{array}\right)~.
\en
Recalling the interpretation of $X^\La$ or $F_\La$ as
coefficients of local sections of $L^{-1}\rightarrow M$,
we read in equation \eqn{nablaXF}  the covariant derivative
of $L^{-1}\rightarrow M$.

\sk
It is also convenient to normalize $\Om$ 
and thus consider the 
(non holomorphic) 
nonvanishing global vector field on 
$\tilde M$ given by 
\eq\label{secVOm}
\VVV=e^{{\cal K}/2}\Om~.
\en
\sk
>From \eqn{horonOm} the covariant derivatives of $\VVV$ are
\eqa
\nabla_i\VVV=e^{\KK/2}\nabla_i\Om\nn~~&,&~~
\bar\nabla_{\bar\imath}\VVV=e^{\KK/2}\bar\nabla_{\bar\imath}\Om=0~,\\[.3em]
\bar\nabla_{\bar\imath}\bar \VVV=e^{\KK/2}
\bar\nabla_{\bar\imath}\bar\Om\nn~~&,&~~
\nabla_{i}\bar \VVV=e^{\KK/2}\nabla_{i}\bar\Om=0~.
\ena
Explicitly we have\footnote{we find also instructive to obtain the covariant derivative 
of the section $\VVV$ via this straighforward calculation that uses $\la{\del\over\del\la}\KK=-1$,
$$
\nabla_i\VVV=\nabla_i(e^{\KK/2}\Om^\xi\mathsf{e}_\xi)=
\hat{\del}_i(e^{\KK/2}\Om^\xi)\mathsf{e}_\xi=
\del_i(e^{\KK/2}\Om^\xi)\mathsf{e}_\xi+
\del_i\KK {\la}{\del\over\del\la}(e^{\KK/2}\Om^\xi)
\mathsf{e}_\xi
=(\partial_iV^\xi+{1\over 2}\del_i {\cal K}_{\,}V^\xi)\,\mathsf{e}_\xi~.
$$}
\eqa\label{nablaVdel}
\nabla_i\VVV=(\partial_iV^\xi+{1\over 2}\del_i {\cal K}_{\,}V^\xi)\,\mathsf{e}_\xi~~&,&~~
\bar\nabla_{\bar\imath}\VVV=(\bar{{\del}}_{\bar\imath}V^\xi-{1\over 2}\bar\del_{\bar\imath}{\cal K} \,V^\xi)\,\mathsf{e}_\xi=0
\label{covhol} \\[.3em]
\bar\nabla_{\bar\imath}\bar \VVV =
(\bar{{\del}}_{\bar\imath}\bar V^\xi+{1\over 2}
\bar{{\del}}_{\bar\imath}{\cal K}\,\bar V^\xi)\,\mathsf{e}_\xi
~~&,&~~
\nabla_{i}\bar \VVV=({{\del}}_{i}\bar V^\xi-{1\over 2}\del_{\bar\imath}{\cal K} \,\bar V^\xi)\,\mathsf{e}_\xi=0~.
\ena
Each coefficient $V^\xi$ of $\VVV$ 
with respect to the $\cc^\times$ 
invariant basis $\mathsf{e}_\xi$ is also a coefficient of a local section of 
the bundle $L^{-1/2}\otimes \bar L^{1/2}
\rightarrow M$.
This bundle has 
connection ${1\over 2}\del_i\KK-{1\over 2}\bar\del_{\bar\imath}\KK$. 
Equation \eqn{nablaVdel} can be 
interpreted as the covariant derivative
of these line bundle local sections.

{}From 
\eqn{ortOm}, and \eqn{ort1}-\eqn{ort3} we have 

\eqa
\label{ortog1}
\le \VVV,\bar \VVV\re&=&-i~~,~\\
\le \VVV,\nabla_i\VVV\re&=&0~~,~ \label{vort1}\\
\le \nabla_i\VVV,\nabla_j\VVV\re&=&0~~,~ \label{vort2}\\
\le \VVV,\bar\nabla_{\bar\imath}\bar \VVV\re&=&0 \label{ortog1bis}~~.
\ena
>From \eqn{lorentianmetric}, or also from 
$[\nabla_j,\bar\nabla_{\bar \imath}]=
-{{\partial}}_j\bar{{\del}}_{\imath}{\cal K}=
-g_{j\bar{\imath}}$ and $\le\bar\nabla_{\bar\imath}\nabla_j\VVV,\bar \VVV\re+\le\nabla_j\VVV,\bar\nabla_{\bar\imath}\bar \VVV\re=0$, we have 
\eq\label{ortog2}
\le \nabla_j \VVV,\bar\nabla_{\bar \imath}\bar \VVV\re=i g_{j\bar\imath}~,
\en
(where $g_{j\bar{\imath}}=\partial_j\bar{{\del}}_{\imath}{\cal K}=-2\pi i\,\pi^* K_{j\bar{\imath}}$ 
is actually $g_{j\bar{\imath}}\!\circ\!\pi$, the pull back via $\pi$ of the positive definite metric on $M$).
If we consider an orthonormal frame
$\{e_I \}$, ($I=1,\ldots \pn$) on $M$, 
\eq\label{eedelg}
e_I=e_I^j\partial_j~~,~~ \partial_j=e_j^I e_I~,~
g_{j\bar\imath}=e_j^I \,\bar e^{\bar J}_{\bar \imath}
\,\delta_{I\bar J}   ~,
\en 
we lift this frame to a frame of horizontal vectors of $T^{(1,0)}\tilde M$, and if we set 
\eq\label{VdelbarVi}
\VVV_M=(\VVV,\bar\nabla_{\bar I}\bar \VVV)~~~,~~~~M=0,1,\ldots \pn~,
\en 
(where 
$\bar \nabla_{\bar I}=\bar e^{\bar\imath}_{\bar I}\,\bar\nabla_{\bar\imath}$),
then relations \eqn{vort1}, \eqn{vort2}, \eqn{ortog1}, \eqn{ortog2} read
\eq
\le  \VVV_M ,\VVV_N\re=0~~,~~~\le  \bar \VVV_M ,\VVV_N\re=i\delta_{MN}~.
\label{N2specG}\en
The index $M$ mixes holomorphic and 
antiholomorphic indices in order to
compensate for the Lorentian signature
of the metric 
$\big({}^{-1}_{\,\,0}{\,}^{\,\,0}_{g_{j\bar\imath}}\big)$ in  \eqn{ortog1}, \eqn{ortog2}.

Explicitly the column vectors of the
 components 
of the sections $\VVV_M=V_{~M}^\xi\mathsf{e}_\xi$ 
are
\eq
\left( V^\xi\right)=\left(
\begin{array}{c}
L^\La  \\
M_\La
\end{array}\right)=e^{{\cal K}/2}\left(
\begin{array}{c}
X^\La  \\
F_\La
\end{array}\right)~~,~~~
\left(\bar\nabla_{\bar I} \bar V^\xi\right) = \left(
\begin{array}{c}
\bar\nabla_{\bar I}L^\La  \\
\bar\nabla_{\bar I}M^\La  
\end{array}\right)~~,~~~
\en
and they can be organized in a $2\nm\times \nm$  matrix
\eq(V^\xi_{~M})=
(V,\bar\nabla_{\bar I}\bar V^\xi)=
\left(
\begin{array}{cc}
L^\La & \bar\nabla_{\bar I}L^\La  \\
M_\La &\, \bar\nabla_{\bar I}M^\La  
\end{array}\right)= \left(
\begin{array}{c}
f_{~M}^\La  \\
h_{\La M}
\end{array}\right)=\left(\begin{array}{c}
f  \\
h
\end{array}\right)~.\label{VLf}
\en
In the last passage we have denoted by $f$ 
(respectively $h$) the 
$\nm\times \nm$ matrix of entries
$f_{~M}^\La$ (respectively  $h_{\La M}$).

The $N=2$ special geometry relations \eqn{N2specG}
are equivalent to 
\eq\label{compsymp1s}
(f^\dagger,h^\dagger) \left(
\begin{array}{cc}
0 & -1\!\!1\\
1\!\!1 & 0
\end{array}\right)
\left(\begin{array}{c}
f  \\
h
\end{array}\right)=
i
 1\!\!1
~~~~~~i.e.~~ -f^\dagger h+h^\dagger f=i 1\!\!1
\en
and
\eq\label{compsymp2s}
(f^t,h^t)\left(
\begin{array}{cc}
0 & -1\!\!1\\
1\!\!1 & 0
\end{array}\right)
\left(\begin{array}{c}
f  \\
h
\end{array}\right)=0
~~~~~~~~~~~~~i.e.~~ -f^th+h^tf=0
\en
These two relations are equivalent to require the real 
matrix
\eq
\left(
\begin{array}{cc}
A & B\\
C & D
\end{array}\right)=\sqrt{2}
\left(\begin{array}{cc}
{\rm Re} f &\,-{\rm Im} f  \\
{\rm Re} h &\,-{\rm Im} h
\end{array}\right)
\en
to be symplectic.
Vice versa any symplectic matrix $\big({}^A_C{}^B_D\big)$
leads to relations \eqn{compsymp1s}, \eqn{compsymp2s} by 
defining
$
\big({}^f_
h\big)=
{1\over \sqrt{2}} \big({}^{A-iB}_{C-iD}
\big)\,.
$
The matrix  
\eq
\VV=\left(\begin{array}{cc}
f  &\bar f\\
h  &\bar h
\end{array}\right)=\left(
\begin{array}{cc}
A & B\\
C & D
\end{array}\right){\cal A}~,
\en
where $
{\cal A}={1\over \sqrt{2}}
\big({}^{~1\!\!1}_{-i1\!\!1}  {}^{\,1\!\!1}_{i1\!\!1}\big)$,
rotates the flat real symplectic frame $\{\mathsf{e_\xi}\}=
\{\mathsf{e}^\La,\mathsf{f}_\La\}$ in the frame 
$\{\VVV_M,\bar \VVV_{\bar M}\}$ that up to a 
rotation by ${\cal A}^{-1}={\cal A}^\dagger$ is also real 
and symplectic (but not flat). This $\{\VVV_M,\bar \VVV_{\bar M}\}$
frame comes from a local coordinate frame on $M$, indeed 
$\bar \VVV_{\bar M}=( e^{\KK/2}\bar\Om, 
e^{\KK/2}e^j_I \hat\partial_j)$. 
The symplectic connection 1-form in this 
frame is simply $\Gamma=\VV^{-1}d\VV$, 
indeed $\nabla \mathsf{e_\xi}=0$ is 
equivalent to 
\eq
d\VV=\VV \Gamma ~.
\en
We can write 
$
\Ga=
\left(
\begin{array}{cc}
\om & {\cal \bar P} \\
{\cal  P}  & \bar\om
\end{array}\right)\,,
$
and see this equation as a condition on 
the Levi-Civita connection $\om$ 
 and the tensor $\cal P$ of $\tilde M$. 
The block decomposition 
$\big({\,}^{\om}_{\cal P} {\,}_{\bar\om}^{\cal\bar P}\,\big)
$
follows by recalling that $\tilde M$ is in 
particular a rigid special \K manifold.
The difference ${\cal P}_\rr=\nabla-D$  between the flat symplectic
connection and the Levi-Civita connection is given by the 
holomorphic symmetric three form $C$  (c.f. \eqn{Cnablaxi})
\eq
C=-\le\nabla\Om,\nabla\nabla\Om\re~.
\en
The properties of $C$ 
previously discussed in the rigid case 
apply also to this projective special 
geometry case.

\sk
\subsection{The $N=2$ theory}
>From the previous section we see that the 
$N=2$ supergravity theories and the higher 
$N$ theories have a similar flat symplectic 
structure. The formalism is  the same, 
indeed since the antisymmetric of the $U(2)$ 
authomorphism group of the $N=2$ 
supersymmetry algebra is a singlet 
we have
\eq
f^\La_{~AB}=f^\La_{~0}\epsilon_{AB}~~,\label{5-70}
h_{\La AB}=h_{\La 0}\epsilon_{AB}~~
\en
where $f^\La_{~0},h_{\La 0}$ are the components of the global section  
$\VVV$, therefore from 
\eqn{VLf} we have as in \eqn{enphasizefh},
\eqa
f &=& (f^\La_{~M}) = (f^\La_{~AB},   \bar f^\La_{~\bar I})~,\nn\\[.2em]
h &=& (h_{\La M}) = (h_{\La AB},   \bar h_{\La\bar I})~,\label{5-71}
\ena
as it should be, the sections 
$\left(^{\bar f^\La_{~\bar I}}
_{\bar h_{\La\bar I}}\right)$ 
have \K weight opposite to the 
$\left(^{f^\La_{~AB}}_{h_{\La I}}\right)$ sections.
\sk
The difference between the $N=2$ cases 
and the $N>2$ cases is that 
the scalar manifold $M$ of the $N=2$ case 
is not in general a coset manifold. The 
flat symplectic bundle 
is therefore not in general a trivial bundle. The 
gauge kinetic term $\N_{\La\Sigma}=h_{\La M}{f^{-1}}^M_{~\Sigma}$
depends on the choice of the flat symplectic frame 
$\{\mathsf{e}_\xi\}=\{\mathsf{e}_\La, \mathsf{f}^\La\}$.
This latter can be defined only locally on
$\tilde M$ (and therefore on $M$). 
In another region we have a different frame
$\{\mathsf{e'}_\xi\}=\{\mathsf{e'}_\La, \mathsf{f'}^\La\}$
and therefore a different gauge kinetic term $\N'_{\La\Sigma}$.
In the common overlapping region the two formulations should give the 
same theory, this is indeed 
the case because the corresponding equations of motion 
are related by a duality rotation.
As a consequence the notion of electric or magnetic charge depends on the 
flat frame chosen. 
In this sense the notion of electric and magnetic charge is not a 
fundamental one.  The symplectic group is a gauge group (where just 
constant gauge transformations are allowed) and only gauge invariant 
quantities are physical.

A related aspect of the comparison between the $N=2$ and the $N>2$ theories is that the 
special \K  structure determines 
the presence of a new geometric quantity, the holomorphic cubic form $C$, 
which physically corresponds to the anomalous magnetic moments 
of the $N=2$ theory. When the special \K manifold 
$M$ is itself a coset manifold \cite{cp},
then the anomalous magnetic moments $C_{ijk}$ are expressible in
terms of the vielbein of ${G}/H$, this is for example the case 
of the $N=2$ theories with scalar manifold 
$G/H= {SU(1,1)\over U(1)}\times {O(6,2)\over {O(6) \times O(2)}}$
and $G/H = {SO^\star (12) \over U(6)}$ 
\cite{cp}. 
\sk
 To complete the analogy
between the $N=2$ theory with $n'$ 
vector multiplets and the higher $N$ 
theories in $D=4$,
we also give the supersymmetry transformation laws, the central and 
matter charges, the differential relations among them and the formula for the potential $\VBH$.

The  supercovariant electric field strength 
$\hat F^\Lambda$ is
\begin{equation}
\hat F^\Lambda  = F^\Lambda + f^\Lambda \bar\psi^A \psi ^B
\epsilon_{AB} -{\rm i} \bar f^\Lambda_{\bar\imath}
\bar \lambda^{\bar\imath}_A\gamma_a\psi_B\epsilon^{AB}V^a + h. c.
\end{equation}
The transformation laws for the chiral gravitino $\psi_A$ and
gaugino $\lambda^{iA}$ fields are:
\begin{equation}
 \delta \psi_{A \mu}={ \nabla}_{\mu}\,\epsilon _A\,+ \epsilon _{AB}
T_{\mu\nu} \gamma^\nu \epsilon^B + \cdots\,, \label{trasfgrav}
 \end{equation}
 \begin{equation}
 \delta\lambda^{iA} = {\rmi} \partial_\mu z^i \gamma^\mu\epsilon^A +
 {\frac{\rmi} 2}
\bar T_{\bar \jmath\mu\nu} \gamma^{\mu \nu}
g^{i\bar\jmath}\epsilon^{AB}\epsilon_B + \cdots\,,
\label{gaugintrasfm}
\end{equation}
where:
\begin{equation}
T =  h_\Lambda F ^{\Lambda}  - f^\Lambda G_\Lambda\,,
\label{T-def}
\end{equation}
\begin{equation}
\bar T_{\bar \imath} =  \bar T_{\bar I} 
\bar e^{\bar I}_{~\bar\imath}\,,\mbox{ with }
\bar T_{\bar I} = \bar h_{\Lambda {\bar I}}  
F^{\Lambda}  - \bar f^\Lambda_{\bar I}
G_\Lambda\,,  \label{G-def}
\end{equation}
are respectively the graviphoton and the matter vectors. In \eqn{trasfgrav},
\eqn{gaugintrasfm} the
position of the $SU(2)$ automorphism index 
$A$ ($A,B=1,2$) is related to chirality, namely $(\psi_A, \lambda^{iA})$ are chiral, $(\psi^A,
 \lambda^{\bar\imath}_A)$ antichiral.
\sk
In order to define the symplectic invariant charges 
let us recall the definition of the
magnetic and electric charges (the moduli independent charges) 
in (\ref{pq}). The central charges and
the matter charges are then defined as the 
integrals over a sphere at spatial infinity 
of the dressed  graviphoton and matter 
vectors \eqn{gravi}, they are given in \eqn{centralcharges1}, \eqn{centralcharges2}:
\begin{equation}
\left(Z_{M}\right)=\left(Z, \bar Z_{\bar I}
\right) = i\overline {\VV(\phi_\infty)}^{-1}Q
\en
where $\phi_{\infty}$ is the value of the scalar fields at spatial infinity.
Because of (\ref{VdelbarVi}) we
get immediately:
\begin{equation}
\nabla_I Z=Z_I\label{fundeq}\,.
\end{equation}
This relation can also be written
 $\nabla_I Z_{AB}=Z_I \epsilon_{AB}$, and 
considering the vielbein 1-form ${\cal P}^I$ dual 
to the frame $e_I$ introduced in \eqn{eedelg}
and setting $\nabla\equiv {\cal P}^I\nabla_I $ 
we obtain 
$\nabla Z_{AB}=Z_I {\cal P}^I \epsilon_{AB}~.$

 The positive definite quadratic invariant $\VBH$ in terms of the charges
  $Z$ and $Z_I$ reads
  \begin{equation}
\VBH= {1\over 2} Z \bar Z +   Z_I \bar Z^I =
  -\frac{1}{2} Q^t \cM(\cN) Q~.
  \label{sumrules}
\end{equation}
Equation (\ref{sumrules}) is obtained by using exactly the same
procedure as in (\ref{mothersumrule}). Invariance of $\VBH$ implies that it 
is a well defined positive function on $M$.
\sk

\section{Duality rotations in Noncommutative Spacetime}

Field theories on noncommutative spaces have received 
renewed interest since their relevance in describing 
Dp-branes effective actions (see \cite{Seiberg:1999vs} and 
references therein).  
Noncommutativity in this context is due to a nonvanishing 
NS background two form on the Dp-brane. 
First space-like (magnetic) backgrounds ($B^{ij}\not= 0$) 
were considered, then NCYM theories also with time 
noncommutativity ($B^{0i}\not=0$) have been studied 
\cite{Seiberg:2000gc}.
The NCYM theories that can be obtained from open strings in the 
decoupling limit $\ap\rightarrow 0$ 
are those with $B$ space-like or light-like (e.g. $B_{0i}=-B_{1i}$),
these were also considered the only theories without unitatrity problems
\cite{Aharony:2000gz}, however by applying a proper perturbative setup
it was shown that also time-space noncommutative field theories can be unitary 
\cite{Bahns}.

Following \cite{Seiberg:1999vs}, gauge 
theory on a Dp-brane with constant two-form $B$  
can be described via a commutative Lagrangian and field strength
$\CL(F+B)$ or via a  noncommutative one  $\widehat{\CL}(\Fh)$,
where $\Fh_{\mu\nu}=\pa_\mu A_\nu - \pa_\nu A_\mu 
-i[ A_\mu\star_{\!\!\!{\textstyle{,}}} A_\nu]$. Here $\star$ is the star 
product,  on coordinates
$[x^\mu\star_{\!\!\!{\textstyle{,}}}x^\nu]=
x^\mu\star x^\nu-x^\nu\star x^\nu=i\th^{\mu\nu}$, where $\th$ depends 
on $B$  and the metric on the Dp-brane.          
The commutative and the noncommutative descriptions are complementary and are related by
Seiberg-Witten map (SW map) \cite{Seiberg:1999vs}, \cite{WessSW, Wessmath}. In the $\ap\rightarrow 0$ 
limit \cite{Seiberg:1999vs} the exact 
effective electromagnetic theory on a Dp-brane is
noncommutative electromagnetism (NCEM), this is equivalent, via SW map, 
to a nonlinear commutative $U(1)$ gauge theory. 

In this section we consider a D3-brane action in the slowly varying field approximation,
we give an explicit expression of this nonlinear $U(1)$ theory and 
we show that it is self-dual when $B$ (or $\th$) is light-like. 
Via SW map solutions of $U(1)$ nonlinear 
electromagnetism are mapped into solutions of NCEM, so that
duality rotations are  also a symmetry of NCEM, i.e., NCEM is self-dual
\cite{Aschieriproc}, \cite{Aschieri:2001zj}.
When $\th$ is space-like we do not have self-duality and the 
S-dual of space-like NCYM is a 
noncommutative open string theory decoupled from closed strings 
\cite{Gopakumar:2000na}. Related work appeared 
in \cite{Rey:2001hh, Lu:2001vv, Alishahiha:2000pu}. We mention that
self-duality  of NCEM was initially studied in
\cite{Ganor:2000my} to first order in $\th$.
On one hand it is per se interesting to provide new
examples of self-dual nonlinear electromagnetism, as the one we
give with the lagrangian  (\ref{L}). 
On the other hand this lagrangian 
is via Seiberg-Witten map, and for slowly varying fields, just NCEM. 
Formally NCEM
resembles  $U(N)$ YM on commutative space, and on tori with rational $\th$
the two theories are $T$-dual \cite{Schwarz:1998qj}.
Self-duality of NCEM then hints to a possible duality symmetry 
property of the equations of motion of $U(N)$ YM.

\sk
\noi{\bf{Self-Duality of the $D3$-brane action}}\\
Consider the $D3$-brane  effective action in a IIB supergravity background 
with constant axion, dilaton NS and RR two-forms. 
The background two-forms can be gauged away in the bulk and we are left
with the field strength $\CF=F+B$ on the D3-brane. 
Here $B$ is defined as the constant part of $\CF$, or
$B=\CF|_{\rm{spatial}\, \infty}$ since 
$F$ vanish at spatial infinity.
{}For slowly varying fields
the Lagrangian, 
in Einstein frame is essentially the Born-Infeld action with axion and dilaton.
We set for simplicity $\N=-i1\!\!1$ and $g_s=1$, where $g_s$ is the string 
coupling constant.  The lagrangian is then
$\LL=\frac{-1}{\ap^2}\sqrt{-{\rm{det}}(g+\ap {\CF})}$.
The explicit expression of $\CK$, is obtained from the definition 
$\CK:={\del \LL\over \del F}$ and is (cf. \eqn{Geasy})  
\eq
{\CK}_{\mu\nu}=
\frac{\CF^*_{\,\mu\nu}+\frac{\ap^2}{4}\CF\CF^*\,\CF_{\mu\nu}}{
\sqrt{1+\frac{\ap^2}{2} \CF^2-\frac{\ap^4}{16}(\CF\CF^*)^2}}~.
\label{Ktransf}
\en
Here $\CF^*_{\,\mu\nu}=\sqrt{g}\epsilon_{\mu\nu\rho\sigma}\CF^{\rho\sigma}$, cf. footnote 2, Section 2.1.
One can then consider a duality rotation by an angle $\gamma$ 
and extract how $B$ 
(the constant part of $\CF$) transforms
\eq
{B'}_{\mu\nu}=\cos\!\gamma\, B_{\mu\nu} - \sin\!\gamma\,
\frac{B^*_{\,\mu\nu}+\frac{\ap^2}{4}B B^*\, B_{\mu\nu}}{
\sqrt{1+\frac{\ap^2}{2} B^2-\frac{\ap^4}{16}(B B^*)^2}}\label{Brot}~.
\en
\sk
\noi{\bf{Open/closed strings and light-like noncommutativity}}
\label{Seiberg-Witten}\\
\noi The open and closed string parameters are related by 
(see \cite{Seiberg:1999vs}, 
the expressions for ${\mathsf{G}}$ and $\th$ first appeared in 
\cite{Chu:1999qz}) 
\eq
\begin{array}{l}
\displaystyle
\frac{1}{g+\ap B}={\mathsf{G}}^{-1}+\frac{\th}{\ap} ~~\nonumber\\[1em]
\displaystyle
g^{-1}=({\mathsf{G}}^{-1}-\th/\ap)\,{\mathsf{G}}\,({\mathsf{G}}^{-1}+\th/\ap)={\mathsf{G}}^{-1}-\ap^{-2}
\th\, {\mathsf{G}}\, \th\nonumber\\[1em]
\displaystyle
\ap B=-({\mathsf{G}}^{-1}-\th/\ap)\,\th/\ap\,({\mathsf{G}}^{-1}+\th/\ap)\nonumber\\[1em]
\displaystyle
{\mathsf{G}}_s=g_s 
\sqrt{\frac{{\rm{det}}{\mathsf{G}}}{\det(g+\ap B)}}=g_s\sqrt{\det {\mathsf{G}}\;
\det{({\mathsf{G}}^{-1}+\th/\ap)}}=g_s\sqrt{\det g^{-1}\;
\det{(g+\ap B)}}
\end{array}
\label{Gsgs}
\en 
The decoupling limit 
$\ap\rightarrow 0$ with ${\mathsf{G}}_s,{\mathsf{G}},\th$ 
nonzero and finite 
\cite{Seiberg:1999vs}  
leads to a well defined field theory only if
$B$ is space-like or
light-like. 
Looking at the closed and open string coupling constants it is 
easy to see why one needs this space-like or light-like condition on 
$B$ in performing this limit.
Consider the coupling constants ratio ${\mathsf{G}}_s/g_s$, that
expanding the 4x4 determinant reads 
(here $B^2=B_{\mu\nu}B_{\rho\sigma}g^{\mu\rho}g^{\nu\sigma}$, 
$\th^2=\th^{\mu\nu}\th^{\rho\sigma}{\mathsf{G}}_{\mu\rho}{\mathsf{G}}_{\nu\sigma}$ and so on)
\eq
\frac{{\mathsf{G}}_s}{g_s}=\sqrt{1+\frac{\ap^{-2}}{2} 
\th^2-\frac{\ap^{-4}}{16}(\th \th^*)^2} 
\;=\;
\sqrt{1+\frac{\ap^2}{2} B^2-\frac{\ap^4}{16}(B B^*)^2}\label{Gs/gs}~.
\en
Both ${\mathsf{G}}_s$ and $g_s$ must be positive; since ${\mathsf{G}}$ and $\th$ are by
definition 
finite for $\ap\rightarrow 0$ this implies 
$\th \th^*=0$ and $\th^2\geq 0$. Now  $\th \th^*=0 \Leftrightarrow 
\det\th=0 \Leftrightarrow \det B=0 \Leftrightarrow B B^*=0$.
In this case from (\ref{Gs/gs}) we also have $\th^2=\ap^4B^2$.
In  conclusion in order for the $\ap\rightarrow 0$ limit defined by
keeping ${\mathsf{G}}_s,{\mathsf{G}},\th$ nonzero and finite \cite{Seiberg:1999vs}, 
to be well defined we need
\eq 
B^2\geq 0~,~~B B^*= 0
~~~~~~~\mbox{i.e.}~~~~~~~\th^2\geq 0~,~~\th \th^*=0
\label{Bth}
\en
This is the condition for $B$ (and 
$\th$) to be space-like or light-like.
Indeed with Minkowski metric and in three vector notation (\ref{Bth}) reads 
${\Bv}^2-\Ev^2\geq 0$ and $\Ev\perp\Bv$. 

If we now require the $\ap\rightarrow 0$ limit to be compatible with 
duality rotations, we immediately see that we have to consider only the 
light-like case $B^2=  B B^*= 0$. Indeed under $U(1)$ rotations 
the electric and magnetic fields mix up, in particular under a $\pi/2$
rotation (\ref{Brot}) a space-like $B$ becomes time-like. 

In the light-like case det$(g+\al' B)\!=\!$ det$(g)$,  relations (\ref{Gsgs}) simplify considerably.
The open and closed string coupling constants coincide, since we set $g_s=1$
we have ${\mathsf{G}}_s=g_s=1$, this also implies 
det$(\mathsf{G})\!=\!$ det$(g)$
so that the hodge dual field $F^\star$ with the $g$ metric equals the one
with the $\mathsf{G}$ metric.
Use of the relations
\eq
\Omega^*_{\mu\rho}{\Omega^*}^{\rho\nu}-
\Omega_{\mu\rho}\Omega^{\rho\nu}=
\frac{1}{2}\Omega^2\,\delta_{\mu}^{~\nu}~\,,\,~~ 
\Omega_{\mu\rho}{\Omega^*}^{\rho\nu}=
{\Omega^*}_{\mu\rho}\Omega^{\rho\nu}=
\frac{-1}{4}\Omega\Omega^*\,\delta_\mu^{~\nu}\label{S}
\en
valid for any antisymmetric tensor $\Omega$, shows that 
any two-tensor at least cubic in $\th$ (or $B$) vanishes.
It follows that $g^{-1}{\mathsf{G}}\,\th=\th$ and that the 
raising or lowering of the  
$\th$ and $B$ indices is independent from the metric used.
We also have
\eq
B_{\mu\nu}=-{\ap}^{-2}\th_{\mu\nu}\label{Btheta}~~.~~ ~
\en
\sk
\noi{\bf{Self-duality of NCBI and NCEM}}

\noi We now study duality rotations for noncommutative 
Born-Infeld (NCBI) theory and its zero slope limit that is NCEM.
The relation between the NCBI and the BI Lagrangians is 
\cite{Seiberg:1999vs}
\eq
\CLh_{BI}(\Fh,{\mathsf{G}},\th,{\mathsf{G}}_s)=\CL_{BI}(F+B,g)+O(\pa F)+\rm{tot.der.}
\label{NCDBI}
\en
where $O(\pa F)$ stands for higher order derivative corrections, 
$\Fh$ is the noncommutative $U(1)$ field strength and we have set 
$g_s=1$.
The NCBI Lagrangian is
\eq
\CLh_{BI}(\Fh,{\mathsf{G}},\th,{\mathsf{G}}_s)=
\frac{-1}{\ap^2 {\mathsf{G}}_s}\sqrt{-{\rm{det}}({\mathsf{G}}+\ap {\Fh})}
+O(\pa\Fh)~.
\en
In the slowly varying field approximation the action of duality 
rotations on $\CLh_{BI}$  
is derived from self-duality of $\CL_{BI}$. 
If $\Fh$ is a solution of the 
$\CLh_{BI}^{{\mathsf{G}}_s,{\mathsf{G}},\th}$ EOM then $\Fh'$ obtained via 
$$\Fh\stackrel{\sma{\rm{SW map}}}{\longleftarrow\!\!\!\longrightarrow}
\CF\stackrel{\sma{\rm{duality\, rot.}}}{\longleftarrow\!\!\longrightarrow}
\CF'
\stackrel{\sma{\rm{SW map}}}{\longleftarrow\!\!\!\longrightarrow}
\Fh'$$ is a solution of the 
$\CLh_{BI}^{{\mathsf{G}}^\prime_{\!s},{\mathsf{G}}',\th'}$ EOM where ${\mathsf{G}}^\prime_s,{\mathsf{G}}',\th'$
are obtained using (\ref{Gsgs}) from $g'$,  $B'$ and $g_s'=g_s=1$. 

In the light-like case we have ${\mathsf{G}}_s=g_s=1$, the
$B$ rotation (\ref{Brot}) simplifies to
\eq
B^\prime_{\mu\nu}=\cos\!\gamma\,B_{\mu\nu}-\sin\!\gamma\,
B^*_{\,\mu\nu}~.\label{Bsrot}
\en
Using (\ref{Bsrot}) the $U(1)$ duality  
action on the open  string variables is 
\eq
{\mathsf{G}}^\prime={\mathsf{G}}~~,~~~\th^{\prime\,\mu\nu}=
\cos\!\gamma\,\th^{\mu\nu}-\sin\!\gamma\,{\th^*}^{\mu\nu}~.
\label{Tsrot}
\en
For $\th$ light-like, solutions $\Fh$ of $\widehat{\CL}^{{\mathsf{G}},\th}$ 
are mapped
into solutions $\Fh'$ of $\widehat{\CL}^{{\mathsf{G}},\th'}$.
Thus we can map solutions of $\widehat{\CL}^{{\mathsf{G}},\th}$ into
solutions of $\widehat{\CL}^{{\mathsf{G}},\th}$,  
therefore the theory described by $\widehat{\CL}^{{\mathsf{G}},\th}$ has 
$U(1)$ duality rotation symmetry.

In order to show self-duality of NCEM we consider the
zero slope limit of (\ref{NCDBI}) and verify that  
the resulting lagrangian
on the r.h.s. of (\ref{NCDBI}) is self-dual.
We
rewrite $\CL_{BI}$ in terms of the open string parameters ${\mathsf{G}},\th$
\eqa
\CL_{BI}&=&\frac{-1}{\ap^2 }\sqrt{-{\rm{det}}(g+\ap\CF)}
=\frac{-\sqrt{{\mathsf{G}}}}{\ap^2 }\sqrt{\frac{{\rm{det}}(g+\ap B+\ap{F})}
{{\rm{det}}(g+\ap B)}}\nonumber\\[1.2em]
&=&\frac{-1}{\ap^2 }\sqrt{-{\rm{det}}({\mathsf{G}}+\ap{F}+{\mathsf{G}}\th F)}
\label{LGth}~.
\ena
The determinant in the last line can be evaluated as sum of
products of traces (Newton-Leverrier formula). Each trace can 
then be rewritten in terms of the six basic Lorentz invariants
$F^2,~F F^*,~F\th,~F\th^*,~\th^2=\th\th^*=0$, explicitly
$$
\begin{array}{l}
{\rm{det}}{{\mathsf{G}}^{-1}}\,{\rm{det}}({\mathsf{G}}+\ap{F}+{\mathsf{G}}\th F)
=(1-\frac{1}{2}\th F)^2+\ap^2[\frac{1}{2}F^2+\frac{1}{4}
\th F^*\;FF^*]-\ap^4(\frac{1}{4}FF^*)^2
\nonumber
\end{array}~\,
$$
Finally we take the $\ap\rightarrow 0$ limit of 
(\ref{LGth}), by dropping the infinite constant and  total derivatives 
the resulting Lagrangian is $\sqrt{\mathsf{G}}$ times
\eq
\frac{-\frac{1}{4}F^2-
\frac{1}{8}\th F^*\,FF^*}{1-
\frac{1}{2}\th F}\label{L}~\,.
\en
We thus have an expression for NCEM in terms of $F$, $\th$ and
${\mathsf{G}}$ (of course $\mathsf{G}_{\mu\nu}$ can be taken $\eta_{\mu\nu}$), $\widehat{\LL}_{\rm{EM}}=\sqrt{\mathsf{G}}\,
\widehat{L}_{\rm{EM}}$,
\eq\label{lagncem}
\widehat{L}_{\rm{EM}}
\equiv-\frac{1}{4}{\Fh}{\Fh}=
\frac{-\frac{1}{4}F^2-
\frac{1}{8}\th F^*\,FF^*}{1-
\frac{1}{2}\th F}
+O(\pa F)+\rm{tot. ~der.}
\en
The Lagrangian (\ref{L}) satisfies the self-duality 
condition (\ref{thecondgen}) with 
$\varphi=\th$, $\kappa=0$, $a=d=0$, $c=-b$ and therefore NCEM is self-dual 
under the $U(1)$ duality 
rotations  (\ref{Tsrot}) and $F'=\cos\!\gamma\,F-\sin\!\gamma\,G$.
The change in $\th\rightarrow \th'$, that is not a dynamical field, 
can be cancelled by a rotation in space so that therefore we can map solution
of the EOM of \eqn{lagncem} into solutions of the EOM of 
\eqn{lagncem} with the same value of $\th$.

\sk
This duality can be enhanced to $Sp(2,\rr)$ by considering also axion 
and dilaton fields; also Higgs fields can be coupled, the coupling is minimal 
in the noncommutative theory. Using this duality 
one can relate space-noncommutative magnetic 
monopoles with a string  (D1-string D3-brane configuration) 
to space-noncommutative electric monopoles (possibly an F-string 
ending on a D3-brane) \cite{Aschieri:2001zj,Aschieri:2001gf}.
\sk
\newpage
\noi{\bf Acknowledgments}\\
We thank  
Laura Andrianopoli, Leonardo Castellani, 
Anna Ceresole, Riccardo D'auria, 
Pietro Fr\'e and Mario Trigiante for valuable discussions.  

The work of P.A. was supported in part by the European Community's Human 
Potential Program under contract MRTN-CT-2004-005104
``\textit{%
Constituents, Fundamental Forces and Symmetries of the Universe''} in
association with University of Alessandria, and in part by the
Italian MIUR under contract PRIN-2005023102.

The work of S.F. was supported in part by the European Community Human
Potential Programme under contract MRTN-CT-2004-503369 ``\textit{%
Constituents, Fundamental Forces and Symmetries of the Universe''} in
association with LNF-INFN and in part by DOE grant DE-FG03-91ER40662, Task C.

The work of B.Z. was supported in part by the
Director, Office of Science, Office of High Energy and Nuclear
Physics, Division of High Energy Physics of the U.S. Department of
Energy under Contract DE-AC03-76SF00098, in part by the National
Science Foundation under grant PHY-0457315.
\sk\sk
\section{Appendix: Symplectic group and transformations}
\subsection{Symplectic group ~($A,B,C,D$ and $f, h$ and $V$ matrices)}
The symplectic group $Sp(2n,\rr)$ is the group of real
$2n\times 2n$ matrices that satisfy
\eq
S^t \left(
\begin{array}{cc}
0 & -1\!\!1\\
1\!\!1 & 0
\end{array}\right) S=\left(
\begin{array}{cc}
0 & -1\!\!1\\
1\!\!1 & 0
\end{array}\right)
\en
Setting $S=\big( {}^A_B{}^C_D\big)$ we explicitly have
\eq\label{sp2nr}
A^tC-C^tA=0~~,~~~B^tD-D^tB=0~~,~~~A^tD-C^tB=1~.
\en
Since the transpose of a symplectic matrix is again symplectic we equivalently have
\eq
AB^t-BA^t=0~~,~~~CD^t-DC^t=0~~,~~~AD^t-BC^t=1~.
\en 
In particular $A^tC, B^tD, CA^{-1}, BD^{-1},A^{-1}B, D^{-1}C, AB^t,DC^t$ are symmetric matrices (in case they exist). 
\sk
If $D$ is invertible we have the factorization
\eq
 \left(
\begin{array}{cc}
A& B\\
C & D
\end{array}\right) =
\left(
\begin{array}{cc}
1\!\!1 & ~BD^{-1}\\
0 & 1\!\!1
\end{array}\right)
\left(
\begin{array}{cc}
{D^t}^{-1} & 0\\
0 & D
\end{array}\right)
\left(
\begin{array}{cc}
1\!\!1 & \,0\\
D^{-1}C &\, 1\!\!1
\end{array}\right)
\label{factorization}
\en
where $A={D^t}^{-1}+BD^{-1}C$ follows from $BD^{-1}={D^t}^{-1}B^t$.
\sk\sk
\noi {\bf The complex basis}

\noi 
It is often convenient to consider the complex basis 
${1\over \sqrt{2}}\big({}^{F+iG}_{F-iG}\big)$
rather than $\big({}^{F}_{G}\big)$.
 The transition from the real to the complex basis is given by the symplectic and unitary matrix $\A^{-1}$, where
\eq
{\cal A}={1\over \sqrt{2}}\left(
\begin{array}{cc}
1\!\!1  & 1\!\!1 \\
-i1\!\!1 & i1\!\!1
\end{array}\right)~~,~~~~{\cal A}^{-1}={\cal A}^\dagger~. \label{defAAm1}
\en
A symplectic matrix $S$, belonging to
the fundamental representation of  $Sp(2n,\rr)$, in the complex basis reads
\eq\label{defAeAt}
U={\cal A}^{-1} S {\cal A}~.
\en
There is a 1-1 correspondence between matrices $U$ as in \eqn{defAeAt} and
complex $2n\times 2n$ matrices belonging to $U(n,n)\cap Sp(2n,\cc)$,
\eq
U^\dagger 
\left(
\begin{array}{cc}
1\!\!1 & 0\\
0 & -1\!\!1
\end{array}\right) 
U=
\left(
\begin{array}{cc}
1\!\!1 & 0\\
0 & -1\!\!1
\end{array}\right) ~~~,~~~~~
U^t\left(
\begin{array}{cc}
0 & -1\!\!1 \\
1\!\!1 &0
\end{array}\right) 
U=
\left(
\begin{array}{cc}
0 & -1\!\!1 \\
1\!\!1 &0
\end{array}\right)~. \label{UtOmU}
\en 
Equations \eqn{UtOmU} define a representation of 
$Sp(2n,\rr)$ on the complex vector space $\cc^{2n}$.
It is the direct sum of the representations $\big({}^\psi_{\bar\psi}\big)$ 
and $\big({}^{~\psi}_{-\bar\psi}\big)$, these are real representations of 
real dimension $2n$. (The representation $\big({}^{\psi^{}}_{\bar\psi}\big)$ is the 
vector space of all linear combinations, with coefficients in $\rr$,
of vectors of the kind $\big({}^\psi_{\bar\psi}\big)$). 
\sk
The maximal compact subgroup of 
$U(n,n)$ is $U(n)\times U(n)$; because of
the second relation in \eqn{UtOmU}
the maximal compact subgroup of
$Sp(2n,\rr)$ is $U(n)$.
The usual embedding of $U(n)$ into 
the complex and the fundamental representations of $Sp(2n,\rr)$ are
respectively \eq
\left(\begin{array}{cc}
u & 0\\
0 &  \bar u
\end{array}\right)~~,~~~~~\left(
\begin{array}{cc}
\rea u &- \im u \\
\im u & \rea u 
\end{array}\right)~~,\label{UinUspSp}
\en
where $u$ belongs to the fundamental of  $U(n)$.
\sk
\sk
\noi{\bf The $f$ and $h$ matrices}

\noi The $f$ and $h$ matrices are $n\times n$ complex matrices that satisfy
the two conditions 
\eq\label{compsymp1}
(f^\dagger,h^\dagger) \left(
\begin{array}{cc}
0 & -1\!\!1\\
1\!\!1 & 0
\end{array}\right)
\left(\begin{array}{c}
f  \\
h
\end{array}\right)=
i
 1\!\!1
~~~~~~i.e.~~ -f^\dagger h+h^\dagger f=i 1\!\!1
\en
and
\eq\label{compsymp2}
(f^t,h^t)\left(
\begin{array}{cc}
0 & -1\!\!1\\
1\!\!1 & 0
\end{array}\right)
\left(\begin{array}{c}
f  \\
h
\end{array}\right)=0
~~~~~~~~~~~~~i.e.~~ -f^th+h^tf=0
\en
These two relations are equivalent to require the real 
matrix
\eq
\left(
\begin{array}{cc}
A & B\\
C & D
\end{array}\right)=\sqrt{2}
\left(\begin{array}{cc}
{\rm Re} f &\,-{\rm Im} f  \\
{\rm Re} h &\,-{\rm Im} h
\end{array}\right)
\en
to be in the fundamental representation of $Sp(2n,\rr)$.
Vice versa any symplectic matrix $\big({}^A_C{}^B_D\big)$
leads to relations \eqn{compsymp1}, \eqn{compsymp2} by 
defining
\eq
\left(\begin{array}{c}
f  \\
h
\end{array}\right)=
{1\over \sqrt{2}} \left(
\begin{array}{c}
A-iB\\
C-iD
\end{array}\right)~.
\label{deffh1} 
\en
In terms of the $f$ and $h$ matrices
we have
\eq
U={\cal A}^{-1} 
\left(
\begin{array}{cc}
A & B\\
C&  D
\end{array}\right) {\cal A}={1\over\sqrt{2}}
\left(
\begin{array}{cc}
f+ih  &\,\, \bar f+i\bar h\\[.2em]
f-ih&\,\, \bar f-i \bar h
\end{array}\right)~.
\label{USfh}
\en

\sk
\noi {\bf The $V$ matrix and its symplectic vectors}

\noi The matrix 

\eq
\VV=\left(
\begin{array}{cc}
A & B\\
C & D
\end{array}\right){\cal A}=\left(\begin{array}{cc}
f  &\bar f\\
h  &\bar h
\end{array}\right)_{}
\en
transforms from the left via the fundamental representation 
of $Sp(2n,\rr)$ and from the right via the complex representation
of $Sp(2n,\rr)$. Since $\A$ is a symplectic matrix we have
that $V$ is a symplectic matrix, 
$
V^t \big({}_{1\!\!1}^0 {}_{~0}^{-1\!\!1}\big) V=
 \big({}_{1\!\!1}^0 {}_{~0}^{-1\!\!1}\big) \,
$, hence also its transpose $V^t$,
$
V \big({}_{1\!\!1}^0 {}_{~0}^{-1\!\!1}\big) V^t=
 \big({}_{1\!\!1}^0 {}_{~0}^{-1\!\!1}\big) ~.
$
The columns of the $V$ matrix are 
therefore mutually symplectic vectors; also the rows 
are mutually symplectic vectors. Explicitly if $V^\xi$ is the vector with components given by the $\xi$-th row of $V$, then
$V^\xi_{~\rho}{\sf\Omega}^{\rho\sigma}
V^\zeta_{~\sigma}={\sf\Omega}^{\xi\zeta}$, 
where ${\sf\Omega}=
\big({}^0_{1\!\!1} {}^{-1\!\!1}_{~0}\big)$.
\sk
\subsection{The coset space $Sp(2n,\rr)/U(n)$ ~($\M$ and $\N$ matrices) }
All positive definite symmetric and 
symplectic matrices ${\cal S}$ are of the form 
\eq
{\cal S}=gg^t~~~~,~~~g\in Sp(2n,\rr)~.
\label{factposdef}
\en
Indeed consider the factorization \eqn{factorization} 
(since ${\cal S}$ is positive definite also its restriction to an
$n$ dimensional subspace is positive definite, therefore $D$
is invertible). The factorization \eqn{factposdef} is 
obtained for example by considering the symplectic matrix
\eq
g=
\left(
\begin{array}{cc}
1\!\!1 & ~BD^{-1}\\
0 & 1\!\!1
\end{array}\right)\left(
\begin{array}{cc}
\sqrt{D^{-1}} & 0\\
0 & \sqrt{D}
\end{array}\right)
~,
\en
where the matrix $\sqrt{D}$ is the unique positive definite 
square root of the symmetric and positive definite matrix $D$.
(Notice that the same proof shows that any 
symmetric and symplectic matrix $\big( {}^A_{B^t}{}^B_D\big)$
with invertible and positive definite matrix $D$ is 
of the form $gg^t$ and therefore is positive definite).

\sk

We can now show that the coset space $Sp(2n,\rr)/U(n)$ is the 
space of all positive definite symmetric and symplectic 
matrices. The maximal compact subgroup 
of $Sp(2n,\rr)$ is $H:=\{g\in Sp(2n,\rr); gg^t=1\!\!1\}$,
and we have seen in \eqn{UinUspSp} that it is $U(n)$.
\sk
We then denote by $gH$ the elements of 
$Sp(2n,\rr)/U(n)$, where $H=U(n)$, and
consider the map 
\eqa
\sigma \,:\, {Sp(2n,\rr)\over U(n)}&~\rightarrow~ &\{{\cal S}\in Sp(2n,\rr); {\cal S}={\cal S}^t {\mbox{ and }} {\cal S}{\mbox{ positive definite}}\}\nn\\
gH &~\mapsto~& gg^t
\ena
This map is well defined  because it does not depend on the representative $g\in Sp(2n,\rr)$ of the equivalence class
$gH$. Formula \eqn{factposdef} shows that this map is surjective. Injectivity is also easily proven:
if $gg^{t}=g'g'^t$ then  
$g'^{-1}g(g'^{-1}g)^{t}=1$, so that $u=g'^{-1}g$ 
is an element of $Sp(2n,\rr)$ that satisfies $uu^{t}=1$. Therefore $u=g'^{-1}g$ belongs to the maximal compact subgroup $H=U(n)$, hence $g$ and $g'$ belong to the same coset. 
\sk\sk
\noi {\bf The $\M$ and $\N$ matrices}

\noi Notice that the $n\times n$ matrices 
$f=(f^\La_{~a})_{\La,a=1,...n}$,  
are invertible. Indeed if the columns 
of $f$ were linearly dependent, say  $f^\La_{~a\,}\psi^a=0$, i.e.
$f\psi=0$, with a nonzero vector $\psi$, then  sandwiching
\eqn{compsymp1} between $\psi^\dagger$ and $\psi$ we would obtain 
\eq
 -(f\psi)^\dagger h\,\psi+\psi^\dagger h^\dagger f\psi=i \psi^\dagger \psi\not=0
\label{fhpsi}
\en
that is  absurd. Similarly also the matrix 
$h=(h_{\La a})$ is invertible.
We can then define the invertible $n\times n$ matrix 
\eq
{\cal N}= hf^{-1}
\label{Nhf}
\en
that is symmetric (cf. \eqn{compsymp2}) and that has negative definite imaginary part  (cf. \eqn{compsymp1}) 
\eq
\N=\N^t~~,~~~~{\rm Im } \,{\cal N}=-{i\over 2}({\cal N}-{\cal N^\dagger})=-{1\over 2}(f{f}^\dagger)^{-1}~,
\label{ImNff}
\en
(while $\N^{-1}$ has positive definite imaginary part $\N^{-1}-\N^{-\dagger}=i(hh^\dagger)^{-1\,}$). Any symmetric matrix with negative definite imaginary part 
is of the form \eqn{Nhf} for some $(f,h)$ satisfying 
\eqn{compsymp1} and \eqn{compsymp2} (just consider any $f$ that satisfyes \eqn{ImNff}).
There is also a 1-1- correspondence between symmetric 
complex matrices $\N$ with negative definite
imaginary part and symmetric negative definite matrices 
$\M$ of $Sp(2n,\rr)$. Given $\N$ 
we consider
\eqa
\M(\N)&=&
\left(\begin{array}{cc}
1\!\!1 & -\rea \N\\
0 & 1\!\!1
\end{array}\right)
\left(\begin{array}{cc}
\im \N & 0  \\
0 & \im \N^{-1}
\end{array}\right)
\left(\begin{array}{cc}
1\!\!1 & 0\\
-\rea \N & 1\!\!1
\end{array}\right)
\nn\\
&=&
\left(\begin{array}{cc}
\im \N +\rea \N \,\im \N^{-1}\, \rea \N &~ -\rea \N \,\im \N^{-1}  \\
-\im \N^{-1}\,\rea \N &~ \im \N^{-1}
\end{array}\right)\nn\\[.2em]
&=&
-i\left(\begin{array}{cc}
0 & -1\!\!1\\
1\!\!1& 0
\end{array}\right)
+\left(\begin{array}{cc}
\N \,\im \N^{-1}\, \N^\dagger &~ -\N \,\im \N^{-1}  \\
-\im \N^{-1}\,\N^\dagger &~ \im \N^{-1}
\end{array}\right)
\nn\\[.2em]
&=&
-i\left(\begin{array}{cc}
0 & -1\!\!1\\
1\!\!1& 0
\end{array}\right)
-2\left(\begin{array}{cc}
hh^\dagger  & -hf^\dagger  \\
-fh^\dagger & ff^\dagger
\end{array}\right)\nn\\[.2em]
&=&
-i\left(\begin{array}{cc}
0 & -1\!\!1\\
1\!\!1& 0
\end{array}\right)
-2\left(\!\begin{array}{cc}
-h\\
f
\end{array}\!\right)
(-h^\dagger~f^\dagger) 
\nn\\[.2em]
&=&
-2\, \rea\!\left[\left(\!\begin{array}{cc}
-h\\
f
\end{array}\!\right)
(-h^\dagger~f^\dagger) \right]\label{Mlast}
\ena
Since symmetric negative  definite matrices 
$\M$ of $Sp(2n,\rr)$ parametrize
the coset space  $Sp(2n,\rr)/U(n)$,
the matrices $\N$ too parametrize this coset space.
\sk
Under symplectic rotations \eqn{uglt} we have 
\eq
\left(\begin{array}{c}
f  \\
h
\end{array}\right)
\rightarrow 
\left(\begin{array}{c}
f  \\
h
\end{array}\right)'
=S
\left(\begin{array}{c}
f  \\
h
\end{array}\right)=
\left(
\begin{array}{cc}
A & B\\
C & D
\end{array}\right)\left(\begin{array}{c}
f  \\
h
\end{array}\right)
\en
and
\eq
{\cal N}\rightarrow {\cal N}'=(C+D{\cal N})(A+B{\cal N})^{-1}~.
\en
The transformation of the imaginary part of $\N$ is
(recall \eqn{ImNff})
\eq
\im \N\rightarrow \im \N' = {(A+B\N)^{-\dagger}}\im \N (A+B\N)^{-1}
\en
The transformation of the corresponding matrix $\M(\N)$
is
\eq
\M(\N)\rightarrow \M(\N ')= {S^t}^{-1}\, \M(\N)\, S^{-1}
\label{Mtransforms}~,
\en
this last relation easily follows from \eqn{Mlast} and 
from 
$\big( {}^{-h}_{\,f}\big)
=\big( {}^{0}_{1\!\!1} {}^{-1\!\!1}_{~0}\big)
\big( {}^{f}_{h}\big)$.

The relation between the negative definite symmetric matrix 
$\M$ defined in \eqn{Mlast} and ${\cal S}$ defined in \eqn{factposdef}
can be obtained from their transformation 
properties under $Sp(2n,\rr)$, 
\eq
\M=-{\cal{S}}^{-1}=\left(
\begin{array}{cc}
0 &-1\!\!1 \\
1\!\!1 & 0
\end{array}\right) 
{\cal S}
\left(
\begin{array}{cc}
0 & -1\!\!1 \\
1\!\!1 & 0
\end{array}\right) ~.\label{MversusS}
\en
We also have
$
{\cal M}= -V^{-\dagger} V^{-1}~.
$
\sk
\subsection{Lie algebra of $Sp(2n,\rr)$ and $U(n)$ ~($a,b,c,d$ matrices)}
\noi If we write $\big({}^A_B {}^C_D\big)=\big({}^{1\!\!1}_0 {}^0_{1\!\!1}\big)+\epsilon\big({}^a_c {}^b_d\big)$ with $\epsilon$ infinitesimal we obtain that
the $2n\times 2n$ matrix
\eq
\left(
\begin{array}{cc}
a & b \\
c & d
\end{array}\right)
\en
belongs to the Lie algebra of $Sp(2n,\rr)$ if $a,b,c,d$ are real $n\times n$ matrices that satisfy the relations
\eq
a^t=-d~,~~b^t=b~,~~c^t=c~.
\en
The Lie algebra of $U(n)$ in this fundamental representation of $Sp(2n,\rr)$ 
is given by the matrices 
$$
\left(
\begin{array}{cc}
a & b \\
-b & \,a
\end{array}\right)
$$ with $b=b^t$, $a=-a^t$.

In the complex basis \eqn{defAeAt} the Lie algebra of $Sp(2n,\rr)$ is given by 
the $2n\times 2n$ matrices 
\eq\label{lieusp}
\left(
\begin{array}{cc}
{\mathsf{a}} & \,{\mathsf{b}} \\
\bar {\mathsf{b}} & \,\bar {\mathsf{a}}
\end{array}\right) 
\en
where ${\mathsf{a}}$ and ${\mathsf{b}}$ are complex $n\times n$ matrices that 
satisfy the relations
\eq\label{LieUsp}
{\mathsf{a}}^\dagger=-{\mathsf{a}}~,~~{\mathsf{b}}^t={\mathsf{b}}~.
\en
The Lie algebra of $U(n)$ in this complex basis is given by the matrices 
$\left(
\begin{array}{cc}
{\mathsf{a}} & 0 \\
0 & \bar {\mathsf{a}}
\end{array}\right) 
$ with ${\mathsf{a}}^\dagger=-{\mathsf{a}}$. 


\section{Appendix: Unilateral Matrix Equations}
The remarkable symmetry property of the trace of the solution 
of the matrix equation \eqn{Qeq} holds for more 
general matrix equations. This trace property and the structure of the solution itself are studied in \cite{ABMZ}, and with a different method in \cite{SchwarzA}; see also  \cite{CerchiaiZumino} for a unified approach based on the generalized 
Bezout theorem, and  \cite{ASVR} for convergence of perturbative solutions
of matrix equations and a new form of the 
noncommutative Lagrange inversion formula.
 
In this appendix we prove the symmetry property of the trace of certain solutions (and their powers) of unilateral matrix equations. These are $N^{\rm th}$ order matrix equations 
for the variable $X$ with matrix coefficients $A_i$ which are all on one
side, e.g. on the left
\beqn
X=A_{0}+A_{1} X +A_{2} X ^{2}+\ldots+ A_{N} X ^{N} .
\label{eqphi}
\eeqn
The matrices are all square and of arbitrary degree. We may equally
consider the $A_i$'s as generators of an associative
algebra, and $ X $ an element of this algebra which satisfies the
above equation.
We consider the formal solution of \eqn{eqphi} obtained as the limit
of the sequence $X_0=0$, $X_{k+1}=A_{0}+A_{1} X_k +A_{2} X_k^{2}+\ldots+ A_{N} X_k^{N} .$
It is convenient to  assign to every matrix 
a dimension $d$ such that $d( X )=-1$. 
Using~\rref{eqphi}, the dimension of the matrix $A_i$ is given by
$d(A_i)=i-1$.

 First note that we can
rewrite equation~\rref{eqphi} as
\[
1-\sum_{i=0}^{N} A_i
~=~
1- X  -
\sum_{k=1}^{N} A_k (1- X ^k)
\]
The right hand side factorizes 
\[
1-\sum_{i=0}^{N} A_i
~=~
(1-\sum_{k=1}^{N} \sum_{m=0}^{k-1} A_k X ^m)(1- X )~.
\]
Under the trace we can use the fundamental property of the
logarithm, even for noncommutative objects, and  obtain
\[
{\rm Tr}\,\log(1-\sum_{i=0}^{N} A_i)
~=~
{\rm Tr}\,\log(1-\sum_{k=1}^{N} \sum_{m=0}^{k-1} A_k X ^m)+
{\rm Tr}\,\log(1- X )~.
\]
Using $d(A_k)=k-1$ and $d( X )=-1$ we have $d(A_k X ^m)=k-m-1$ and
we see that all the words in the argument of the first logarithm on the right hand
side have semi-positive dimension.  
Since all the words in the expansion of the second term have negative
dimension we obtain
\beqn
{\rm Tr}\, \log (1- X )={\rm Tr}\, \log(1-\sum_{i=0}^{N} A_i)
\Big|_{d<0}~.
\label{genr}
\eeqn
On the right hand side of~\rref{genr} one must expand the logarithm
and restrict the sum to words of negative dimension. Since
$d( X ^r)= -r$\, 
by extracting the dimension $d=-r$ terms from the right hand side 
of~\rref{genr} we obtain
\beqn
{\rm Tr}
\,\phi^r\,
~=~r
\sum_{
\stackrel{\{a_i\}}
{\sum{(i-1)a_i=-r}}
}
\frac{\left( \sum_{i=0}^{N} a_i - 1 \right)!}{a_0!a_1! \ldots a_N!}
~{\rm Tr}~
{\cal S}(A_{0}^{a_0}A_{1}^{a_1} \ldots A_{N}^{a_N})~.
\label{trphir}
\eeqn  
The relevant  point is that all the terms in the expansion of 
${\rm Tr}\, \log(1-\sum_{i=0}^{N} A_i)$
are automatically symmetrized, this explains the symmetrization operator
${\cal S}$ in the $A_0,A_1,...A_N$ matrix coefficients. 
\sk

If the coefficient $A_N$ is unity, we have 
the following identity  for the symmetrization operators of $N+1$ and of $N$
coefficients (words)
\[
{\cal S}(A_{0}^{a_0}A_{1}^{a_1} \ldots A_{N}^{a_N})|_{A_N = 1}
=
{\cal S}(A_{0}^{a_0}A_{1}^{a_1} \ldots A_{N-1}^{a_{N-1}})~.
\]
This is obviously true up to normalization; the normalization  
can be checked in the commutative case.

The trace of the solution of~\rref{Qeq} can now be obtained
from~\rref{trphir} by considering $r=1$ and  $N=2$ and by setting $A_2$ to unity.

\end{document}